\documentclass[prd,showpacs,amsmath,amssymb,twocolumn]{revtex4}

\usepackage{graphicx}% Include figure files
\usepackage{dcolumn}% Align table columns on decimal point
\usepackage{bm}% bold math
\usepackage{psfrag}
\usepackage{subfigure}

\newcommand{\be}{\begin{eqnarray}}
\newcommand{\ee}{\end{eqnarray}}
\newcommand{\bn}{\begin{eqnarray*}}
\newcommand{\en}{\end{eqnarray*}}

\newcommand{\nn}{\nonumber \\}
\newcommand{\nl}{\\}

\renewcommand{\vec}[1]{\mbox{\boldmath$#1$}}
\renewcommand{\d}{\mbox{\rm d}}

\renewcommand{\th}{\ensuremath{\theta}}

\newcommand{\ph}{\ensuremath{\phi}}

\newcommand{\al}{\ensuremath{\alpha}}
\newcommand{\bt}{\ensuremath{\beta}}
\newcommand{\sg}{\ensuremath{\sigma}}
\newcommand{\gm}{\ensuremath{\gamma}}
\newcommand{\dl}{\ensuremath{\delta}}
\newcommand{\lm}{\ensuremath{\lambda}}
\newcommand{\Lm}{\ensuremath{\Lambda}}

\newcommand{\Dl}{\ensuremath{\Delta}}

\newcommand{\Gm}{\ensuremath{\Gamma}}

\newcommand{\OmK}{\ensuremath{\Omega_{\rm K}}}

\newcommand{\Cmu}{\ensuremath{\hat{\mu}}}
\newcommand{\Cnu}{\ensuremath{\hat{\nu}}}
\newcommand{\Cal}{\ensuremath{\hat{\al}}}
\newcommand{\Cbt}{\ensuremath{\hat{\bt}}}

\newcommand{\ze}{\ensuremath{\hat{0}}}
\newcommand{\on}{\ensuremath{\hat{1}}}
\newcommand{\tw}{\ensuremath{\hat{2}}}
\newcommand{\tr}{\ensuremath{\hat{3}}}

\newcommand{\lt}{\ensuremath{\left}}
\newcommand{\rt}{\ensuremath{\right}}

\renewcommand{\d}{\mbox{\rm d}}

\begin{document}

\pagenumbering{arabic}

\title{Perturbation Method for Classical Spinning Particle Motion:  II. Vaidya Space-Time}

\author{Dinesh Singh}
\email{dinesh.singh@uregina.ca}
\affiliation{%
Department of Physics, University of Regina \\
Regina, Saskatchewan, S4S 0A2, Canada
}%
\date{\today}

\begin{abstract}
This paper describes an application of the Mathisson-Papapetrou-Dixon (MPD) equations in analytic perturbation form
to the case of circular motion around a radially accreting or radiating black hole described by the Vaidya metric.
Based on the formalism presented earlier, this paper explores the effects of mass accretion or loss
of the central body on the overall dynamics of the orbiting spinning particle.
This includes changes to its squared mass and spin magnitude due to the classical analog of radiative corrections
from spin-curvature coupling.
Various quantitative consequences are explored when considering orbital motion near the black hole's event horizon.
An analysis on the orbital stability properties due to spin-curvature interactions is examined briefly, with conclusions
in general agreement with previous work performed for the case of circular motion around a Kerr black hole.
\end{abstract}


\maketitle

\section{Introduction}
\label{sec:1}

The Mathisson-Papapetrou-Dixon (MPD) equations \cite{Mathisson,Papapetrou,Dixon1,Dixon2} represent a well-known description of
classical spinning particle motion in the presence of a curved space-time background.
They comprise the ``pole-dipole approximation'' for the dynamics of extended bodies with spin angular momentum
in the vicinity of black holes, neutron stars, or other sources of space-time curvature where a strong
gravitational field is generated.
This includes sources which themselves are time-varying for the duration of a spinning particle's motion along
its worldline.
As a consequence, the spin-curvature coupling term in the MPD equations, which generates an external force and torque to act on the spinning
particle, also becomes time-varying, leading to potentially very interesting dynamical effects experienced by the particle.

One particularly interesting space-time background with an explicit time dependence is known as the Vaidya metric,
which describes space-time curvature due to a spherically symmetric compact source that either radially accretes
surrounding radiation, or radiates away its central mass.
Given that most astrophysical sources have at least some orbital or spin angular momentum during their formation,
it is unlikely to find candidate sources in the night sky that carry the properties exactly described by the Vaidya metric.
However, because of its relative simplicity compared to the Kerr metric to describe rotating black holes, while also
having a time-dependent central mass, the Vaidya metric nonetheless provides an ideal testing ground for understanding
subtle properties of the MPD equations for an orbiting spinning particle that is sensitive to a time-varying
gravitational field.
A recent paper \cite{Singh1} presents an extensive numerical investigation of the MPD equations in a Vaidya
background, modelling a point dipole in circular orbit around a much heavier non-rotating black hole
described by a monotonically increasing central mass function in terms of known functions.
This paper also shows, using only the quadrupole moment formula, that the dynamical background due to a growing central mass
can influence the shape and frequency of gravitational waveforms generated by the spinning particle for a sufficiently large
mass accretion rate, with potentially useful implications for low-frequency gravitational wave astronomy via the
space-based LISA observatory \cite{LISA}.

Although a numerical treatment of the MPD equations in a Vaidya background is undoubtedly a useful exercise,
an analytical exploration of the same problem is definitely beneficial in many respects.
For example, knowing the explicit time-dependence of the mass function within an analytical expression of the
MPD equations allows for the study of conditions where instabilities in the dynamical system most likely will occur.
It can also potentially give useful insight for knowing when a mass increase or loss will lead to macroscopic changes in
the particle's orbit for a predetermined mass accretion or loss rate.
Furthermore, the general results obtained from such a study can provide clues for how a spinning particle may respond due to
a more realistic time-dependent source than one described by the Vaidya metric, such as a pulsating star,
particularly on determining the most dominant contribution to its response.

A recent development on the study of the MPD equations involves a linear perturbative approach first introduced by
Chicone, Mashhoon, and Punsly (CMP) \cite{Chicone}, with an application by Mashhoon and Singh \cite{Mashhoon2}
for determining a first-order perturbation of a circular orbit around a Kerr black hole due to spin-curvature coupling.
This first approach was more recently generalized by Singh \cite{Singh0} to accommodate for higher-order contributions
in powers of $s/(m \, r) \ll 1$, where $\rho = s/m$ is the M{\o}ller radius \cite{Mashhoon2,Moller} in terms of the
particle's spin magnitude $s$ and mass $m$, and $r$ is the radial distance from the background mass source to the particle's location.
This generalization can be applied to formally infinite order in the perturbation expansion parameter and makes no reference to
any particular space-time metric or symmetries therein.
A detailed application of the generalized CMP approximation to the MPD equations was just presented
for the case of circular motion around a Kerr black hole.
For future reference, this recent paper is now identified as ``Paper~I'' \cite{Singh2}.
It would be very interesting to perform the same investigation as found in Paper~I, but this time applied to the Vaidya metric.
%For the purpose of comparison with Paper~I, the outcome of this analysis is identified as ``Paper~II.''

The purpose of this paper is to apply the generalized CMP approximation of the MPD equations to describe circular motion around
a static compact object described by the Vaidya metric, and incorporate both mass accretion from null radiation and
outgoing radiation within the formalism.
This paper begins with a brief review of the MPD equations and the generalized CMP approximation \cite{Mashhoon2,Singh0,Singh2},
found in Sec.~\ref{sec:2}.
An introduction to the Vaidya metric \cite{Singh1} and its application to the generalized CMP approximation is then presented in Sec.~\ref{sec:3}.
Following this, Sec.~\ref{sec:4} describes the main results for the case of circular motion around the central body
to second order in the perturbation expansion parameter, including
the ``radiative corrections'' of the squared mass and spin magnitudes predicted within the underlying formalism \cite{Singh0,Singh2}.
Afterwards, a discussion of the obtained results is given in Sec.~\ref{sec:5}, followed by a brief conclusion.
Consistent with Paper~I, the Riemann and Ricci tensors follow the conventions of MTW \cite{MTW} with signature $+2$,
and assuming geometric units of $G = c = 1$.

\section{Mathisson-Papapetrou-Dixon (MPD) Equations and the Generalized CMP Approximation}
\label{sec:2}

\subsection{MPD Equations}
\label{sec:2.1}

Given the dynamical degrees of freedom $P^\mu(\tau)$ and $S^{\al \bt}(\tau)$ for the spinning particle's
linear four-momentum and spin tensor, respectively, the MPD equations are
\begin{subequations}
\label{MPD-equations}
\be
{DP^\mu \over \d \tau} & = & - {1 \over 2} \, R^\mu{}_{\nu \al \bt} \, u^\nu \, S^{\al \bt} \, ,
\label{MPD-momentum}
\nl
\nn
{DS^{\al \bt} \over \d \tau} & = & P^\al \, u^\bt - P^\bt \, u^\al \, ,
\label{MPD-spin}
\ee
\end{subequations}
where $R_{\mu \nu \al \bt}$ is the Riemann curvature tensor and
$u^\mu(\tau) = \d x^\mu(\tau)/\d \tau$ is the four-velocity with affine parametrization $\tau$.
While $\tau$ can be chosen to satisfy $u^\mu \, u_\mu = -1$ to describe proper time, it is not
necessary to impose this particular constraint if desired.
The combined force equation (\ref{MPD-momentum}) and torque equation (\ref{MPD-spin})
infer that the particle's four-momentum precesses around the centre-of-mass worldline,
giving rise to non-trivial motion away from time-like geodesic motion.

The MPD equations presented in (\ref{MPD-equations}) are underdetermined, and require
supplementary equations to specify the system.
A commonly accepted constraint is to impose orthogonality between the particle's linear and spin
angular momenta, following Dixon's approach \cite{Dixon1,Dixon2}, such that
\be
S^{\alpha \beta} \, P_\beta & = & 0 \, .
\label{spin-condition}
\ee
As well, the mass and spin parameters $m$ and $s$ are identified by the constraint equations
\begin{subequations}
\label{m-s-magnitude}
\be
m^2 & = & -P_\mu \, P^\mu \, ,
\label{mass}
\nl
\nn
s^2 & = & {1 \over 2} \, S_{\mu \nu} \, S^{\mu \nu} \, ,
\label{spin}
\ee
\end{subequations}
which become {\em constants of the motion} \cite{Chicone} when (\ref{spin-condition}) is implemented within
the MPD equations.
It is also well-known that the four-velocity $u^\mu$ can be expressed in terms of $P^\mu$ and $S^{\al \bt}$ within
the MPD formalism \cite{Tod}, leading to
\be
u^\mu & = & -{P \cdot u \over m^2} \lt[P^\mu
+ {1 \over 2} \, {S^{\mu \nu} \, R_{\nu \gm \al \bt} \, P^\gm \, S^{\al \bt} \over
m^2 + {1 \over 4} \, R_{\al \bt \rho \sg} \, S^{\al \bt} \, S^{\rho \sg}} \rt],
\label{MPD-velocity}
\ee
where specification of $P \cdot u$ determines the parametrization constraint for $\tau$.
Clearly, (\ref{MPD-velocity}) shows that the spin-curvature coupling creates a displacement of the particle's
four-velocity away from geodesic motion.

\subsection{Generalized CMP Approximation}
\label{sec:2.2}

While a more detailed account of the generalized CMP approximation can be found in Paper~I, it is useful to
briefly summarize the main points of this approach to the MPD equations.
This is a perturbation approach based on the assumption that
\begin{subequations}
\label{P-S-approx}
\be
P^\mu(\varepsilon) & \equiv & \sum_{j = 0}^\infty \varepsilon^j \, P_{(j)}^\mu \, , % \ = \ P_{(0)}^\mu + \varepsilon \, P_{(1)}^\mu + \varepsilon^2 \, P_{(2)}^\mu + \cdots \, ,
\label{P-approx-def}
\nl
\nn
S^{\mu \nu} (\varepsilon) & \equiv & \varepsilon \sum_{j = 0}^\infty \varepsilon^j \, S_{(j)}^{\mu \nu}
\ = \ \sum_{j = 1}^\infty \varepsilon^j \, S_{(j-1)}^{\mu \nu} \, ,
\label{S-approx-def}
\ee
\end{subequations}
where $P_{(j)}^\mu$ and $S_{(j-1)}^{\mu \nu}$ are the respective jth-order contributions of the linear momentum and
spin angular momentum in $\varepsilon$, an expansion parameter associated with $s$.
In addition, the four-velocity is described as
\be
u^\mu(\varepsilon) & \equiv & \sum_{j = 0}^\infty \varepsilon^j \, u_{(j)}^\mu \, .
\label{MPD-velocity-e0}
\ee
The zeroth-order expressions in $\varepsilon$ then correspond to a spinless particle in geodesic motion,
while higher-order contributions are identified with spin-curvature coupling.

The main idea to the generalized CMP approximation is to substitute (\ref{P-S-approx}) and (\ref{MPD-velocity-e0})
into both the MPD equations (\ref{MPD-equations}) and the exact expression for $u^\mu$ according to (\ref{MPD-velocity}),
expand these equations with respect to $\varepsilon$, and solve for each order of the perturbation expansion iteratively.
It follows that the jth-order expressions of the MPD equations are
\begin{subequations}
\label{MPD-j}
\be
{DP_{(j)}^\mu \over \d \tau} & = & - \frac{1}{2} \, R^\mu{}_{\nu \alpha \beta} \sum_{k = 0}^{j - 1} u_{(j-1-k)}^\nu \, S_{(k)}^{\alpha \beta} \, ,
\label{DP-j}
\nl
\nn
{DS_{(j-1)}^{\alpha \beta} \over \d \tau} & = &  2 \sum_{k = 0}^{j-1} P_{(j-1-k)}^{[\alpha} \, u_{(k)}^{\beta]} \, ,
\label{DS-j}
\ee
\end{subequations}
where $j = 0$ implies that
\be
{D P_{(0)}^\mu \over \d \tau} & = & 0 \, ,
\label{DP-0}
\ee
while $j = 1$, corresponding to the CMP approximation \cite{Chicone,Singh0,Singh2}, is
\begin{subequations}
\label{MPD-1}
\be
{D P_{(1)}^\mu \over \d \tau} & = & -{1 \over 2} \, R^\mu{}_{\nu \al \bt} \, u_{(0)}^\nu \, S_{(0)}^{\al \bt} \, ,
\label{DP-1}
\nl
\nn
{D S_{(0)}^{\al \bt} \over \d \tau} & = & 0 \, .
\label{DS-0}
\ee
\end{subequations}
In addition to (\ref{MPD-equations}), the supplementary spin condition equation (\ref{spin-condition})
and constraint equations (\ref{m-s-magnitude}) for the squared mass and spin magnitudes need to be incorporated
within this formalism.
For the spin condition, it is straightforward to show that
\be
P_\mu^{(0)} \, S_{(j)}^{\mu \nu} & = & - \sum_{k=1}^j P_\mu^{(k)} \, S_{(j-k)}^{\mu \nu} \, , \qquad j \geq 1
\label{s.p=0}
\ee
for the (j+1)th-order contribution, where
\be
P_\mu^{(0)} \, S_{(0)}^{\mu \nu} & = & 0 \,
\label{s.p=0-e1}
\ee
for the first-order perturbation in $\varepsilon$.
As for the squared mass and spin magnitude constraint equations, it is possible to identify a {\em bare mass} $m_0$ and {\em bare spin}
$s_0$ according to
\begin{subequations}
\label{m0-s0-sq}
\be
m_0^2 & \equiv & -P^{(0)}_\mu \, P_{(0)}^\mu \, ,
\label{m0-sq}
\nl
s_0^2 & \equiv & {1 \over 2} \, S_{\mu \nu}^{(0)} \, S_{(0)}^{\mu \nu} \, ,
\label{s0-sq}
\ee
\end{subequations}
such that
\begin{subequations}
\label{m-s-total-sq}
\be
m^2 (\varepsilon) & = & m_0^2 \lt(1 + \sum_{j=1}^\infty \varepsilon^j \, \bar{m}_j^2 \rt),
\label{m-sq}
\nl
s^2 (\varepsilon) & = & \varepsilon^2 \, s_0^2 \lt(1 + \sum_{j=1}^\infty \varepsilon^j \, \bar{s}_j^2 \rt),
\label{s-sq}
\ee
\end{subequations}
where
\begin{subequations}
\label{m-s-shift-sq}
\be
\bar{m}_j^2 & = & - {1 \over m_0^2} \, \sum_{k=0}^j P_\mu^{(j-k)} \, P_{(k)}^\mu \, ,
\label{m-bar-sq}
\nl
\bar{s}_j^2 & = & {1 \over s_0^2} \, \sum_{k=0}^j S_{\mu \nu}^{(j-k)} \, S_{(k)}^{\mu \nu} \, ,
\label{s-bar-sq}
\ee
\end{subequations}
are dimensionless jth-order ``radiative corrections'' to $m_0^2$ and $s_0^2$, respectively, due to spin-curvature coupling.
Each expression of (\ref{m-s-shift-sq}) satisfies
\be
{D \bar{s}_j^2 \over \d \tau} & = & {D \bar{m}_j^2 \over \d \tau} \ = \ 0 \, .
\label{Ds/dt-Dm/dt=0}
\ee

Solving for the four-velocity (\ref{MPD-velocity-e0}) requires specifying the parametrization constraint within (\ref{MPD-velocity}).
Following Paper~I, the particularly useful choice of
\be
P \cdot u & \equiv & -m(\varepsilon),
\label{P.u}
\ee
leads to
\begin{widetext}
\be
u^\mu(\varepsilon) & = & \sum_{j = 0}^\infty \varepsilon^j \, u_{(j)}^\mu
\ = \ {P_{(0)}^\mu \over m_0} + \varepsilon \lt[{1 \over m_0} \lt(P_{(1)}^\mu - {1 \over 2} \, \bar{m}_1^2 \, P_{(0)}^\mu\rt) \rt]
\nn
\nn
\nn
&   & {} + \varepsilon^2 \lt\{ {1 \over m_0} \lt[P_{(2)}^\mu - {1 \over 2} \, \bar{m}_1^2 \, P_{(1)}^\mu
         - {1 \over 2} \lt(\bar{m}_2^2 - {3 \over 4} \, \bar{m}_1^4\rt) P_{(0)}^\mu \rt]
%\rt.
%\nn
%\nn
%\nn
%&   & {} + \lt.
+ {1 \over 2 m_0^3} \, S_{(0)}^{\mu \nu} \, R_{\nu \gm \al \bt}  \, P_{(0)}^\gm \, S_{(0)}^{\al \bt} \rt\}
\nn
\nn
\nn
&   & {} + \varepsilon^3 \lt\{ {1 \over m_0} \lt[P_{(3)}^\mu - {1 \over 2} \, \bar{m}_1^2 \, P_{(2)}^\mu
         - {1 \over 2} \lt(\bar{m}_2^2 - {3 \over 4} \, \bar{m}_1^4\rt) P_{(1)}^\mu \rt.
%\rt.
%\nn
%\nn
%\nn
%&   & {} - \lt.
- {1 \over 2} \lt(\bar{m}_3^2 - {3 \over 2} \, \bar{m}_1^2 \, \bar{m}_2^2 + {5 \over 8} \, \bar{m}_1^6 \rt) P_{(0)}^\mu \rt]
\nn
\nn
&   & {} + \lt.
{1 \over 2 m_0^3} \, R_{\nu \gm \al \bt} \lt[\sum_{n=0}^1 S_{(1-n)}^{\mu \nu} \sum_{k=0}^n P_{(n-k)}^\gm \, S_{(k)}^{\al \bt}
- {3 \over 2} \, \bar{m}_1^2 \, S_{(0)}^{\mu \nu} \, P_{(0)}^\gm \, S_{(0)}^{\al \bt} \rt] \rt\}
%\nn
%\nn
%\nn
%&   & {}
+ O(\varepsilon^4) \, ,
\label{MPD-velocity-explicit}
\ee
\end{widetext}
where
\be
u_\mu(\varepsilon) \, u^\mu(\varepsilon) & = & -1 + O(\varepsilon^4) \, ,
\label{u.u}
\ee
implying that $u^\mu$ is indeed the four-velocity with unit normal to third-order in $\varepsilon$.
It is not necessarily true, however, that $u \cdot u = -1$ applied to all orders of $\varepsilon$
corresponds to (\ref{P.u}).
Extending (\ref{u.u}) to fourth-order in $\varepsilon$ and higher requires a more general approach,
where
\be
P \cdot u & \equiv & \sum_{j=0}^\infty \lt(P \cdot u\rt)_{(j)} \, \varepsilon^j
\label{P.u-general}
\ee
and the choice for each $\lt(P \cdot u\rt)_{(j)}$ is determined from constraint equations
for each order of $\varepsilon$ as required.

\subsection{Summary of the Linear Momentum and Spin Angular Momentum Expansion Components}
\label{sec:2.3}

The approach adopted to solve for the linear momentum and spin tensor expansion components in the generalized
CMP approximation is to use the tetrad formalism and work in Fermi normal co-ordinates.
Full details for obtaining these expressions are shown in Paper~I, but it is worthwhile to give a brief outline
of the procedure.
Suppose that an orthonormal tetrad frame $\lm^\mu{}_{\hat{\al}}$ satisfying
\be
\eta_{\hat{\al} \hat{\bt}} & = & g_{\mu \nu} \, \lm^\mu{}_{\hat{\al}} \, \lm^\nu{}_{\hat{\bt}} \,
\label{ortho-tetrad}
\ee
and parallel transport $({D \lm^\mu{}_{\hat{\al}} / \d \tau} = 0)$
describes a projection of space-time curvature described by general space-time co-ordinates $\mu$ onto a locally flat tangent space,
denoted by $\hat{\al}$.
The Fermi co-ordinates are described by $X^{\hat{\al}}$ in the local neighbourhood about the spinning particle's
centre-of-mass worldline, while general space-time co-ordinates are denoted by $X^\mu$.
As usual, $\lm^\mu{}_{\hat{0}} = u_{(0)}^\mu$.
Furthermore, the Riemann curvature tensor in the Fermi frame is then given by
\be
{}^F{}R_{\hat{\al} \hat{\bt} \hat{\gm} \hat{\dl}} & = & R_{\mu \nu \rho \sg} \,
\lm^\mu{}_{\hat{\al}} \, \lm^\nu{}_{\hat{\bt}} \, \lm^\rho{}_{\hat{\gm}} \, \lm^\sg{}_{\hat{\dl}} \, ,
\label{F-Riemann}
\ee
and that for $j \geq 0$,
\begin{subequations}
\label{Pj-Sj=}
\be
P_{(j)}^\mu & = & \lm^\mu{}_{\hat{\al}} \, P_{(j)}^{\hat{\al}} \, ,
\label{Pj=}
\nl
S_{(j)}^{\mu \nu} & = & \lm^\mu{}_{\hat{\al}} \, \lm^\nu{}_{\hat{\bt}} \, S_{(j)}^{\hat{\al} \hat{\bt}} \, .
\label{Sj=}
\ee
\end{subequations}
Following the approach taken in Paper~I and elsewhere \cite{Singh0,Singh2}, it is shown from (\ref{Pj-Sj=}) for $j = 0$ that
\begin{subequations}
\label{P0-S0=}
\be
P_{(0)}^\mu & = & \lm^\mu{}_{\hat{\al}} \, P_{(0)}^{\hat{\al}} \ = \ m_0 \, \lm^\mu{}_{\hat{0}} \, ,
\label{P0=}
\nl
S_{(0)}^{\mu \nu} & = & \lm^\mu{}_{\hat{\imath}} \, \lm^\nu{}_{\hat{\jmath}} \, S_{(0)}^{\hat{\imath} \hat{\jmath}} \, ,
\label{S0=}
\ee
\end{subequations}
satisfying the first-order spin condition (\ref{s.p=0-e1}), where $P_{(0)}^{\hat{\al}} =  m_0 \, \dl^{\hat{\al}}{}_{\hat{0}}$
and $S_{(0)}^{\hat{\imath} \hat{\jmath}}$ is a constant-valued spatial antisymmetric tensor determined from initial conditions.

For $j = 1$, the linear momentum is straightforwardly determined to be
\be
P_{(1)}^\mu & = & -{1 \over 2} \, \lm^\mu{}_{\hat{k}}
\int \lt({}^F{}R^{\hat{k}}{}_{\hat{0} \hat{\imath} \hat{\jmath}} \, S_{(0)}^{\hat{\imath} \hat{\jmath}}\rt) \d \tau \, ,
\label{P1-general}
\ee
while
\be
S_{(1)}^{\mu \nu} & = & \lm^\mu{}_{\hat{\al}} \, \lm^\nu{}_{\hat{\bt}} \, S_{(1)}^{\hat{\al} \hat{\bt}} \,
\nn
& = & \lt[{1 \over 4} \, \bar{s}_1^2 \, \lm^\mu{}_{\hat{\imath}} \, \lm^\nu{}_{\hat{\jmath}}
- {2 \over m_0} \, \lm^{[\mu}{}_{\hat{0}} \, \lm^{\nu]}{}_{\hat{\jmath}} \, P^{(1)}_{\hat{\imath}} \rt] S_{(0)}^{\hat{\imath} \hat{\jmath}}
 \, ,
\label{S1-tetrad=}
\ee
subject to
\be
{D S_{(1)}^{\mu \nu} \over \d \tau} & = & 0 \, .
\label{DS1/dt=0}
\ee
Contracting (\ref{P1-general}) into $P^{(0)}_\mu$ shows that the first-order mass shift contribution is
identically
\be
\bar{m}_1^2 & = & 0 \, .
\label{m1-bar^2=0}
\ee
However, the expression for first-order spin shift is still formally {\em undetermined} based on (\ref{S1-tetrad=}) alone, and while
it is tempting to set $\bar{s}_1^2 = 0$ in analogy with (\ref{m1-bar^2=0}), this is not justified given that
$\bar{s}_1^2$ only needs to be {\em covariantly constant} according to (\ref{Ds/dt-Dm/dt=0}), and not necessarily zero.
To obtain an expression for $\bar{s}_1^2$ requires the direct solving of (\ref{DS1/dt=0}), the details of which are given in Paper~I
and are presented in Appendix~\ref{appendix:s1-bar^2} of this paper.

Solving for the $j = 2$ expressions for both $P^\mu$ and $S^{\al \bt}$ is straightforward, such that
\begin{widetext}
\be
P_{(2)}^\mu & = & -{1 \over 2} \, \lm^\mu{}_{\hat{\al}}
\int \lt({1 \over m_0} \, {}^F{}R^{\hat{\al}}{}_{\hat{\bt} \hat{k} \hat{l}} \, P_{(1)}^{\hat{\bt}} \, S_{(0)}^{\hat{k} \hat{l}}
+ {}^F{}R^{\hat{\al}}{}_{\hat{0} \hat{\gm} \hat{\bt}} \, S_{(1)}^{\hat{\gm} \hat{\bt}}\rt) \d \tau
\nn
& \approx & -{1 \over 2} \, \lm^\mu{}_{\hat{\al}}
\int \lt({1 \over m_0} \, {}^F{}R^{\hat{\al}}{}_{\hat{\bt} \hat{k} \hat{l}} \, P_{(1)}^{\hat{\bt}}
+ {1 \over 4} \, \lt\langle \bar{s}_1^2 \rt\rangle \, {}^F{}R^{\hat{\al}}{}_{\hat{0} \hat{k} \hat{l}}
- {2 \over m_0} \, {}^F{}R^{\hat{\al}}{}_{\hat{0} \hat{0} \hat{l}} \, P^{(1)}_{\hat{k}} \rt) S_{(0)}^{\hat{k} \hat{l}} \, \d \tau
\,
\label{P2-general}
\ee
\end{widetext}
for the linear momentum, where
\be
\lt\langle \bar{s}_j^2 \rt\rangle & = & {1 \over T} \, \int_0^T \bar{s}_j^2(\tau) \, \d \tau
\label{<sj^2>}
\ee
is the time-averaged jth-order correction to the squared spin magnitude.
The corresponding expression for the spin tensor is given by
\be
S_{(2)}^{\mu \nu} & = & {1 \over m_0} \, \lm^{[\mu}{}_{\hat{0}} \, \lm^{\nu]}{}_{\hat{\imath}}
\int S_{(0)}^{\hat{\imath} \hat{\jmath}} \, {}^F{}R_{\hat{\jmath} \hat{0} \hat{k} \hat{l}} \, S_{(0)}^{\hat{k} \hat{l}} \, \d \tau \, ,
\label{S2-general}
\ee
the solution to
\be
{D S_{(2)}^{\mu \nu} \over \d \tau} & = & {1 \over m_0^3} \, P_{(0)}^{[\mu} \, S_{(0)}^{\nu]\sg} \, R_{\sg \gm \al \bt} \, P_{(0)}^\gm \,
S_{(0)}^{\al \bt} \, ,
\label{DS2/dt=0}
\ee
after substituting $u_{(1)}^\mu$ from (\ref{MPD-velocity-explicit}).

%%%%%%%%%%%%%%%%%%%%%%%%%%%%%%%%%%%%%%%%%%%%%%%

\subsection{Perturbations of the M{\o}ller Radius}

As noted in Paper~I, the M{\o}ller radius $\rho = s/m$ is closely identified with the strength of spin-curvature coupling
experienced by the spinning particle.
Previous studies of chaotic dynamics in the Kerr background \cite{Suzuki1,Suzuki2,Hartl1,Hartl2} indicate the possibility
that perturbations of $\rho$ may reveal the conditions where a transition from stable to chaotic motion can appear for a spinning
particle in a general space-time background.
The perturbation expression for the M{\o}ller radius is then formally given by
\be
{s(\varepsilon) \over m(\varepsilon)}
& = & \varepsilon \, {s_0 \over m_0} \lt\{1 + \varepsilon \lt[{1 \over 2} \lt(\bar{s}_1^2 - \bar{m}_1^2\rt) \rt] \rt.
\nn
&  &{} + \varepsilon^2 \lt[{1 \over 2} \lt(\bar{s}_2^2 - \bar{m}_2^2\rt) - {1 \over 4} \, \bar{s}_1^2 \, \bar{m}_1^2
- {1 \over 8} \lt(\bar{s}_1^4 - 3 \, \bar{m}_1^4\rt)\rt]
\nn
&  &{} + \lt. O(\varepsilon^3) \rt\},
\label{s/m}
\ee
where the contributions due to $\bar{m}_1^2$ are retained for completeness' sake.
It is of particular interest to see how (\ref{s/m}) behaves for the Vaidya metric, which can then be compared
directly with the results obtained in Paper~I.

\section{Generalized CMP Approximation in Vaidya Space-Time}
\label{sec:3}

With the formalism of the generalized CMP approximation presented, it is possible to now
develop the framework for applications to motion in a Vaidya space-time background.
The most immediate challenge is to derive the orthonormal tetrad frame $\lm^\mu{}_{\hat{\al}}$
for application of the formalism just outlined.
It is very surprising to note that, while the Vaidya metric is much simpler in form compared to the Kerr metric
used in Paper~I, the relevant computations are technically much more involved, leading to much greater complexity
than first anticipated.
This is because the Vaidya metric is effectively {\em time-dependent}, since the mass function is no longer static, but either
grows or shrinks monotonically along null rays.
Ultimately, this property must be incorporated within the structure of the orthonormal tetrad.

The Vaidya metric in $\lt(\xi, r, \th, \ph\rt)$ co-ordinates is described in general form as \cite{Singh1,Vaidya,Carmeli}
\be
\d s^2 & = & -\lt(1 - {2 \, M(\xi) \over r} \rt) \d \xi^2 + 2 \, \al \, \d \xi \, \d r
\nn
&  &{} + r^2 \lt(\d \th^2 + \sin^2 \th \, \d \ph^2\rt) \, ,
\label{Vaidya=}
\ee
where $\xi$ is a generalized null co-ordinate denoting time development and
$\al$ is a dimensionless parameter chosen such that
\be
\xi & = & \nu \, , \qquad \al = 1
\label{xi=nu}
\ee
for ingoing radiation \cite{Singh1} along the advanced null co-ordinate $\nu$, while
\be
\xi & = & \mu \, , \qquad \al = -1
\label{xi=mu}
\ee
corresponding to outgoing radiation \cite{Carmeli} along the retarded null co-ordinate $\mu$.
For the Vaidya metric, the central mass function $M(\xi)$ is a monotonically increasing or decreasing function
of $\xi(\tau)$ for a given choice of $\al$ to satisfy the weak energy condition, but is otherwise an arbitrary function.
The mass function can also be defined as
\be
M(\xi) & = & M_0 + \Dl M(\xi) \, ,
\label{M=}
\ee
where $M_0$ is the static mass for a Schwarzschild black hole and $\Dl M(0) = 0$.
It will prove useful to express (\ref{Vaidya=}) in terms of $\lt(t, r, \th, \ph\rt)$ co-ordinates, where
$\xi$ is described by the tortoise co-ordinate condition \cite{Singh1}
\be
\xi & = & t + \al \lt[r + 2 M_0 \, \ln \lt({r \over 2M_0} - 1 \rt) \rt].
\label{xi}
\ee
This leads to the Vaidya metric expressed as
\be
\d s^2 & = & - \lt[\lt(1 - {2M_0 \over r}\rt) - {2 \Dl M \over r} \rt] \d t^2
\nn
&  &{} + \lt[4 \, \al \lt(1 - {2M_0 \over r}\rt)^{-1} \, {\Dl M \over r}\rt] \d t \, \d r
\nn
& &{} + \lt[\lt(1 - {2M_0 \over r}\rt)^{-1} + 2 \lt(1 - {2M_0 \over r}\rt)^{-2} \, {\Dl M \over r} \rt] \d r^2
\nn
&  &{} + r^2 \, \lt(\d \th^2 + \sin^2 \th \, \d \ph^2 \rt) \, ,
\label{Vaidya-metric}
\ee
%
%where $\al$ is $+1$ for infalling radiation and $-1$ for outgoing radiation described by the central mass $M = M_0 + \Dl M$
%defined in terms of a mass function $\Dl M \equiv \Dl M \lt(\xi(\tau)\rt)$ that monotonically increases
%$(\al = +1)$ or decreases $(\al = -1)$ along the null co-ordinate.
%In the limit as $\Dl M \rightarrow 0$, (\ref{Vaidya-metric}) reduces to the Schwarzschild metric expressed in terms of $M_0$.
%
which reduces to the Schwarzschild metric as $\Dl M (\xi)~\rightarrow~0$.

While it is mathematically acceptable to leave $\Dl M (\xi)$ unspecified, it creates computational obstacles
for an exact treatment of the problem.
Therefore, a simplifying assumption adopted is to let $\Dl M/M_0 \ll 1$, which is well-justified on physical grounds,
since the Eddington luminosity limit \cite{Heyl} imposes an upper bound mass accretion rate of
\be
{\d (\Dl M) \over \d t} & = & 3 \times 10^{-22} \, {\lt(1 - \gm\rt) \over \gm} \, \lt({M_0 \over M_\odot}\rt) \ll 1 \, ,
\label{Mdot}
\ee
where $M_\odot$ is one solar mass and $\gamma \approx 0.1$ is the energy release efficiency of the
outgoing photon flux.
This allows for a derivation of the Vaidya tetrad frame $\lm^\mu{}_{\hat{\al}}$ in terms of a linear perturbation about
$\lm^\mu{}_{\hat{\al} \, (\rm Sch)}$, the Schwarzchild tetrad frame for circular motion, which is presented below.

Consider the Schwarzschild orthonormal tetrad frame \cite{Mashhoon2} for circular motion with fixed radius $r > 2M_0$, such that
\begin{subequations}
\label{tetrad-sch}
\be
\lm^\mu{}_{\hat{0} \, (\rm Sch)} & = & \lt({E \over A^2}, 0, 0, {L \over r^2 \, \sin \th} \rt) \ = \ u^\mu_{(\rm Sch)},
\label{tetrad-0-sch}
\nl
\lm^\mu{}_{\hat{1} \, (\rm Sch)} & = & \lt(-{L \over r \, A} \, \sin \lt(\OmK \, \tau \rt), \, A \, \cos \lt(\OmK \, \tau \rt), \, 0, \, \rt.
\nn
&  &{} \lt. -{E \over r \, A \, \sin \th} \, \sin \lt(\OmK \, \tau \rt) \rt),
\label{tetrad-1-sch}
\nl
\lm^\mu{}_{\hat{2} \, (\rm Sch)} & = & \lt(0, 0, {1 \over r}, 0\rt),
\label{tetrad-2-sch}
\nl
\lm^\mu{}_{\hat{3} \, (\rm Sch)} & = & \lt({L \over r \, A} \, \cos \lt(\OmK \, \tau \rt), \, A \, \sin \lt(\OmK \, \tau \rt), \, 0, \, \rt.
\nn
&  &{} \lt. {E \over r \, A \, \sin \th} \, \cos \lt(\OmK \, \tau \rt) \rt),
\label{tetrad-3-sch}
\ee
\end{subequations}
where
\be
\OmK & = & \sqrt{M_0 \over r^3}
\label{Om-Kepler}
\ee
is the Keplerian frequency of the orbit,
\be
N & = & \sqrt{1 - {3 M_0 \over r}} \, , \qquad A \ = \ \sqrt{1 - {2M_0 \over r}} \, ,
\label{N,A=}
\ee
and the energy $E$ and orbital angular momentum $L$ for the orbit are
\be
E & = & {A^2 \over N} \, , \qquad
L \ = \ {r^2 \OmK \over N} \, .
\label{E-L}
\ee
The boundary conditions are determined such that $t = \ph = 0$ at $\tau = 0$.
Furthermore, given that the Vaidya metric is spherically symmetric, the Cartesian axis centred on
the black hole is oriented such that orbital motion is confined to the plane defined by $\th = \pi/2$
with respect to an assigned $z$-axis.

The next step is to derive the Vaidya orthonormal tetrad in the form
\be
\lm^\mu{}_{\hat{\al}} & \approx & \lm^\mu{}_{\hat{\al} \, (\rm Sch)} + \Dl \lm^\mu{}_{\hat{\al}} \, ,
\label{Vaidya-tetrad-formal}
\ee
where $\Dl \lm^\mu{}_{\hat{\al}}$ is the linear perturbation proportional to $\Dl M$.
While an exact treatment within this perturbation approach is given in Appendices~\ref{appendix:tetrad-frame} and~\ref{appendix:zeroth-component},
the outcome is considerably more complicated than for the exact orthonormal tetrad in the Kerr background \cite{Mashhoon2}.
Therefore, another simplifying assumption is introduced, in the form of a series expansion with respect
to inverse powers of $N$, since any deviations away from the Schwarzschild contribution will only be potentially identifiable
when the spinning particle approaches the nearest (photon) orbit of $r \rightarrow 3 \, M_0$, corresponding to $N \rightarrow 0$.
In addition, an expression for $\Dl M$ needs to be chosen in terms of $\tau$ that is consistent with both
the properties of the metric (\ref{Vaidya-metric}) and the mass accretion rate upper bound (\ref{Mdot}).
This leads to the choice of
\be
\Dl M(\tau) & \approx & {\al \over A} \, \lt|\d \lt(\Dl M\rt) \over \d \xi \rt| \, \tau \, ,
\label{dM-approx}
\ee
where the prefactor of $\al$ in (\ref{dM-approx}) accounts for the direction of radiation flow, and $\lt|\d \lt(\Dl M\rt)/\d \xi \rt| \ll 1$.
With these further assumptions incorporated, it can be shown that the Vaidya orthonormal tetrad frame components
for a particle in orbit near the event horizon $(N \rightarrow 0)$ at $\th = \pi/2$ are
\begin{subequations}
\label{tetrad-0-approx}
\be
\lm^0{}_{\hat{0}} & \approx & {1 \over N} + {\al \over 2 \, N^7} \, \lt|\d \lt(\Dl M\rt) \over \d \xi \rt|
\, (r \, \OmK) \, C\lt(r, \OmK \, \tau\rt) \, ,
%\lt[{r \, \OmK \over N^7} \, \sin \lt(2 \, \OmK \tau\rt) + {1 \over 2 \, N^6} \lt(1 - 2 \, r \, \OmK \rt)
%\lt[\sin \lt(2 \, \OmK \, \tau \rt) - \OmK \, \tau  \rt] \rt] \, ,
\nn
\label{tetrad-00-approx}
\nl
\lm^1{}_{\hat{0}} & \approx & - {2 \, \al \over N^3} \lt|\d \lt(\Dl M\rt) \over \d \xi \rt| \sin^2 \lt(\OmK \, \tau \rt) \, , \qquad
\label{tetrad-01-approx}
\nl
\lm^2{}_{\hat{0}} & = & 0 \, , \qquad
\label{tetrad-02-approx}
\nl
\lm^3{}_{\hat{0}} & = & \OmK \, \lm^0{}_{\hat{0}} \, ,
\label{tetrad-03-approx}
\ee
\end{subequations}
for $\lm^\mu{}_{\hat{0}}$,
\begin{subequations}
\label{tetrad-1-approx}
\be
\lm^0{}_{\hat{1}} & \approx & - r \, \OmK \lt[{1 \over N \, A}
+ {\al \over 2 \, N^7} \, \lt|\d \lt(\Dl M\rt) \over \d \xi \rt| \, C\lt(r, \OmK \, \tau\rt)\rt]
\nn
&  &{} \times \sin \lt(\OmK \, \tau \rt) \, ,
\label{tetrad-10-approx}
\nl
\lm^1{}_{\hat{1}} & \approx & A \, \cos \lt(\OmK \, \tau \rt)
+ {2 \, \al \over N^3} \lt|\d \lt(\Dl M\rt) \over \d \xi \rt| \sin^3 \lt(\OmK \, \tau \rt) \, , \qquad
\nl
\label{tetrad-11-approx}
\lm^2{}_{\hat{1}} & = & 0 \, ,
\label{tetrad-12-approx}
\nl
\lm^3{}_{\hat{1}} & = & - \lt[{A \over r \, N}
+ {\al \, \OmK \over 2 \, N^7} \, \lt|\d \lt(\Dl M\rt) \over \d \xi \rt| \, (r \, \OmK) \, C\lt(r, \OmK \, \tau\rt)\rt]
\nn
&  &{} \times \sin \lt(\OmK \, \tau \rt) \, ,
\label{tetrad-13-approx}
\ee
\end{subequations}
for $\lm^\mu{}_{\hat{1}}$,
\begin{subequations}
\label{tetrad-2-approx}
\be
\lm^0{}_{\hat{2}} & = & 0 \, , \qquad
\lm^1{}_{\hat{2}} \ = \ 0 \, , \qquad
\label{tetrad-20-21-approx}
\nl
\lm^2{}_{\hat{2}} & = & {1 \over r} \, , \qquad
\lm^3{}_{\hat{2}} \ = \ 0 \, ,
\label{tetrad-22-23-approx}
\ee
\end{subequations}
for $\lm^\mu{}_{\hat{2}}$, and
\begin{subequations}
\label{tetrad-3-approx}
\be
\lm^0{}_{\hat{3}} & \approx & r \, \OmK \lt[{1 \over N \, A}
+ {\al \over 2 \, N^7} \, \lt|\d \lt(\Dl M\rt) \over \d \xi \rt| \, C\lt(r, \OmK \, \tau\rt)\rt]
\nn
&  &{} \times \cos \lt(\OmK \, \tau\rt) \, ,
\label{tetrad-30-approx}
\nl
\lm^1{}_{\hat{3}} & \approx & A \, \sin \lt(\OmK \, \tau \rt)
\nn
&  &{} - {2 \, \al \over N^3} \lt|\d \lt(\Dl M\rt) \over \d \xi \rt| \sin^2 \lt(\OmK \, \tau \rt) \, \cos \lt(\OmK \, \tau \rt) \, , \qquad
\nl
\label{tetrad-31-approx}
\lm^2{}_{\hat{3}} & = & 0 \, ,
\label{tetrad-32-approx}
\nl
\lm^3{}_{\hat{3}} & = & \lt[{A \over r \, N}
+ {\al \, \OmK \over 2 \, N^7} \, \lt|\d \lt(\Dl M\rt) \over \d \xi \rt| \, (r \, \OmK) \, C\lt(r, \OmK \, \tau\rt)\rt]
\nn
&  &{} \times \cos \lt(\OmK \, \tau \rt) \, ,
\label{tetrad-33-approx}
\ee
\end{subequations}
for $\lm^\mu{}_{\hat{3}}$, where
\be
C\lt(r, \OmK \, \tau\rt) & \equiv & 2 \, \sin \lt(2 \, \OmK \tau\rt)
\nn
&  &{} + {N \over r \, \OmK} \lt[\lt(1 - 2 \, r \, \OmK\rt) \sin \lt(2 \, \OmK \, \tau\rt) - 2 \, \OmK \, \tau \rt] \, .
\nn
\label{C1}
\ee

Given (\ref{tetrad-0-approx})--(\ref{tetrad-3-approx}), it is now possible to obtain the Riemann tensor components in the Fermi frame.
While the exact expressions for ${}^F{}R_{\Cmu \Cnu \Cal \Cbt}$ are found in Appendix~\ref{appendix:riemann-frame-components},
for the special case of $N \rightarrow 0$ and
$\th = \pi/2$ considered in this paper, the dominant nonzero components are
\begin{widetext}
\begin{subequations}
\label{Riemann-Fermi}
\be
{}^F{}R_{\ze \on \ze \on} & \approx & -{\OmK^2 \over N^2} \lt[2 \, A^2 + r^2 \, \OmK^2
+ {3 \, \al \over N^6} \, \lt|\d \lt(\Dl M\rt) \over \d \xi \rt| \, (r^3 \, \OmK^3) \, C\lt(r, \OmK \, \tau\rt) \rt] \cos^2 \lt(\OmK \, \tau\rt)
\ = \ - {}^F{}R_{\tw \tr \tw \tr} \, ,
\nl
{}^F{}R_{\ze \on \ze \tr} & \approx & -{\OmK^2 \over N^2} \lt[2 \, A^2 + r^2 \, \OmK^2
+ {3 \, \al \over N^6} \, \lt|\d \lt(\Dl M\rt) \over \d \xi \rt| \, (r^3 \, \OmK^3) \, C\lt(r, \OmK \, \tau\rt) \rt]
\sin \lt(\OmK \, \tau \rt) \, \cos \lt(\OmK \, \tau \rt)
\nn
& = & - {}^F{}R_{\on \tw \tw \tr} \, ,
\nl
{}^F{}R_{\ze \on \on \tr} & \approx & {3 \, \OmK^2 \over N^2} \lt[A \, (r \, \OmK)
+ {\al \over N^6} \, \lt|\d \lt(\Dl M\rt) \over \d \xi \rt| \, (r^3 \, \OmK^3) \, C\lt(r, \OmK \, \tau\rt) \rt] \cos \lt(\OmK \, \tau \rt)
\ = \ - {}^F{}R_{\ze \tw \tw \tr} \, ,
\nl
{}^F{}R_{\ze \tw \ze \tw} & \approx & {\OmK^2 \over N^2} \lt[1
+ {3 \, \al \over N^6} \, \lt|\d \lt(\Dl M\rt) \over \d \xi \rt| \, (r^3 \, \OmK^3) \, C\lt(r, \OmK \, \tau\rt) \rt]
\ = \ - {}^F{}R_{\on \tr \on \tr} \, ,
\nl
{}^F{}R_{\ze \tw \on \tw} & \approx & -{3 \, \OmK^2 \over N^2} \lt[A \, (r \, \OmK)
+ {\al \over N^6} \, \lt|\d \lt(\Dl M\rt) \over \d \xi \rt| \, (r^3 \, \OmK^3) \, C\lt(r, \OmK \, \tau\rt) \rt]
\sin \lt(\OmK \, \tau \rt)
%\nn
%& = &
\ = \ - {}^F{}R_{\ze \tr \on \tr} \, ,
\nl
\nn
\nn
{}^F{}R_{\ze \tr \ze \tr} & \approx & -{\OmK^2 \over N^2} \lt[2 \, A^2 + r^2 \, \OmK^2
+ {3 \, \al \over N^6} \, \lt|\d \lt(\Dl M\rt) \over \d \xi \rt| \, (r^3 \, \OmK^3) \, C\lt(r, \OmK \, \tau\rt) \rt] \sin^2 \lt(\OmK \, \tau \rt)
\nn
& = & - {}^F{}R_{\on \tw \on \tw} \, .
\ee
\end{subequations}
\end{widetext}
It is straightforward to confirm that (\ref{Riemann-Fermi}) agrees with (55) of Paper~I in the Schwarzschild limit when $\al = 0$.
As well, it is interesting to note that the Vaidya contribution to the curvature
becomes significant when $\lt|\d \lt(\Dl M\rt)/\d \xi \rt| \sim N^6$, which sets an appropriate scale for the mass accretion or loss rate
in the analysis to follow in this paper.

\section{Application to Circular Motion in the Vaidya Background}
\label{sec:4}

Having now obtained the orthonormal tetrad for the Vaidya background in the Fermi frame, it is possible
to make use of the generalized CMP approximation for the MPD equations.
This first requires evaluation of the unperturbed orbit from $P_{(0)}^\mu(\tau)$, which then gets integrated
with respect to $\tau$ to eventually obtain $X_{(0)}^\mu(\tau)$.
Because of the time-dependence in the Vaidya metric due to the evolving central mass function,
it is clear that the unperturbed orbit will not be truly circular.
However, since $\lt|\d \lt(\Dl M\rt)/\d \xi \rt| \ll 1$, the deviation from circular motion is minimal.
Based on (\ref{P0=}) and (\ref{tetrad-0-approx}), the unperturbed four-momentum components are determined to be
\begin{subequations}
\label{P0-components}
\be
P_{(0)}^0(\tau) & = & m_0 \lt[{1 \over N} + {\al \over 2 \, N^7} \, \lt|\d \lt(\Dl M\rt) \over \d \xi \rt|
r \, \OmK \, C\lt(r, \OmK \, \tau\rt)\rt] \, ,
\nn
\label{P0-0}
\nl
P_{(0)}^1(\tau) & = &  - 2 \, m_0 \, {\al \over N^3} \lt|\d \lt(\Dl M\rt) \over \d \xi \rt| \sin^2 \lt(\OmK \, \tau \rt) \, ,
\label{P0-1}
\nl
P_{(0)}^2(\tau) & = & 0 \, , \qquad
\label{P0-2}
\nl
P_{(0)}^3(\tau) & = & m_0 \, \OmK \lt[{1 \over N} \rt.
\nn
&  &{} + \lt. {\al \over 2 \, N^7} \, \lt|\d \lt(\Dl M\rt) \over \d \xi \rt| r \, \OmK \, C\lt(r, \OmK \, \tau\rt)\rt] \, .
\label{P0-3}
\ee
\end{subequations}
It is clear from (\ref{P0-1}) that the radial component of the four-momentum is directed inwards for infalling radiation $(\al = 1)$
and outwards for outflowing radiation $(\al = -1)$.
This makes physical sense because a growing central mass creates stronger curvature that gives rise to a stronger inward force
felt by the orbiting particle, and vice versa for a dissipating central mass.
Integrating (\ref{P0-components}) over $\tau$ in the form
\be
X_{(0)}^\mu(\tau) & = & {1 \over m_0} \, \int_0^\tau P_{(0)}^\mu(\tau') \, \d \tau' + X_{(0)}^\mu(0) \,
\label{X0=}
\ee
leads to the unperturbed orbit, where the initial position is $X_{(0)}^\mu(0) = \lt(0, \, r, \, \pi/2, \, 0\rt)$
to correspond with the $x$-axis.
Evaluation of (\ref{X0=}) results in
\begin{subequations}
\label{X0-components}
\be
X_{(0)}^0(\tau) & = & {\tau \over N} + {\al \over 2 \, N^7} \, \lt|\d \lt(\Dl M\rt) \over \d \xi \rt|
\nn
&  &{} \times r \, \OmK \int_0^\tau C\lt(r, \OmK \, \tau'\rt) \, \d \tau' \, ,
\label{X0-0}
\nl
X_{(0)}^1(\tau) & = &  - {\al \over \OmK \, N^3} \lt|\d \lt(\Dl M\rt) \over \d \xi \rt|
\nn
&  &{} \times \lt[\OmK \, \tau - {1 \over 2} \, \sin \lt(2 \, \OmK \, \tau\rt)\rt] \, ,
\label{X0-1}
\nl
X_{(0)}^2(\tau) & = & {\pi \over 2} \, , \qquad
\label{X0-2}
\nl
X_{(0)}^3(\tau) & = & {\OmK  \, \tau \over N} + {\al \over 2 \, N^7} \, \lt|\d \lt(\Dl M\rt) \over \d \xi \rt|
\nn
&  &{} \times r \, \OmK^2 \int_0^\tau C\lt(r, \OmK \, \tau'\rt) \, \d \tau' \, ,
\label{X0-3}
\ee
\end{subequations}
where
\be
\lefteqn{\int_0^\tau C\lt(r, \OmK \, \tau'\rt) \, \d \tau' \ = \
{1 \over 2 \, r \, \OmK^2} \lt\{2 \, r \, \OmK \lt[1 - \cos \lt(2 \, \OmK \, \tau\rt) \rt] \rt. }
\nn
&&{} + \lt. N \lt[\lt(1 - 2 \, r \, \OmK \rt)\lt[1 - \cos \lt(2 \, \OmK \, \tau\rt) \rt] - 2 \, \OmK^2 \, \tau^2 \rt] \rt\} \, .
\nn
\label{C-int=}
\ee
It is clear from (\ref{X0-components}) and (\ref{C-int=}) that a spinless particle in the Vaidya background
experiences a quasi-circular orbit with an overall growth or decay of its radial position over proper time, plus some
non-trivial oscillatory structure embedded within its time development.

\subsection{First-Order Perturbations in $\varepsilon$}
\label{sec:4.1}

As with the computation in Paper~I, progressing to the first-order perturbation (CMP approximation) in the Vaidya background
is conceptually straightforward.
However, the outcome is analytically more complicated than for its counterpart in the Kerr background.
For the spinning particle initially positioned on the $x$-axis of the Cartesian frame,
the initial spin orientation $(\hat{\th}, \hat{\ph})$ for $S_{(0)}^{\mu \nu}$ is chosen \cite{Mashhoon2,Singh2} to agree with the
standard definition of $(\th, \ph)$ for the spherical co-ordinates with respect to the Cartesian frame's $z$-axis.
This leads to
\begin{subequations}
\label{S0-proj}
\be
S_{(0)}^{\tw \tr} & = & s_0 \, \sin \hat{\th} \, \cos \hat{\ph} \, ,
\label{S0-23-proj}
\nl
S_{(0)}^{\tr \on} & = & -s_0 \, \cos \hat{\th} \, ,
\label{S0-31-proj}
\nl
S_{(0)}^{\on \tw} & = & s_0 \, \sin \hat{\th} \, \sin \hat{\ph} \, ,
\label{S0-12-proj}
\ee
\end{subequations}
with the outcome that
\begin{widetext}
\begin{subequations}
\label{S0}
\be
S_{(0)}^{01}(\tau) & = & -m_0 \, r \, \lt(s_0 \over m_0 \, r \rt) \lt[ {(r \, \OmK) \over N} \, \cos \hat{\th}
+ {\al \, A \over N^7} \, \lt|\d \lt(\Dl M\rt) \over \d \xi \rt| \rt.
\nn
&  &{} \times \lt.
\lt\{ {(r \, \OmK) \over 2} \lt[\sin (2 \, \OmK \, \tau + \hat{\th}) + \sin (2 \, \OmK \, \tau - \hat{\th}) \rt]
- N \, \cos \hat{\th} \, (\OmK \, \tau) \rt\} \rt]
 \, ,
\label{S0-01}
\nl
S_{(0)}^{02}(\tau) & = & -m_0 \lt(s_0 \over m_0 \, r \rt) \lt[{(r \, \OmK) \over 2 \, N \, A} \,
\, \lt[\sin (\OmK \, \tau + \hat{\th} - \hat{\ph}) - \sin (\OmK \, \tau - \hat{\th} - \hat{\ph})\rt] \rt.
\nn
\nn
\nn
&  &{} - {\al \over 4 \, N^7} \, \lt|\d \lt(\Dl M\rt) \over \d \xi \rt| \, (r \, \OmK)
\lt\{\lt[\cos (3 \, \OmK \, \tau + \hat{\th} - \hat{\ph}) - \cos (3 \, \OmK \, \tau - \hat{\th} - \hat{\ph}) \rt] \rt.
\nn
&  &{}
+ \lt[\cos (\OmK \, \tau + \hat{\th} + \hat{\ph}) - \cos (\OmK \, \tau - \hat{\th} + \hat{\ph}) \rt]
\nn
&  &{} + \lt. \lt.
{2 \, N \over (r \, \OmK)} \, \lt[\sin (\OmK \, \tau + \hat{\th} - \hat{\ph}) - \sin (\OmK \, \tau - \hat{\th} - \hat{\ph}) \rt]
(\OmK \, \tau) \rt\} \rt]
 \, ,
\label{S0-02}
\nl
S_{(0)}^{03}(\tau) & = & 0 \, ,
\label{S0-03}
\nl
S_{(0)}^{12}(\tau) & = & m_0 \, \lt(s_0 \over m_0 \, r \rt) \lt[{A \over 2}
\lt[\cos (\OmK \, \tau + \hat{\th} - \hat{\ph}) - \cos (\OmK \, \tau - \hat{\th} - \hat{\ph})\rt] \rt.
\nn
\nn
\nn
&  &{} - {\al \over 4 \, N^3} \, \lt|\d \lt(\Dl M\rt) \over \d \xi \rt|
\lt\{\lt[\sin (3 \, \OmK \, \tau + \hat{\th} - \hat{\ph}) - \sin (3 \, \OmK \, \tau - \hat{\th} - \hat{\ph}) \rt] \rt.
\nn
&  &{} + \lt. \lt.
\lt[\sin (\OmK \, \tau + \hat{\th} + \hat{\ph}) - \sin (\OmK \, \tau - \hat{\th} + \hat{\ph}) \rt]
- 2 \, \lt[\sin (\OmK \, \tau + \hat{\th} - \hat{\ph}) - \sin (\OmK \, \tau - \hat{\th} - \hat{\ph}) \rt] \rt\} \rt]
\, ,
\label{S0-12}
\nl
S_{(0)}^{23}(\tau) & = & {m_0 \over r} \, \lt(s_0 \over m_0 \, r \rt) \lt[{A \over 2 \, N}
\lt[\sin (\OmK \, \tau + \hat{\th} - \hat{\ph}) - \sin (\OmK \, \tau - \hat{\th} - \hat{\ph})\rt] \rt.
\nn
\nn
\nn
&  &{} - {\al \over 4 \, N^7} \, \lt|\d \lt(\Dl M\rt) \over \d \xi \rt| \, (r^2 \, \OmK^2)
\lt\{\lt[\cos (3 \, \OmK \, \tau + \hat{\th} - \hat{\ph}) - \cos (3 \, \OmK \, \tau - \hat{\th} - \hat{\ph}) \rt] \rt.
\nn
&  &{}
+ \lt[\cos (\OmK \, \tau + \hat{\th} + \hat{\ph}) - \cos (\OmK \, \tau - \hat{\th} + \hat{\ph}) \rt]
\nn
&  &{} + \lt. \lt.
{2 \, N \over (r \, \OmK)} \, \lt[\sin (\OmK \, \tau + \hat{\th} - \hat{\ph}) - \sin (\OmK \, \tau - \hat{\th} - \hat{\ph}) \rt]
(\OmK \, \tau) \rt\} \rt]
 \, ,
\label{S0-23}
\nl
S_{(0)}^{31}(\tau) & = & -m_0 \, \lt(s_0 \over m_0 \, r \rt) \lt[{A^2 \over N} \cos \hat{\th}
+ {\al \, A \over N^7} \, \lt|\d \lt(\Dl M\rt) \over \d \xi \rt| \rt.
\nn
&  &{} \times \lt. (r \, \OmK)
\lt\{{(r \, \OmK) \over 2} \lt[\sin (2 \, \OmK \, \tau + \hat{\th}) + \sin (2 \, \OmK \, \tau - \hat{\th}) \rt]
- N \, \cos \hat{\th} \, (\OmK \, \tau) \rt\} \rt]
\, .
\label{S0-31}
\ee
\end{subequations}
\end{widetext}
As noted earlier in Paper~I, a complicated beat structure in the sinusoidal functions exists in (\ref{S0}), due to
the initial spin orientation angles.
It is also confirmed that the leading-order spin tensor agrees with its Paper~I counterpart in the Schwarzschild limit
when $\al = 0$.

The first-order perturbation of the linear momentum is determined from (\ref{P1-general}).
A straightforward evaluation leads to the expressions
\begin{widetext}
\begin{subequations}
\label{P1-components}
\be
P_{(1)}^0(\tau) & = & m_0 \, \lt(s_0 \over m_0 \, r \rt) \lt[ {3 \over 2} \, {(r^3 \, \OmK^3) \over N^3} \,
\lt[\cos (\OmK \, \tau + \hat{\th}) + \cos (\OmK \, \tau - \hat{\th}) - 2 \, \cos \hat{\th} \rt] \rt.
\nn
\nn
\nn
&  &{} + {\al \over N^9} \,  \lt|\d \lt(\Dl M\rt) \over \d \xi \rt| \, (r^3 \, \OmK^3)
\lt\{{3 \over 4} \, A \lt[\sin (3 \, \OmK \, \tau + \hat{\th}) + \sin (3 \, \OmK \, \tau - \hat{\th}) \rt] \rt.
\nn
&  &{} - {1 \over 2 \, A} \, \lt(3 \, A^2 - 2 \, r^2 \, \OmK^2\rt)
\lt[\sin (2 \, \OmK \, \tau + \hat{\th}) + \sin (2 \, \OmK \, \tau - \hat{\th}) \rt]
\nn
&  &{} + {1 \over 4 \, A} \, \lt(3 \, A^2 - 8 \, r^2 \, \OmK^2\rt)
\lt[\sin (\OmK \, \tau + \hat{\th}) + \sin (\OmK \, \tau - \hat{\th}) \rt]
\nn
&  &{} - \lt. \lt. {3 \, N \, A \over 2 \, (r \, \OmK)} \, \lt[\cos (\OmK \, \tau + \hat{\th}) + \cos (\OmK \, \tau - \hat{\th})
- 2 \, \cos \hat{\th}\rt](\OmK \, \tau) + {6 \, N \, (r \, \OmK) \over A} \, \cos \hat{\th} \, (\OmK \, \tau)
\rt\} \rt]
\, ,
\label{P1-0}
\nl
\nn
\nn
P_{(1)}^1(\tau) & = &  m_0 \, \lt(s_0 \over m_0 \, r \rt) \lt[ {3 \over 2} \, {A^2 \, (r^2 \, \OmK^2) \over N^2} \,
\lt[\sin (\OmK \, \tau + \hat{\th}) + \sin (\OmK \, \tau - \hat{\th}) \rt] \rt.
\nn
\nn
\nn
&  &{} - {2 \, \al \, A \over N^8} \,  \lt|\d \lt(\Dl M\rt) \over \d \xi \rt| \, (r^4 \, \OmK^4)
\nn
&  &{} \times \lt.
\lt\{\lt[\cos (2 \, \OmK \, \tau + \hat{\th}) + \cos (2 \, \OmK \, \tau - \hat{\th}) \rt]
- \lt[\cos (\OmK \, \tau + \hat{\th}) + \cos (\OmK \, \tau - \hat{\th}) \rt] \rt\} \rt]
\, ,
\label{P1-1}
\nl
\nn
\nn
P_{(1)}^2(\tau) & = & {m_0 \over r} \, \lt(s_0 \over m_0 \, r \rt) \lt[ {3 \over 2} \, {A \, (r^2 \, \OmK^2) \over N^2} \,
\lt[\cos (\OmK \, \tau + \hat{\th} - \hat{\ph}) - \cos (\OmK \, \tau - \hat{\th} - \hat{\ph})
+ \cos (\hat{\th} + \hat{\ph}) - \cos (\hat{\th} - \hat{\ph}) \rt] \rt.
\nn
\nn
\nn
&  &{} + {\al \over 2 \, N^8} \,  \lt|\d \lt(\Dl M\rt) \over \d \xi \rt| \, (r^4 \, \OmK^4)
\lt\{\lt[\sin (3 \, \OmK \, \tau + \hat{\th} - \hat{\ph}) - \sin (3 \, \OmK \, \tau - \hat{\th} - \hat{\ph}) \rt] \rt.
\nn
&  &{} + 3 \lt[\sin (\OmK \, \tau + \hat{\th} + \hat{\ph}) - \sin (\OmK \, \tau - \hat{\th} + \hat{\ph}) \rt]
- 4 \lt[\sin (\hat{\th} + \hat{\ph}) + \sin (\hat{\th} - \hat{\ph})\rt]
\nn
&  &{} \lt. \lt.
- {6 \, N \over (r \, \OmK)}
\lt[\cos (\OmK \, \tau + \hat{\th} - \hat{\ph}) - \cos (\OmK \, \tau - \hat{\th} - \hat{\ph}) \rt] (\OmK \, \tau) \rt\} \rt]
\, ,
\label{P1-2}
\nl
\nn
\nn
P_{(1)}^3(\tau) & = & {m_0 \over r} \, \lt(s_0 \over m_0 \, r \rt)
\lt[{3 \over 2} \, {A^2 \, (r^2 \, \OmK^2) \over N^3} \,
\lt[\cos (\OmK \, \tau + \hat{\th}) + \cos (\OmK \, \tau - \hat{\th}) - 2 \, \cos \hat{\th} \rt] \rt.
\nn
\nn
\nn
&  &{} + {\al \over N^9} \,  \lt|\d \lt(\Dl M\rt) \over \d \xi \rt| \, A \, (r^4 \, \OmK^4)
\lt\{{3 \over 4} \, \lt[\sin (3 \, \OmK \, \tau + \hat{\th}) + \sin (3 \, \OmK \, \tau - \hat{\th}) \rt] \rt.
\nn
&  &{} - {1 \over 2} \, \lt[\sin (2 \, \OmK \, \tau + \hat{\th}) + \sin (2 \, \OmK \, \tau - \hat{\th}) \rt]
- {5 \over 4} \, \lt[\sin (\OmK \, \tau + \hat{\th}) + \sin (\OmK \, \tau - \hat{\th}) \rt]
\nn
&  &{} - \lt. \lt. {3 \, N \over 2 \, (r \, \OmK)} \, \lt[\cos (\OmK \, \tau + \hat{\th}) + \cos (\OmK \, \tau - \hat{\th})
- 6 \, \cos \hat{\th}\rt](\OmK \, \tau) \rt\} \rt]
 \, .
\label{P1-3}
\ee
\end{subequations}
\end{widetext}
It is useful to note the ratio between the azimuthal component of the linear momentum to its time component,
\be
{P_{(1)}^3(\tau) \over P_{(1)}^0(\tau)} & \approx &
{E \over L} \lt[1  - {\al \over N^4 \, A} \, \lt|\d \lt(\Dl M\rt) \over \d \xi \rt| \rt.
\nn
&  &{} \times \lt. \lt\{ \sin (2 \, \OmK \, \tau) - {N \over (r \, \OmK)} \,  (\OmK \, \tau) \rt\} \rt] \, . \qquad
\label{P1-3-P1-0}
\ee
Comparison between (\ref{P1-3-P1-0}) and its counterpart (61) in Paper~I shows an important distinction between
the two expressions, where (\ref{P1-3-P1-0}) records a predominantly oscillatory {\em time variation} in the ratio,
while the expression in Paper~I gives a strictly {\em time-independent} ratio of $E/L$.

\subsection{Higher-Order Perturbations in $\varepsilon$}
\label{sec:4.2}

While computations for the second-order perturbation quantities are straightforward to perform, in the Vaidya
background they become prohibitively long.
Therefore, expressions for the second-order linear momentum and spin tensor components
are presented in numerical form, contained in the next section.
In similar fashion to that in Paper~I, the ``radiative corrections'' to the squared mass
and spin magnitudes (\ref{m-bar-sq}) and (\ref{s-bar-sq}) exist in relatively compact form, necessary
to evaluate the perturbed M{\o}ller radius (\ref{s/m}).

It is first necessary to introduce the notation for beat functions in the form
\begin{subequations}
\label{Q-cs}
\be
Q_{\rm c}^\pm (n_1 \, \OmK \, \tau \, ,  n_2 \, \hat{\th} \, , \, n_3 \, \hat{\ph}) & \equiv &
\cos (n_1 \OmK \, \tau + n_2 \, \hat{\th} - n_3 \, \hat{\ph})
\nn
& \pm & \cos (n_1 \OmK \, \tau - n_2 \, \hat{\th} - n_3 \, \hat{\ph}) \, ,
\nn
\label{Q-c}
\nl
Q_{\rm s}^\pm (n_1 \, \OmK \, \tau \, ,  n_2 \, \hat{\th} \, , \, n_3 \, \hat{\ph}) & \equiv &
\sin (n_1 \OmK \, \tau + n_2 \, \hat{\th} - n_3 \, \hat{\ph})
\nn
& \pm & \sin (n_1 \OmK \, \tau - n_2 \, \hat{\th} - n_3 \, \hat{\ph}) \, ,
\nn
\label{Q-s}
\ee
\end{subequations}

Then the first-order spin shift in the squared spin magnitude in the Vaidya background is
\be
\bar{s}_1^2(\tau) & = & {1 \over N^2} \, \lt(s_0 \over m_0 \, r \rt) \lt[
\tilde{s}_{1a}^2(\tau) + {\al \over N^6} \, \lt|\d \lt(\Dl M\rt) \over \d \xi \rt| \, \tilde{s}_{1b}^2(\tau) \rt] \, ,
\nn
\label{s1-bar^2=}
\ee
where
\begin{widetext}
\begin{subequations}
\label{s1ab-bar^2=}
\be
\tilde{s}_{1a}^2(\tau) & = &
{3 \over 16} \, r^3 \, \OmK^3
\lt\{\lt(2 \, \OmK \, \tau\rt) \lt[Q_{\rm s}^+ (2 \, \OmK \, \tau \, ,  3 \, \hat{\th} \, , 2 \, \hat{\ph})
- Q_{\rm s}^+ (2 \, \OmK \, \tau \, , \hat{\th} \, , 2 \, \hat{\ph}) \rt] \rt.
\nn
&  &{} + 3 \lt[Q_{\rm c}^+ (2 \, \OmK \, \tau \, ,  3 \, \hat{\th} \, , 2 \, \hat{\ph})
- Q_{\rm c}^+ (2 \, \OmK \, \tau \, , \hat{\th} \, , 2 \, \hat{\ph}) \rt]
- Q_{\rm c}^+ (\OmK \, \tau \, ,  3 \, \hat{\th} \, , 2 \, \hat{\ph})
\nn
&  &{} + Q_{\rm c}^+ (\OmK \, \tau \, ,  \hat{\th} \, , 2 \, \hat{\ph})
- Q_{\rm c}^+ (\OmK \, \tau \, , 3 \, \hat{\th} \, , 0) + Q_{\rm c}^+ (\OmK \, \tau \, , \hat{\th} \, , 0)
\nn
&  &{} - \lt. 2 \lt[Q_{\rm c}^+ (0 \, ,  3 \, \hat{\th} \, , 2 \, \hat{\ph}) - Q_{\rm c}^+ (0 \, , \hat{\th} \, , 2 \, \hat{\ph}) \rt]
+ 2 \lt[\cos (3 \, \hat{\th}) - \cos \hat{\th} \rt] \rt\} \, ,
\label{s1a-bar^2=}
\nl
\nn
\nn
\tilde{s}_{1b}^2(\tau) & = &
{1 \over 256} \, (r^4 \, \OmK^4) \lt[24 \, (\OmK^2 \, \tau^2) \lt\{2 \lt[Q_{\rm s}^+ (4 \, \OmK \, \tau \, , 3 \, \hat{\th} \, , 2 \, \hat{\ph})
- Q_{\rm s}^+ (4 \, \OmK \, \tau \, , \hat{\th} \, , 2 \, \hat{\ph})\rt]  \rt. \rt.
\nn
&  &{} + \lt.
3 \lt[Q_{\rm s}^+ (2 \, \OmK \, \tau \, , 3 \, \hat{\th} \, , 0) - Q_{\rm s}^+ (2 \, \OmK \, \tau \, , \hat{\th} \, , 0) \rt]
- Q_{\rm s}^+ (0 \, , 3 \, \hat{\th} \, , 2 \, \hat{\ph}) + Q_{\rm s}^+ (0 \, , \hat{\th} \, , 2 \, \hat{\ph})
\rt\}
\nn
&  &{} + (\OmK \, \tau) \lt\{12 \lt[Q_{\rm c}^+ (4 \, \OmK \, \tau \, , 3 \, \hat{\th} \, , 2 \, \hat{\ph})
- Q_{\rm c}^+ (4 \, \OmK \, \tau \, , \hat{\th} \, , 2 \, \hat{\ph})\rt] \rt.
\nn
&  &{} + {12 \over A} \, \lt(5 \, A + 8 \, r \, \OmK\rt) \lt[Q_{\rm c}^+ (2 \, \OmK \, \tau \, , 3 \, \hat{\th} \, , 0)
- Q_{\rm c}^+ (2 \, \OmK \, \tau \, , \hat{\th} \, , 0) \rt]
\nn
&  &{} + \lt. {48 \over A} \, \lt(A + 6 \, r \, \OmK\rt) \lt[Q_{\rm c}^+ (0 \, , 3 \, \hat{\th} \, , 2 \, \hat{\ph})
- Q_{\rm c}^+ (0 \, , \hat{\th} \, , 2 \, \hat{\ph}) \rt]\rt\}
\nn
&  &{} - {1 \over A} \, \lt(15 \, A - 8 \, r \, \OmK\rt) \lt[Q_{\rm s}^+ (4 \, \OmK \, \tau \, , 3 \, \hat{\th} \, , 2 \, \hat{\ph})
- Q_{\rm s}^+ (4 \, \OmK \, \tau \, , \hat{\th} \, , 2 \, \hat{\ph})\rt]
\nn
&  &{} - {1 \over A} \, \lt(27 \, A - 328 \, r \, \OmK\rt) \, Q_{\rm s}^+ (2 \, \OmK \, \tau \, , 3 \, \hat{\th} \, , 0)
+ {3 \over A} \, \lt(9 \, A + 232 \, r \, \OmK\rt) \, Q_{\rm s}^+ (2 \, \OmK \, \tau \, , \hat{\th} \, , 0)
\nn
&  &{} + {1 \over A} \, \lt(3 \, A - 40 \, r \, \OmK\rt) \lt[Q_{\rm s}^+ (\OmK \, \tau \, , 3 \, \hat{\th} \, , 2 \, \hat{\ph})
- Q_{\rm s}^+ (\OmK \, \tau \, , 3 \, \hat{\th} \, , 0) - Q_{\rm s}^+ (\OmK \, \tau \, , \hat{\th} \, , 2 \, \hat{\ph})
+ Q_{\rm s}^+ (\OmK \, \tau \, , \hat{\th} \, , 0) \rt]
\nn
&  &{} + \lt. {4 \over A} \, \lt(3 \, A + 8 \, r \, \OmK\rt) \lt[Q_{\rm s}^+ (0 \, , 3 \, \hat{\th} \, , 2 \, \hat{\ph})
- Q_{\rm s}^+ (0 \, , \hat{\th} \, , 2 \, \hat{\ph})\rt] \rt] \, .
\label{s1b-bar^2=}
\ee
\end{subequations}
\end{widetext}
When integrated over a cycle defined by the Keplerian frequency, the time-averaged expression for (\ref{s1-bar^2=}) is
determined with respect to
\begin{subequations}
\label{s1-tilde^2-time-avg}
\be
\lt\langle \tilde{s}_{1a}^2 \rt\rangle & \equiv & {\OmK \over 2 \, \pi} \int_0^{2 \pi/\OmK} \tilde{s}_{1a}^2(\tau) \, \d \tau
\nn
& = & {3 \over 2} \, \lt(r^3 \, \OmK^3\rt) \,
\sin^2 \hat{\th} \, \cos \hat{\th} \lt[3 \, \cos (2 \, \hat{\th}) - 1 \rt],
\label{s1a-tilde^2-time-avg=}
\nl
\lt\langle \tilde{s}_{1b}^2 \rt\rangle & = & {(r^4 \, \OmK^4) \over 16 \, A} \, \sin^2 \hat{\th} \, \cos \hat{\th} \,
\nn
&  &{} \times
\lt[{1 \over 2} \lt(21 \, A - 32 \, \pi^2 \, A + 32 \, r \, \OmK \rt) \sin (2 \, \hat{\ph}) \rt.
\nn
&  &{} - \lt. 12 \, \pi \lt(A + 12 \, r \, \OmK\rt) \cos (2 \, \hat{\ph}) + 36 \, \pi \, A \rt]. \qquad
\label{s1b-tilde^2-time-avg=}
\ee
\end{subequations}
It is clear from (\ref{s1-tilde^2-time-avg}) that $\lt\langle \bar{s}_1^2 \rt\rangle$ is well-behaved for the full range of
$\hat{\th}$ and $\hat{\ph}$.

For the second-order mass shift in the squared mass magnitude, it is shown that
\begin{widetext}
\be
\bar{m}_2^2(\tau) & = & \lt(s_0 \over m_0 \, r \rt)^2 \lt[ {9 \over 8} \, {A^2 \over N^4} \,  (r^4 \, \OmK^4) \rt.
\nn
&  &{} \times \lt\{Q_{\rm c}^+ (2 \, \OmK \, \tau \, , 2 \, \hat{\th} \, , 2 \, \hat{\ph}) - Q_{\rm c}^+ (0 \, , 2 \, \hat{\th} \, , 2 \, \hat{\ph})
- 2 \lt[ \cos (2 \, \OmK \, \tau - 2 \, \hat{\ph}) - 2 \, \cos (2 \, \hat{\ph})\rt] \rt\}
\nn
&  &{} + {3 \over 4} \, {\al \, A \over N^{10}} \, \lt|\d \lt(\Dl M\rt) \over \d \xi \rt| (r^6 \, \OmK^6)
\lt\{ Q_{\rm s}^+ (4 \, \OmK \, \tau \, , 2 \, \hat{\th} \, , 2 \, \hat{\ph}) - 2 \, \sin (4 \, \OmK \, \tau - 2 \, \hat{\ph})
- 2 \, Q_{\rm s}^+ (2 \, \OmK \, \tau \, , 2 \, \hat{\th} \, , 0) \rt.
\nn
&  &{} - \lt. \lt. 12 \, \sin(2 \, \OmK \, \tau) - Q_{\rm s}^+ (0 \, , 2 \, \hat{\th} \, , 2 \, \hat{\ph})
- 2 \, \sin (2 \, \hat{\ph}) \rt\} \rt]
\, ,
\label{m2-bar^2=}
\ee
\end{widetext}
where the corresponding time-averaged expression is
\be
\lt\langle \bar{m}_2^2 \rt\rangle & = & \lt(s_0 \over m_0 \, r \rt)^2 \, (r^4 \, \OmK^4) \lt[
{9 \over 2} \, {A^2 \over N^4} \, \, \sin^2 \hat{\th} \, (2 \, \cos^2 \hat{\ph} - 1) \rt.
\nn
&  &{} - \lt. {6 \, \al \over N^{10}} \, \lt|\d \lt(\Dl M\rt) \over \d \xi \rt| \, A \, (r^2 \, \OmK^2) \,
\sin^2 \hat{\th} \, \sin \hat{\ph} \, \cos \hat{\ph} \rt]
\, .
\nn
\label{m2-bar^2-time-avg=}
\ee
In similar fashion, the second-order spin shift is determined to be
\begin{widetext}
\be
\bar{s}_2^2(\tau) & = & -{3 \over 8 \, N^4} \, \lt(s_0 \over m_0 \, r \rt) \, A^2 (r^3 \, \OmK^3)
\lt[Q_{\rm c}^+ (4 \, \OmK \, \tau \, , 3 \, \hat{\th} \, , 2 \, \hat{\ph}) - Q_{\rm c}^+ (4 \, \OmK \, \tau \, , \hat{\th} \, , 2 \, \hat{\ph}) \rt.
\nn
&  &{} - Q_{\rm c}^+ (3 \, \OmK \, \tau \, , 3 \, \hat{\th} \, , 2 \, \hat{\ph})
+ 3 \, Q_{\rm c}^+ (3 \, \OmK \, \tau \, , 3 \, \hat{\th} \, , 0) + Q_{\rm c}^+ (3 \, \OmK \, \tau \, , \hat{\th} \, , 2 \, \hat{\ph})
+ 5 \, Q_{\rm c}^+ (3 \, \OmK \, \tau \, , \hat{\th} \, , 0)
\nn
&  &{} + Q_{\rm c}^+ (\OmK \, \tau \, , 3 \, \hat{\th} \, , -2 \, \hat{\ph})
- 3 \, Q_{\rm c}^+ (\OmK \, \tau \, , 3 \, \hat{\th} \, , 0) - Q_{\rm c}^+ (\OmK \, \tau \, , \hat{\th} \, , -2 \, \hat{\ph})
- 5 \, Q_{\rm c}^+ (\OmK \, \tau \, , \hat{\th} \, , 0)
\nn
&  &{} - \lt. Q_{\rm c}^+ (0 \, , 3 \, \hat{\th} \, , 2 \, \hat{\ph}) + Q_{\rm c}^+ (0 \, , \hat{\th} \, , 2 \, \hat{\ph}) \rt]
\nn
&  &{} + {3 \, \al \over 2 \, N^{10}} \, \lt(s_0 \over m_0 \, r \rt)^2 \, \lt|\d \lt(\Dl M\rt) \over \d \xi \rt| \, A \, (r^6 \, \OmK^6)
\lt[Q_{\rm s}^+ (4 \, \OmK \, \tau \, , 4 \, \hat{\th} \, , 2 \, \hat{\ph}) - Q_{\rm s}^+ (4 \, \OmK \, \tau \, , 2 \, \hat{\th} \, , 2 \, \hat{\ph}) \rt.
\nn
&  &{} - Q_{\rm s}^+ (3 \, \OmK \, \tau \, , 4 \, \hat{\th} \, , 2 \, \hat{\ph})
+ Q_{\rm s}^+ (3 \, \OmK \, \tau \, , 4 \, \hat{\th} \, , 0)
+ Q_{\rm s}^+ (3 \, \OmK \, \tau \, , 2 \, \hat{\th} \, , 2 \, \hat{\ph}) + Q_{\rm c}^+ (3 \, \OmK \, \tau \, , 2 \, \hat{\th} \, , 0)
+ 4 \, \sin(3 \, \OmK \, \tau)
\nn
&  &{} + 2 \, Q_{\rm s}^+ (2 \, \OmK \, \tau \, , 4 \, \hat{\th} \, , 0)
+ 2 \, Q_{\rm s}^+ (2 \, \OmK \, \tau \, , 2 \, \hat{\th} \, , 0) + 8 \, \sin(2 \, \OmK \, \tau)
\nn
&  &{} - 4 \, Q_{\rm s}^+ (\OmK \, \tau \, , 4 \, \hat{\th} \, , 2 \, \hat{\ph})
+ 3 \, Q_{\rm s}^+ (\OmK \, \tau \, , 4 \, \hat{\th} \, , -2 \, \hat{\ph})
- 7 \, Q_{\rm s}^+ (\OmK \, \tau \, , 4 \, \hat{\th} \, , 0)
\nn
&  &{} + 4 \, Q_{\rm s}^+ (\OmK \, \tau \, , 2 \, \hat{\th} \, , 2 \, \hat{\ph})
- 3 \, Q_{\rm s}^+ (\OmK \, \tau \, , 2 \, \hat{\th} \, , -2 \, \hat{\ph})
- 7 \, Q_{\rm s}^+ (\OmK \, \tau \, , 2 \, \hat{\th} \, , 0) - 28 \, \sin (\OmK \, \tau)
\nn
&  &{} + \lt. 7 \, Q_{\rm s}^+ (0 \, , 4 \, \hat{\th} \, , 2 \, \hat{\ph}) - 7 \, Q_{\rm s}^+ (0 \, , 2 \, \hat{\th} \, , 2 \, \hat{\ph}) \rt]
%\nn
%&  &{}
+ {1 \over 4 \, N^{10}} \, \lt(s_0 \over m_0 \, r \rt)^2 \, \lt|\d \lt(\Dl M\rt) \over \d \xi \rt| \, \tilde{s}_1^2(\tau) \, \tilde{s}_2^2(\tau) \, ,
\label{s2-bar^2=}
\ee
\end{widetext}
whose time-averaged expression is
\newpage
\begin{widetext}
\be
\lt\langle \bar{s}_2^2 \rt\rangle & = & - {3 \, A^2 \over N^4} \, \lt(s_0 \over m_0 \, r \rt)  \, (r^3 \, \OmK^3) \,
\sin^2 \hat{\th} \, \cos \hat{\th} \, (2 \, \cos^2 \hat{\ph} - 1)
\nn
&  &{} + {1 \over N^{10}} \lt(s_0 \over m_0 \, r \rt)^2 \, \lt|\d \lt(\Dl M\rt) \over \d \xi \rt|
\lt\{ 84 \, \al \, A \, (r^6 \, \OmK^6) \, \sin^2 \hat{\th} \, (4 \, \cos^2 \hat{\th} - 1) \, \sin \hat{\ph} \, \cos \hat{\ph}
+ {1 \over 4} \, \lt\langle \tilde{s}_{1a}^2 \, \tilde{s}_{1b}^2 \rt\rangle
\rt\}
\, .
\label{s2-bar^2-time-avg=}
\ee
\end{widetext}

\section{Numerical Analysis}
\label{sec:5}

Based on the analytic expressions presented in this paper, it is useful to explore some numerical analysis
for the main results obtained.
For the same reasons as given in Paper I, the purpose for taking this route is to visually determine the consequences of increasing
the order of the perturbation expansion in the generalized CMP approximation, when applied to the Vaidya background.
It may be possible to identify a correspondence between this approach and that of a purely numerical treatment of the MPD equations.
This numerical treatment of the generalized CMP approximation is given in terms of plots found in Appendix~\ref{appendix:plots}.

As with Paper I, the emphasis for this analysis is to identify the stability properties of the
spinning particle's motion in the Vaidya background, assuming $r = 6 M$, $\hat{\th} = \hat{\ph} = \pi/4$,
and $\lt|\d \lt(\Dl M\rt) / \d \xi \rt| = 10^{-4}$.
While it is understood from (\ref{Mdot}) that this choice for the mass accretion rate $(\al = 1)$
is too large given the Eddington luminosity limit, it nonetheless provides a useful means to directly compare with
the corresponding set of plots found in Paper I.
As for the mass loss rate $(\al = -1)$, there is apparently no reason to exclude this choice for
$\lt|\d \lt(\Dl M\rt) / \d \xi \rt|$ on physical grounds.

Throughout this paper, the numerical analysis assumes that $\mu \equiv s_0/(m_0 \, r) = 10^{-2}$ and $\mu = 10^{-1}$,
where $m_0 = 10^{-2} M$.
To determine the magnitude for a realistic spin, it is first shown that
\be
s_0 & = & \lt(10^2 \, {r \, \mu \over M} \rt) \, m_0^2 \,
\label{s0-realistic}
\ee
for given $r$ and $m_0$, which suggests that
\be
\mu & \lesssim & 10^{-2} \, {M \over r} \,
\label{mu-realistic}
\ee
to accommodate for a realistic spin of $s_0 \lesssim m_0^2$, corresponding to solar mass black holes and neutron stars \cite{Hartl1,Cook}
in orbit around supermassive black holes.
Since $\mu = 10^{-2}$ and $\mu = 10^{-1}$ lead to unrealistically large values \cite{Hartl1,Hartl2} for $s_0$,
while agreeing with the choice given previously \cite{Suzuki1,Suzuki2} to explore chaotic behaviour for the MPD equations,
it follows that any chaotic effects determined in this paper occur outside of astrophysically realistic conditions,
as noted in Paper~I.

\begin{figure*}
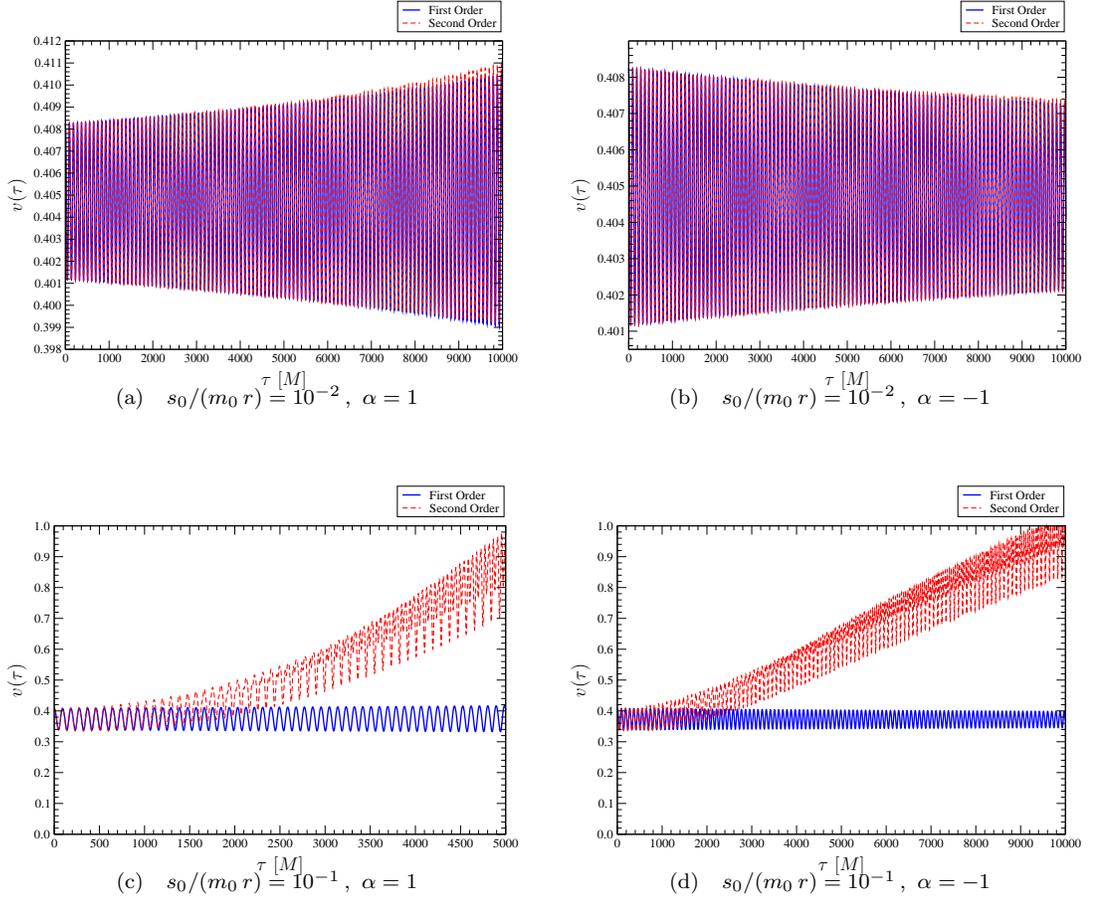

\psfrag{T}[tc][][1.8][0]{\Large $\tau \ [M]$}
\psfrag{v}[bc][][1.8][0]{\Large $v(\tau)$}
\begin{minipage}[t]{0.3 \textwidth}
\centering
\subfigure[\hspace{0.2cm} $s_0/(m_0 \, r) = 10^{-2} \, , \ \al = 1$]{
\label{fig:v-Vaidya-r=6-T=025-P=025-dMdx=1e-04-alpha=+1-mu=1e-2}
\rotatebox{0}{\includegraphics[width = 6.6cm, height = 5.0cm, scale = 1]{1a}}}
%\vspace{0.5cm}
\end{minipage}%
\hspace{2.0cm}
\begin{minipage}[t]{0.3 \textwidth}
\centering
\subfigure[\hspace{0.2cm} $s_0/(m_0 \, r) = 10^{-2} \, , \ \al = -1$]{
\label{fig:v-Vaidya-r=6-T=025-P=025-dMdx=1e-04-alpha=-1-mu=1e-2}
\rotatebox{0}{\includegraphics[width = 6.6cm, height = 5.0cm, scale = 1]{1b}}}
%\vspace{0.5cm}
\end{minipage} \\
\vspace{0.8cm}
\begin{minipage}[t]{0.3 \textwidth}
\centering
\subfigure[\hspace{0.2cm} $s_0/(m_0 \, r) = 10^{-1} \, , \ \al = 1$]{
\label{fig:v-Vaidya-r=6-T=025-P=025-dMdx=1e-04-alpha=+1-mu=1e-1}
\rotatebox{0}{\includegraphics[width = 6.6cm, height = 5.0cm, scale = 1]{1c}}}
%\vspace{0.5cm}
\end{minipage}%
\hspace{2.0cm}
\begin{minipage}[t]{0.3 \textwidth}
\centering
\subfigure[\hspace{0.2cm} $s_0/(m_0 \, r) = 10^{-1} \, , \ \al = -1$]{
\label{fig:v-Vaidya-r=6-T=025-P=025-dMdx=1e-04-alpha=-1-mu=1e-1}
\rotatebox{0}{\includegraphics[width = 6.6cm, height = 5.0cm, scale = 1]{1d}}}
%\vspace{0.5cm}
\end{minipage}
\caption{\label{fig:v-Vaidya-r=6-T=025-P=025-dMdx=1e-04} Co-ordinate speed $v(\tau)$ of the spinning particle for circular motion around
a black hole described by the Vaidya metric, for $r = 6 M$ and $\hat{\th} = \hat{\ph} = \pi/4$.
Fig.~\ref{fig:v-Vaidya-r=6-T=025-P=025-dMdx=1e-04-alpha=+1-mu=1e-2} shows a gradual increase in amplitude for $s_0/(m_0 \, r) = 10^{-2}$
with no significant difference due to second-order contributions in $\varepsilon$.
For $\al = -1$, Fig.~\ref{fig:v-Vaidya-r=6-T=025-P=025-dMdx=1e-04-alpha=-1-mu=1e-2} shows a corresponding gradual decrease in amplitude.
In contrast, Figs.~\ref{fig:v-Vaidya-r=6-T=025-P=025-dMdx=1e-04-alpha=+1-mu=1e-1} and \ref{fig:v-Vaidya-r=6-T=025-P=025-dMdx=1e-04-alpha=-1-mu=1e-1}
indicate the existence of an instability in the orbit when $s_0/(m_0 \, r) = 10^{-1}$, due to the rise of $v(\tau)$
from the second-order contribution in $\varepsilon$.}
\end{figure*}

Figure~\ref{fig:v-Vaidya-r=6-T=025-P=025-dMdx=1e-04} lists the plots of the particle's co-ordinate speed
\be
v(\varepsilon) & = & \sqrt{g_{ij} \, V^i(\varepsilon) \, V^j(\varepsilon)} \,
\label{v-defn}
\ee
as a function of $\tau$, where
\be
V^i(\varepsilon) & \equiv & {u^i (\varepsilon) \over u^0 (\varepsilon)} \,
\label{v-comp-defn}
\ee
for $u^\mu(\varepsilon)$ according to (\ref{MPD-velocity-explicit}), to first- and second-order
in $\varepsilon$, and with the restriction of $0 \leq v < 1$.
As with the result of Paper I, Figs.~\ref{fig:v-Vaidya-r=6-T=025-P=025-dMdx=1e-04-alpha=+1-mu=1e-2} and
\ref{fig:v-Vaidya-r=6-T=025-P=025-dMdx=1e-04-alpha=-1-mu=1e-2} show that the expression to first-order in $\varepsilon$
is almost exclusively responsible for $v$ due to $s_0/(m_0 \, r) = 10^{-2}$.
Unlike the corresponding set of plots due to the Kerr background, however, the range of co-ordinate speed changes
with $\tau$ in accordance with the choice for $\al$.
In Fig.~\ref{fig:v-Vaidya-r=6-T=025-P=025-dMdx=1e-04-alpha=+1-mu=1e-2}, the range for $v$ grows steadily for $\al = 1$,
while the opposite is true for $\al = -1$, as shown in Fig.~\ref{fig:v-Vaidya-r=6-T=025-P=025-dMdx=1e-04-alpha=-1-mu=1e-2}.
This outcome suggests that the eccentricity of the particle's orbit increases or decreases accordingly.
When $s_0/(m_0 \, r) = 10^{-1}$, it is clear from Figs.~\ref{fig:v-Vaidya-r=6-T=025-P=025-dMdx=1e-04-alpha=+1-mu=1e-1} and
\ref{fig:v-Vaidya-r=6-T=025-P=025-dMdx=1e-04-alpha=-1-mu=1e-1} that the particle's motion becomes unstable due to
the $O(\varepsilon^2)$ expression, and the co-ordinate speed rapidly approaches $v = 1$.
For Fig.~\ref{fig:v-Vaidya-r=6-T=025-P=025-dMdx=1e-04-alpha=+1-mu=1e-1} with $\al = 1$ this occurs at around $\tau = 1500 M$,
while Fig.~\ref{fig:v-Vaidya-r=6-T=025-P=025-dMdx=1e-04-alpha=-1-mu=1e-1} with $\al = -1$ denotes this outcome
at around $\tau = 2000 M$.

\begin{figure*}
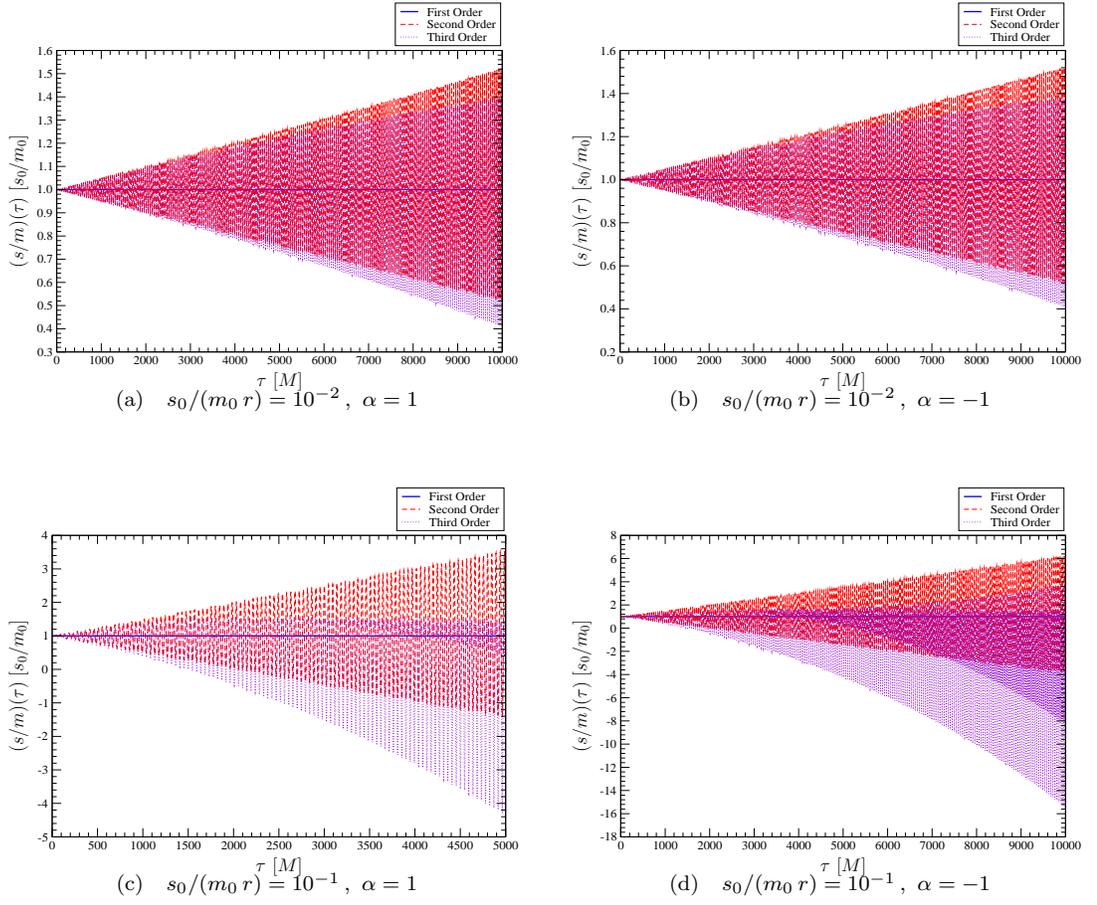

\psfrag{T}[tc][][1.8][0]{\Large $\tau \ [M]$}
\psfrag{s/m}[bc][][1.8][0]{\Large $(s/m)(\tau) \ [s_0/m_0]$}
\begin{minipage}[t]{0.3 \textwidth}
\centering
\subfigure[\hspace{0.2cm} $s_0/(m_0 \, r) = 10^{-2} \, , \ \al = 1$]{
\label{fig:moller-Vaidya-r=6-T=025-P=025-dMdx=1e-04-alpha=+1-mu=1e-2}
\rotatebox{0}{\includegraphics[width = 6.6cm, height = 5.0cm, scale = 1]{2a}}}
%\vspace{0.5cm}
\end{minipage}%
\hspace{2.0cm}
\begin{minipage}[t]{0.3 \textwidth}
\centering
\subfigure[\hspace{0.2cm} $s_0/(m_0 \, r) = 10^{-2} \, , \ \al = -1$]{
\label{fig:moller-Vaidya-r=6-T=025-P=025-dMdx=1e-04-alpha=-1-mu=1e-2}
\rotatebox{0}{\includegraphics[width = 6.6cm, height = 5.0cm, scale = 1]{2b}}}
%\vspace{0.5cm}
\end{minipage} \\
\vspace{0.8cm}
\begin{minipage}[t]{0.3 \textwidth}
\centering
\subfigure[\hspace{0.2cm} $s_0/(m_0 \, r) = 10^{-1} \, , \ \al = 1$]{
\label{fig:moller-Vaidya-r=6-T=025-P=025-dMdx=1e-04-alpha=+1-mu=1e-1}
\rotatebox{0}{\includegraphics[width = 6.6cm, height = 5.0cm, scale = 1]{2c}}}
%\vspace{0.5cm}
\end{minipage}%
\hspace{2.0cm}
\begin{minipage}[t]{0.3 \textwidth}
\centering
\subfigure[\hspace{0.2cm} $s_0/(m_0 \, r) = 10^{-1} \, , \ \al = -1$]{
\label{fig:moller-Vaidya-r=6-T=025-P=025-dMdx=1e-04-alpha=-1-mu=1e-1}
\rotatebox{0}{\includegraphics[width = 6.6cm, height = 5.0cm, scale = 1]{2d}}}
%\vspace{0.5cm}
\end{minipage}
\caption{\label{fig:moller-Vaidya-r=6-T=025-P=025-dMdx=1e-04} M{\o}ller radius $\rho(\tau) = (s/m)(\tau)$ in the Vaidya background
for $r = 6 M$ and $\hat{\th} = \hat{\ph} = \pi/4$, in units of $s_0/m_0$.
Figs.~\ref{fig:moller-Vaidya-r=6-T=025-P=025-dMdx=1e-04-alpha=+1-mu=1e-2} and \ref{fig:moller-Vaidya-r=6-T=025-P=025-dMdx=1e-04-alpha=-1-mu=1e-2}
show that while the higher-order contributions in $\varepsilon$
lead to a slowly increasing amplitude in $\rho$, Figs.~\ref{fig:moller-Vaidya-r=6-T=025-P=025-dMdx=1e-04-alpha=+1-mu=1e-1}
and \ref{fig:moller-Vaidya-r=6-T=025-P=025-dMdx=1e-04-alpha=-1-mu=1e-1} indicate a much larger amplitude increase as
$s_0/(m_0 \, r) = 10^{-1}$, where the second- and third-order contributions in $\varepsilon$ become distinctive.}
\end{figure*}

Since roughly the same behaviour occurs for the Kerr background in Paper I, this provides further evidence that
a sufficiently large choice for $s_0/(m_0 \, r)$ can trigger the transition from stable to unstable orbital motion.
This motivates a similar examination of the M{\o}ller radius $\rho(\tau) = (s/m)(\tau)$, as given by (\ref{s/m}),
and expressed by Figure~\ref{fig:moller-Vaidya-r=6-T=025-P=025-dMdx=1e-04} for the same set of initial conditions,
to third-order in $\varepsilon$.
When compared with Figure~\ref{fig:v-Vaidya-r=6-T=025-P=025-dMdx=1e-04}, and in general agreement with the corresponding
plots in Paper I, there is further confirmation that the spin-curvature interaction due to $\rho(\tau)$ induces
the respective kinematic outcome for $v(\tau)$.
As well, the expression to third-order in $\varepsilon$ implies that the mass shift contribution $\bar{m}_2^2$ and the second-order
spin shift term $\bar{s}_2^2$ listed in (\ref{s/m}) results in a downward shift in the range of oscillation
for Figure~\ref{fig:moller-Vaidya-r=6-T=025-P=025-dMdx=1e-04}.
Given $s_0/(m_0 \, r) = 10^{-2}$, comparison of Fig.~\ref{fig:moller-Vaidya-r=6-T=025-P=025-dMdx=1e-04-alpha=+1-mu=1e-2} for $\al = 1$,
with Fig.~\ref{fig:moller-Vaidya-r=6-T=025-P=025-dMdx=1e-04-alpha=-1-mu=1e-2} for $\al = -1$
shows relatively little difference between them.
Not surprisingly, the situation changes considerably when $s_0/(m_0 \, r) = 10^{-1}$, as shown in
Figs.~\ref{fig:moller-Vaidya-r=6-T=025-P=025-dMdx=1e-04-alpha=+1-mu=1e-1} and \ref{fig:moller-Vaidya-r=6-T=025-P=025-dMdx=1e-04-alpha=-1-mu=1e-1}.
It is very interesting to note that, when compared with Figs.~\ref{fig:v-Vaidya-r=6-T=025-P=025-dMdx=1e-04-alpha=+1-mu=1e-1} and
\ref{fig:v-Vaidya-r=6-T=025-P=025-dMdx=1e-04-alpha=-1-mu=1e-1},
the sudden increase in $v$ for both cases corresponds precisely with the condition that
$\rho(\tau) < 0$ for $\tau > 1500 M$ and $\tau > 2000 M$, respectively.
This provides more evidence suggesting that the M{\o}ller radius must be strictly positive-valued to
correspond with stable orbital motion, roughly agreeing with a similar set of conditions noted in Paper I for the Kerr background.

\begin{figure*}
\psfrag{Th}[cc][][1.8][0]{\Large $\hat{\th}$}
\psfrag{Ph}[cc][][1.8][0]{\Large $\hat{\ph}$}
\psfrag{s/m}[bc][][1.8][90]{\Large $\lt\langle s/m \rt\rangle \ [s_0/m_0]$}
\begin{minipage}[t]{0.3 \textwidth}
\centering
\subfigure[\hspace{0.2cm} $\al = 1 \, , \ O(\varepsilon^2)$]{
\label{fig:avg-Moller-Vaidya-alpha=+1-e2}
\rotatebox{0}{\includegraphics[width = 5.0cm, height = 3.5cm, scale = 1]{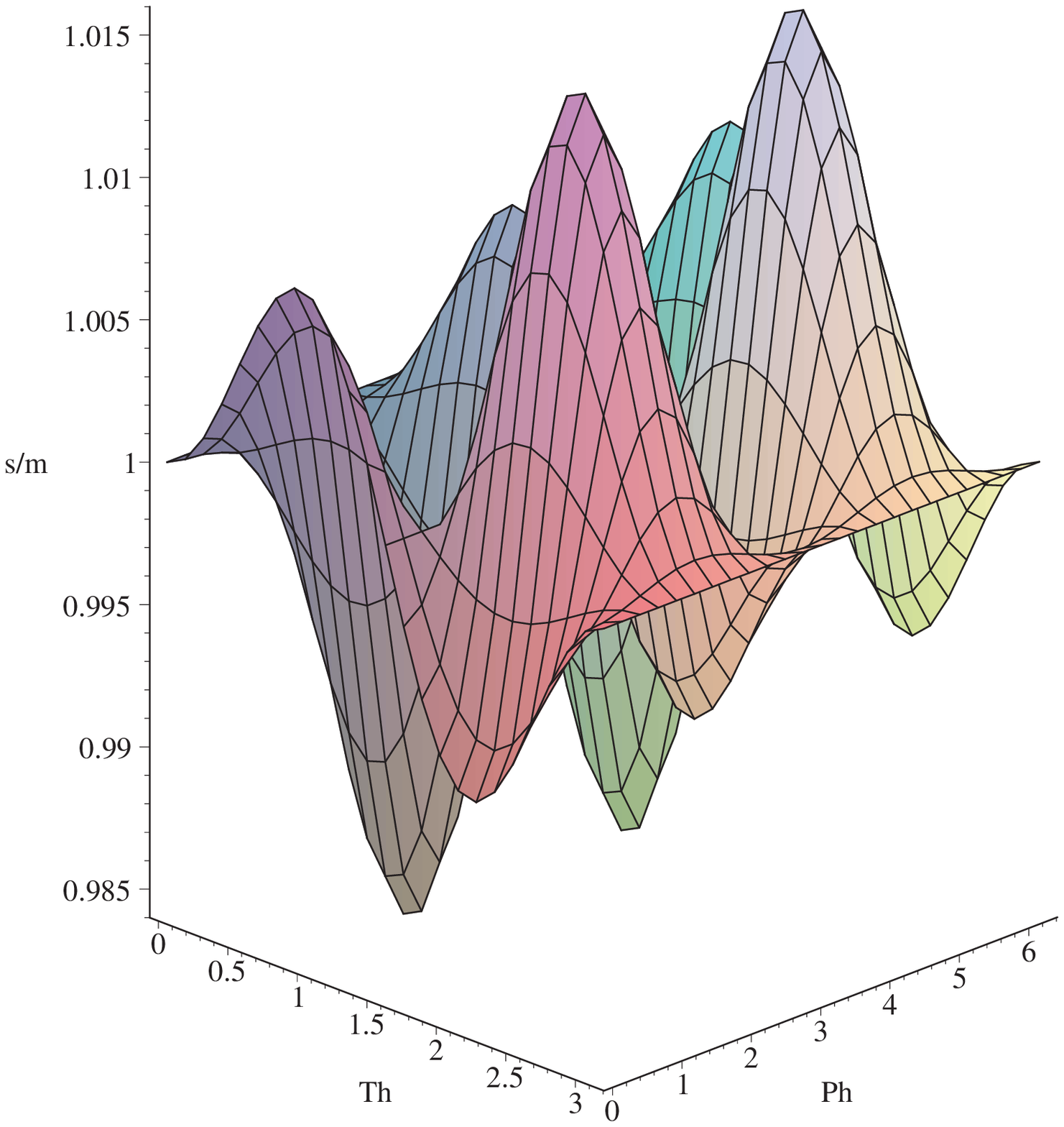}}}
\end{minipage}%
\hspace{0.5cm}
\begin{minipage}[t]{0.3 \textwidth}
\centering
\subfigure[\hspace{0.2cm} $\al = 1 \, , \ O(\varepsilon^3)$]{
\label{fig:avg-Moller-Vaidya-alpha=+1-e3}
\rotatebox{0}{\includegraphics[width = 5.0cm, height = 3.5cm, scale = 1]{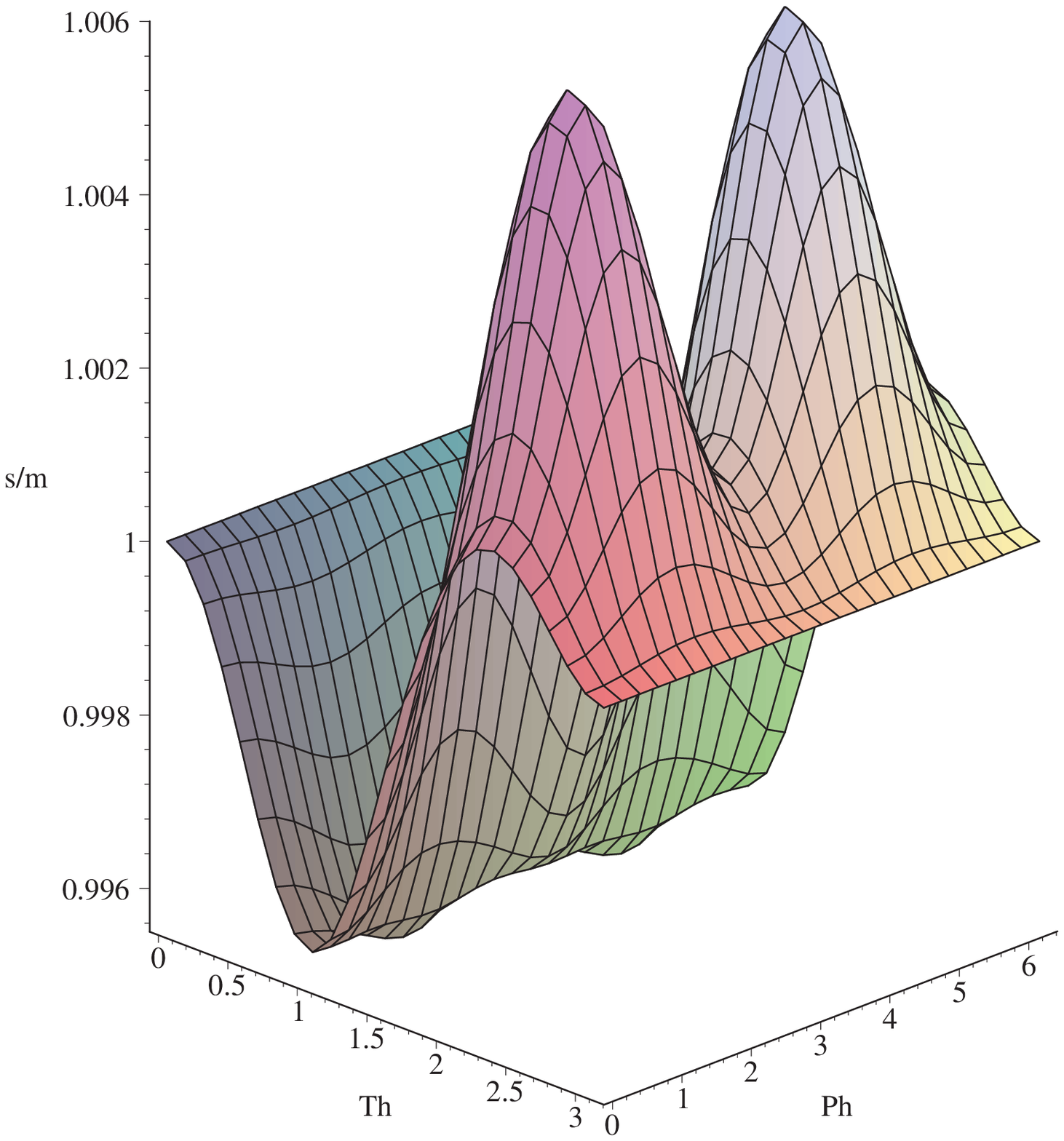}}}
\end{minipage}%
\hspace{0.5cm}
\begin{minipage}[t]{0.3 \textwidth}
\centering
\vspace{-3.5cm}
\caption{\label{fig:avg-Moller-Vaidya-alpha=+1} Three-dimensional plot of the time-averaged M{\o}ller radius
$\lt\langle \rho \rt\rangle = \lt\langle s/m \rt\rangle$ as a function of $\hat{\th}$ and $\hat{\ph}$
for $r = 6 M$ and $\al = 1$.
Fig.~\ref{fig:avg-Moller-Vaidya-alpha=+1-e2} shows a complicated peak and valley structure to $\lt\langle \rho \rt\rangle$
that simplifies somewhat in Fig.~\ref{fig:avg-Moller-Vaidya-alpha=+1-e3}.
\vspace{1mm}}
\end{minipage}
\end{figure*}

\begin{figure*}
\psfrag{Th}[cc][][1.8][0]{\Large $\hat{\th}$}
\psfrag{Ph}[cc][][1.8][0]{\Large $\hat{\ph}$}
\psfrag{s/m}[bc][][1.8][90]{\Large $\lt\langle s/m \rt\rangle \ [s_0/m_0]$}
\begin{minipage}[t]{0.3 \textwidth}
\centering
\subfigure[\hspace{0.2cm} $\al = -1 \, , \ O(\varepsilon^2)$]{
\label{fig:avg-Moller-Vaidya-alpha=-1-e2}
\rotatebox{0}{\includegraphics[width = 5.0cm, height = 3.5cm, scale = 1]{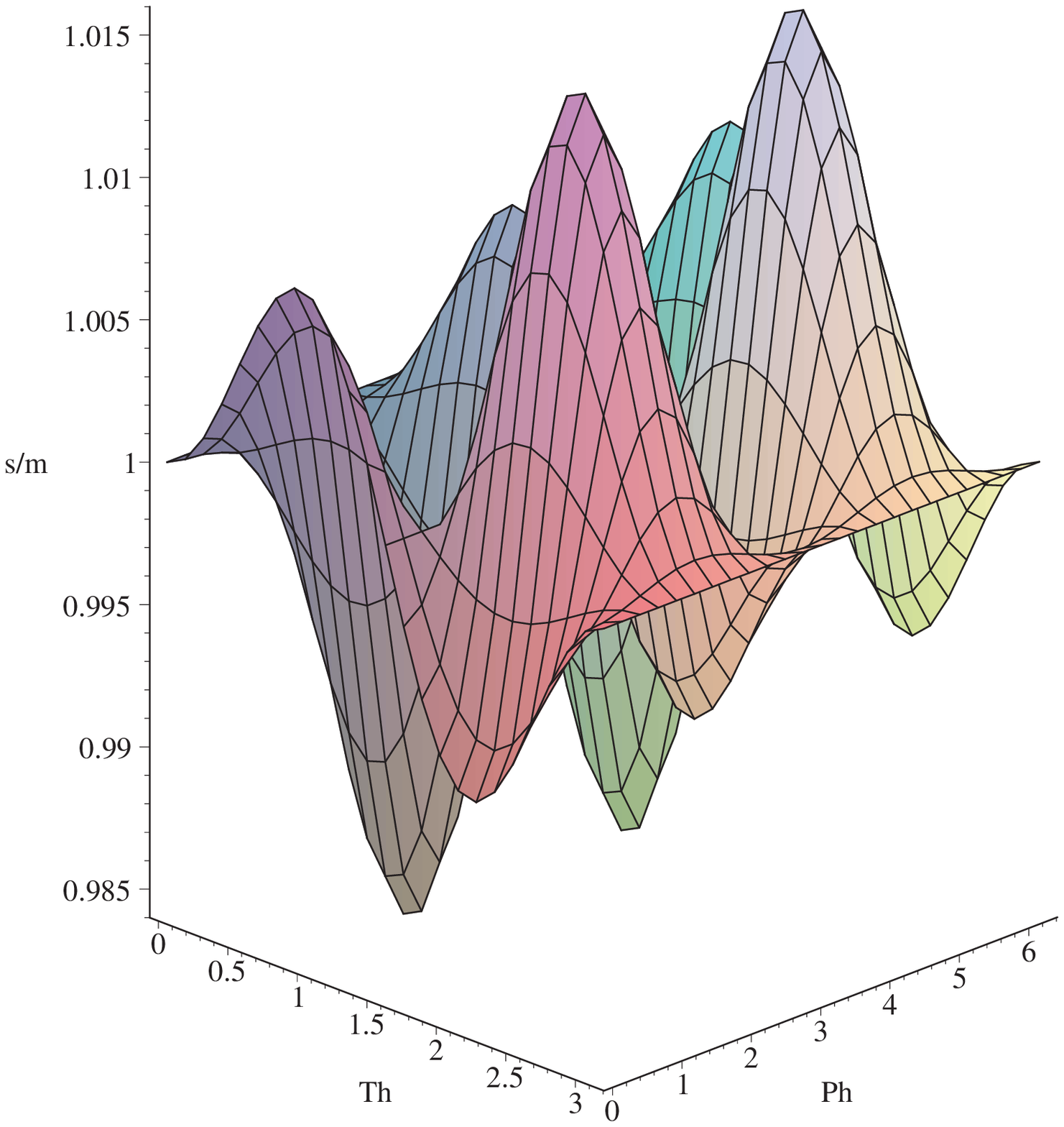}}}
\end{minipage}%
\hspace{0.5cm}
\begin{minipage}[t]{0.3 \textwidth}
\centering
\subfigure[\hspace{0.2cm} $\al = -1 \, , \ O(\varepsilon^3)$]{
\label{fig:avg-Moller-Vaidya-alpha=-1-e3}
\rotatebox{0}{\includegraphics[width = 5.0cm, height = 3.5cm, scale = 1]{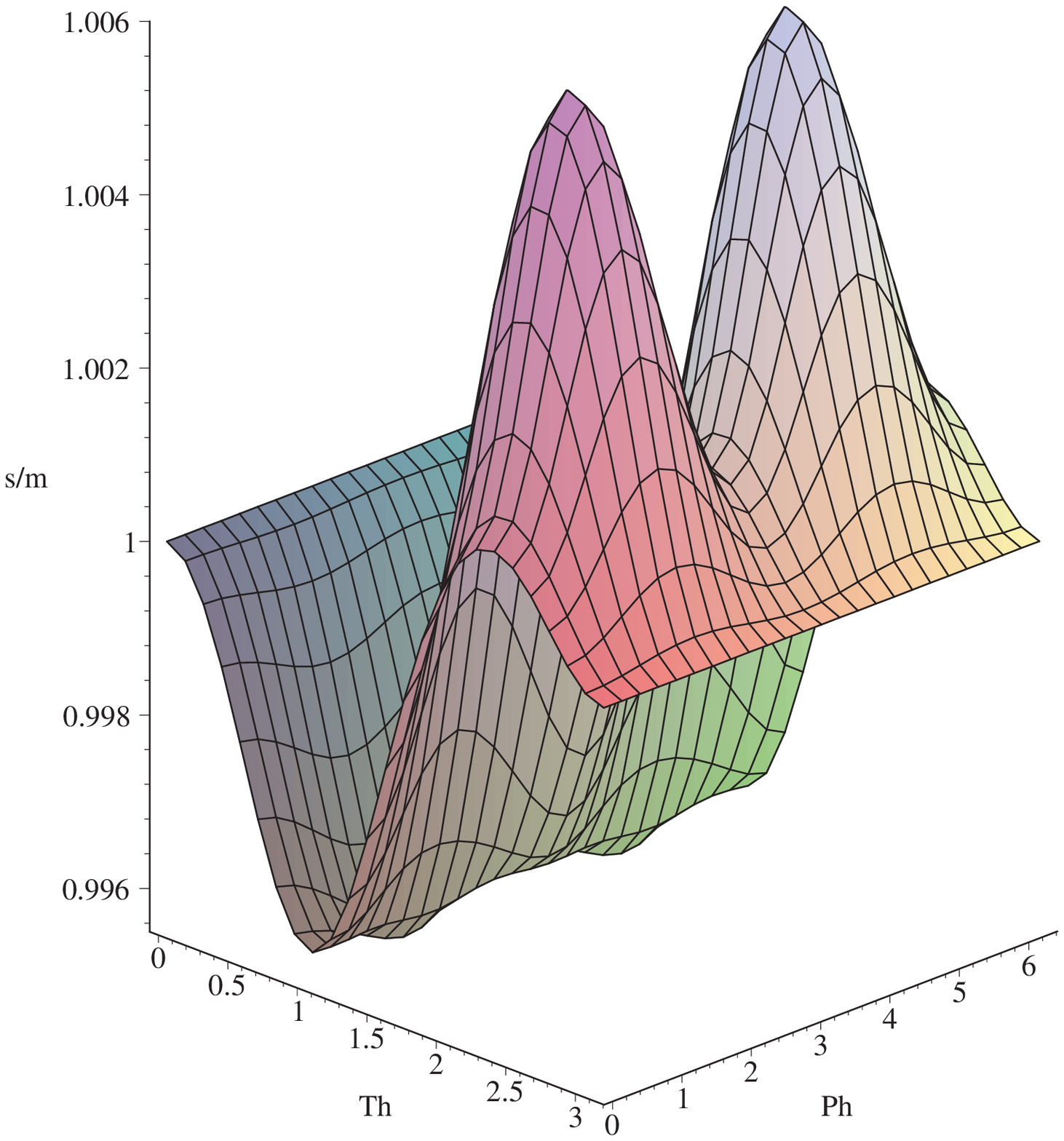}}}
\end{minipage}%
\hspace{0.5cm}
\begin{minipage}[t]{0.3 \textwidth}
\centering
\vspace{-3.5cm}
\caption{\label{fig:avg-Moller-Vaidya-alpha=-1} Time-averaged M{\o}ller radius as a function of $\hat{\th}$ and $\hat{\ph}$
for $r = 6 M$ and $\al = -1$.
The plots are essentially indistinguishable when compared to Figure~\ref{fig:avg-Moller-Vaidya-alpha=+1}, as $\al = 1$ goes to
$\al = -1$.
\vspace{1mm}}
\end{minipage}
\end{figure*}

Following the treatment given in Paper I, the time-averaged value for the M{\o}ller radius
$\lt\langle \rho \rt\rangle = \lt\langle s/m \rt\rangle$ as a function of the initial spin orientation angles $\hat{\th}$ and $\hat{\ph}$
is considered, leading to three-dimensional plots given by Figures~\ref{fig:avg-Moller-Vaidya-alpha=+1} and \ref{fig:avg-Moller-Vaidya-alpha=-1}
for $s_0/(m_0 \, r) = 10^{-1}$.
In similar fashion to that shown in Paper I, these set of plots identify an even function symmetry according to $\hat{\ph} = \pi$.
As denoted by Figs.~\ref{fig:avg-Moller-Vaidya-alpha=+1-e2} and \ref{fig:avg-Moller-Vaidya-alpha=-1-e2},
there exists a non-trivial peak and valley structure in $\lt\langle \rho \rt\rangle$ that agrees in form with the
corresponding set in Paper I, while Figs.~\ref{fig:avg-Moller-Vaidya-alpha=+1-e3} and \ref{fig:avg-Moller-Vaidya-alpha=-1-e3}
indicate a loss of structure in $0 \leq \hat{\th} < \pi$, with two peaks that remain.
However, unlike the plots given in Paper I, this set of plots also numerically agree with each other, irrespective
of the choice for $\al$, which suggests that the relevant terms are either small compared to terms not coupled to $\al$,
or they integrate to zero entirely.

\begin{figure*}
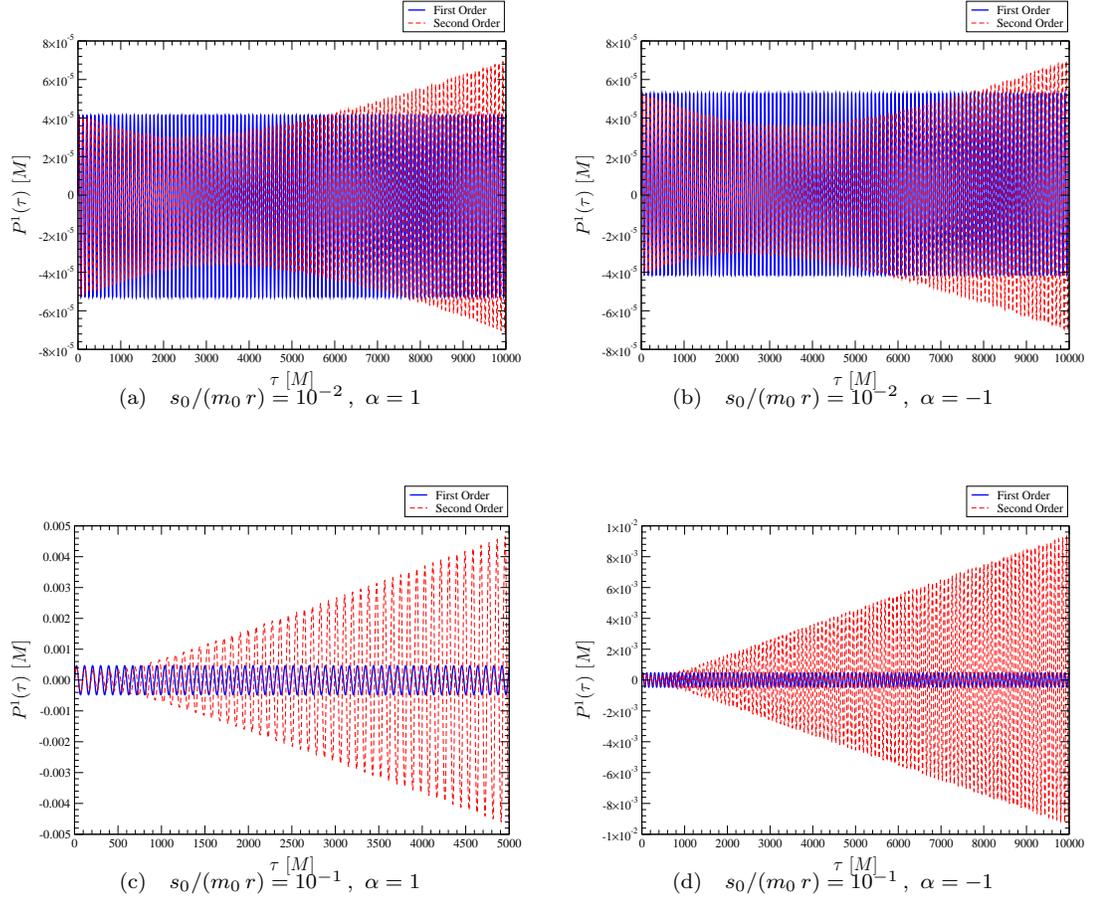

\psfrag{T}[tc][][1.8][0]{\Large $\tau \ [M]$}
\psfrag{P1}[bc][][1.8][0]{\Large $P^1(\tau) \ [M]$}
\begin{minipage}[t]{0.3 \textwidth}
\centering
\subfigure[\hspace{0.2cm} $s_0/(m_0 \, r) = 10^{-2} \, , \ \al = 1$]{
\label{fig:P1-Vaidya-r=6-T=025-P=025-dMdx=1e-04-alpha=+1-mu=1e-2}
\rotatebox{0}{\includegraphics[width = 6.6cm, height = 5.0cm, scale = 1]{5a}}}
%\vspace{0.5cm}
\end{minipage}%
\hspace{2.0cm}
\begin{minipage}[t]{0.3 \textwidth}
\centering
\subfigure[\hspace{0.2cm} $s_0/(m_0 \, r) = 10^{-2} \, , \ \al = -1$]{
\label{fig:P1-Vaidya-r=6-T=025-P=025-dMdx=1e-04-alpha=-1-mu=1e-2}
\rotatebox{0}{\includegraphics[width = 6.6cm, height = 5.0cm, scale = 1]{5b}}}
%\vspace{0.5cm}
\end{minipage} \\
\vspace{0.8cm}
\begin{minipage}[t]{0.3 \textwidth}
\centering
\subfigure[\hspace{0.2cm} $s_0/(m_0 \, r) = 10^{-1} \, , \ \al = 1$]{
\label{fig:P1-Vaidya-r=6-T=025-P=025-dMdx=1e-04-alpha=+1-mu=1e-1}
\rotatebox{0}{\includegraphics[width = 6.6cm, height = 5.0cm, scale = 1]{5c}}}
%\vspace{0.5cm}
\end{minipage}%
\hspace{2.0cm}
\begin{minipage}[t]{0.3 \textwidth}
\centering
\subfigure[\hspace{0.2cm} $s_0/(m_0 \, r) = 10^{-1} \, , \ \al = -1$]{
\label{fig:P1-Vaidya-r=6-T=025-P=025-dMdx=1e-04-alpha=-1-mu=1e-1}
\rotatebox{0}{\includegraphics[width = 6.6cm, height = 5.0cm, scale = 1]{5d}}}
%\vspace{0.5cm}
\end{minipage}
\caption{\label{fig:P1-Vaidya-r=6-T=025-P=025-dMdx=1e-04} Radial component $P^1(\tau)$ of the linear momentum in the Vaidya background for
$r = 6 M$ and $\hat{\th} = \hat{\ph} = \pi/4$.
When adding the second-order contribution in $\varepsilon$ for $s_0/(m_0 \, r) = 10^{-2}$,
Figs.~\ref{fig:P1-Vaidya-r=6-T=025-P=025-dMdx=1e-04-alpha=+1-mu=1e-2} and \ref{fig:P1-Vaidya-r=6-T=025-P=025-dMdx=1e-04-alpha=-1-mu=1e-2}
indicate a slight increase in the amplitude, while
Figs.~\ref{fig:P1-Vaidya-r=6-T=025-P=025-dMdx=1e-04-alpha=+1-mu=1e-1} and \ref{fig:P1-Vaidya-r=6-T=025-P=025-dMdx=1e-04-alpha=-1-mu=1e-1}
show a much stronger amplitude increase as $s_0/(m_0 \, r) = 10^{-1}$.}
\end{figure*}

\begin{figure*}
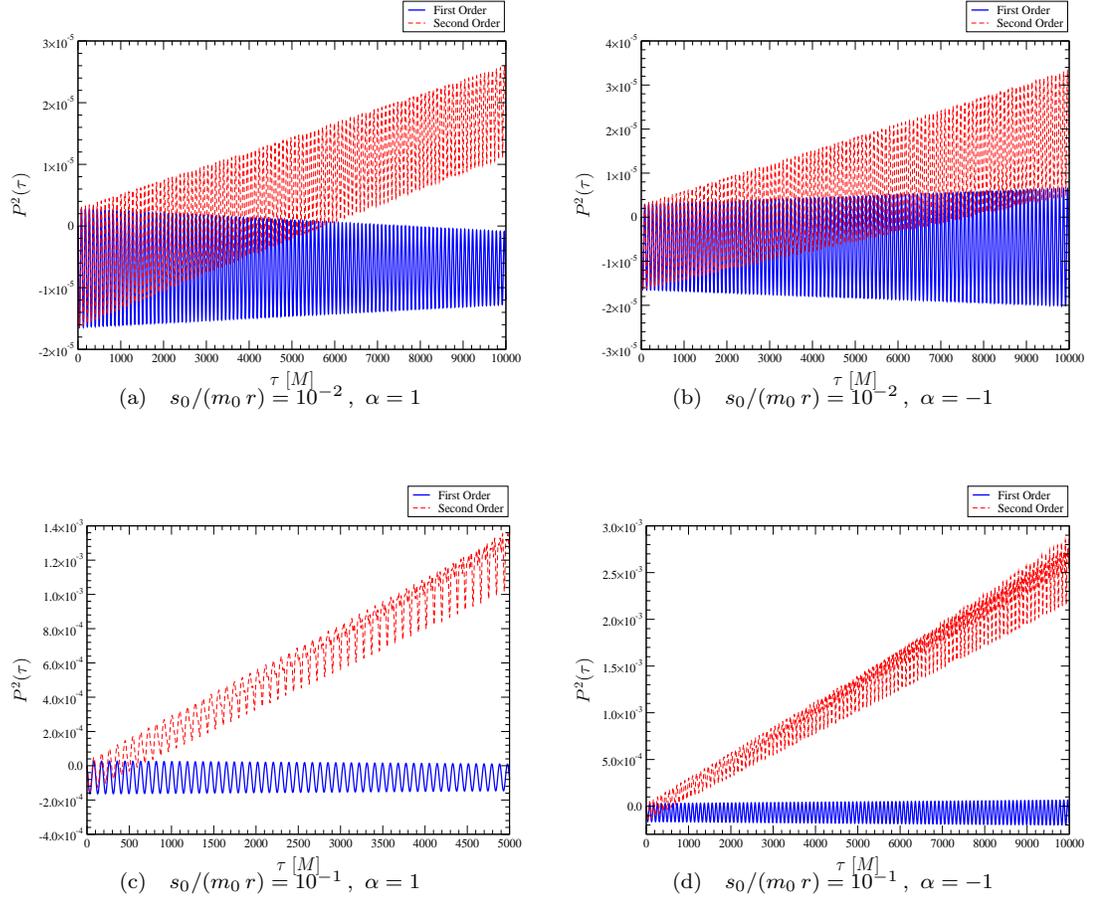

\psfrag{T}[tc][][1.8][0]{\Large $\tau \ [M]$}
\psfrag{P2}[bc][][1.8][0]{\Large $P^2(\tau)$}
\begin{minipage}[t]{0.3 \textwidth}
\centering
\subfigure[\hspace{0.2cm} $s_0/(m_0 \, r) = 10^{-2} \, , \ \al = 1$]{
\label{fig:P2-Vaidya-r=6-T=025-P=025-dMdx=1e-04-alpha=+1-mu=1e-2}
\rotatebox{0}{\includegraphics[width = 6.6cm, height = 5.0cm, scale = 1]{6a}}}
%\vspace{0.5cm}
\end{minipage}%
\hspace{2.0cm}
\begin{minipage}[t]{0.3 \textwidth}
\centering
\subfigure[\hspace{0.2cm} $s_0/(m_0 \, r) = 10^{-2} \, , \ \al = -1$]{
\label{fig:P2-Vaidya-r=6-T=025-P=025-dMdx=1e-04-alpha=-1-mu=1e-2}
\rotatebox{0}{\includegraphics[width = 6.6cm, height = 5.0cm, scale = 1]{6b}}}
%\vspace{0.5cm}
\end{minipage} \\
\vspace{0.8cm}
\begin{minipage}[t]{0.3 \textwidth}
\centering
\subfigure[\hspace{0.2cm} $s_0/(m_0 \, r) = 10^{-1} \, , \ \al = 1$]{
\label{fig:P2-Vaidya-r=6-T=025-P=025-dMdx=1e-04-alpha=+1-mu=1e-1}
\rotatebox{0}{\includegraphics[width = 6.6cm, height = 5.0cm, scale = 1]{6c}}}
%\vspace{0.5cm}
\end{minipage}%
\hspace{2.0cm}
\begin{minipage}[t]{0.3 \textwidth}
\centering
\subfigure[\hspace{0.2cm} $s_0/(m_0 \, r) = 10^{-1} \, , \ \al = -1$]{
\label{fig:P2-Vaidya-r=6-T=025-P=025-dMdx=1e-04-alpha=-1-mu=1e-1}
\rotatebox{0}{\includegraphics[width = 6.6cm, height = 5.0cm, scale = 1]{6d}}}
%\vspace{0.5cm}
\end{minipage}
\caption{\label{fig:P2-Vaidya-r=6-T=025-P=025-dMdx=1e-04} Polar component $P^2(\tau)$ of the linear momentum for
$r = 6 M$ and $\hat{\th} = \hat{\ph} = \pi/4$ in the Vaidya background.
All three plots show that the second-order contribution in $\varepsilon$ results in a net non-zero magnitude for the polar component.
While Figs.~\ref{fig:P2-Vaidya-r=6-T=025-P=025-dMdx=1e-04-alpha=+1-mu=1e-2} and \ref{fig:P2-Vaidya-r=6-T=025-P=025-dMdx=1e-04-alpha=-1-mu=1e-2}
yield a modest non-zero effect for $s_0/(m_0 \, r) = 10^{-2}$,
Figs.~\ref{fig:P2-Vaidya-r=6-T=025-P=025-dMdx=1e-04-alpha=+1-mu=1e-1} and \ref{fig:P2-Vaidya-r=6-T=025-P=025-dMdx=1e-04-alpha=-1-mu=1e-1}
show a significantly more pronounced effect for $s_0/(m_0 \, r) = 10^{-1}$.}
\end{figure*}

\begin{figure*}
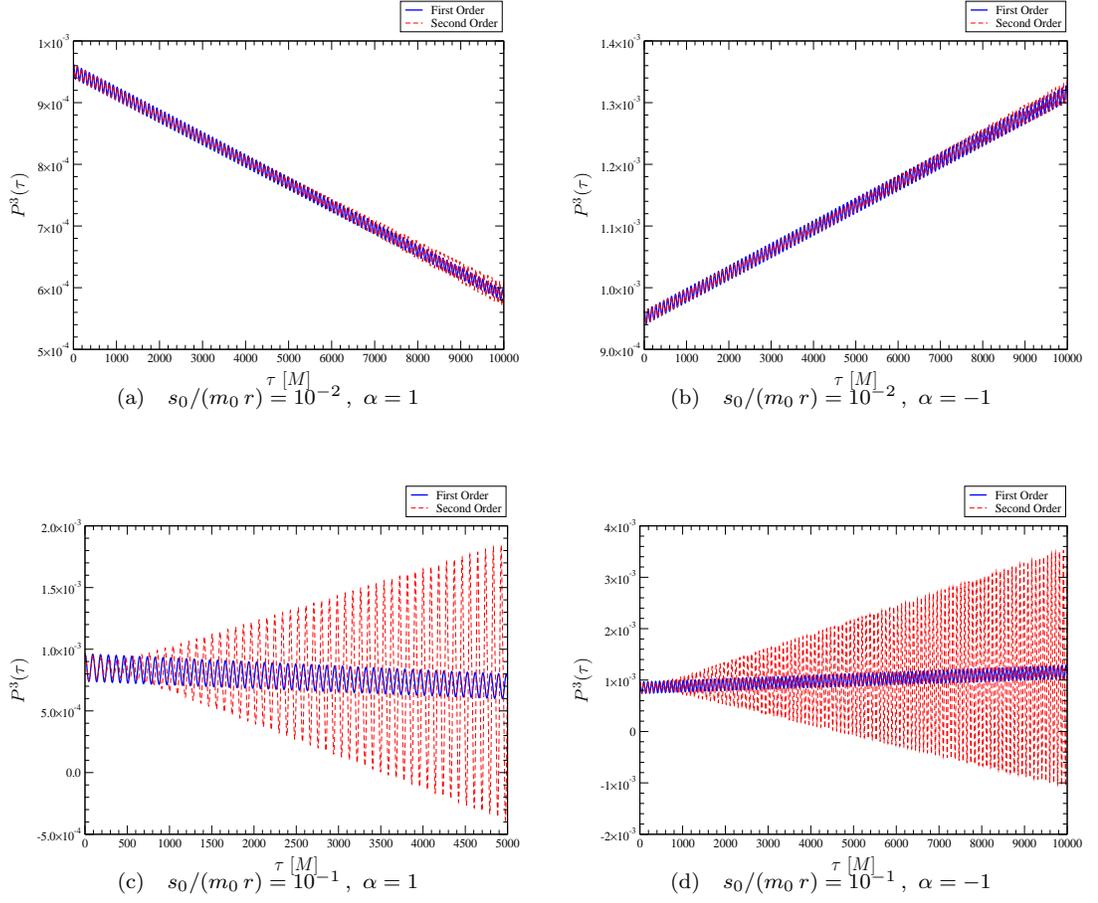

\psfrag{T}[tc][][1.8][0]{\Large $\tau \ [M]$}
\psfrag{P3}[bc][][1.8][0]{\Large $P^3(\tau)$}
\begin{minipage}[t]{0.3 \textwidth}
\centering
\subfigure[\hspace{0.2cm} $s_0/(m_0 \, r) = 10^{-2} \, , \ \al = 1$]{
\label{fig:P3-Vaidya-r=6-T=025-P=025-dMdx=1e-04-alpha=+1-mu=1e-2}
\rotatebox{0}{\includegraphics[width = 6.6cm, height = 5.0cm, scale = 1]{7a}}}
%\vspace{0.5cm}
\end{minipage}%
\hspace{2.0cm}
\begin{minipage}[t]{0.3 \textwidth}
\centering
\subfigure[\hspace{0.2cm} $s_0/(m_0 \, r) = 10^{-2} \, , \ \al = -1$]{
\label{fig:P3-Vaidya-r=6-T=025-P=025-dMdx=1e-04-alpha=-1-mu=1e-2}
\rotatebox{0}{\includegraphics[width = 6.6cm, height = 5.0cm, scale = 1]{7b}}}
%\vspace{0.5cm}
\end{minipage} \\
\vspace{0.8cm}
\begin{minipage}[t]{0.3 \textwidth}
\centering
\subfigure[\hspace{0.2cm} $s_0/(m_0 \, r) = 10^{-1} \, , \ \al = 1$]{
\label{fig:P3-Vaidya-r=6-T=025-P=025-dMdx=1e-04-alpha=+1-mu=1e-1}
\rotatebox{0}{\includegraphics[width = 6.6cm, height = 5.0cm, scale = 1]{7c}}}
%\vspace{0.5cm}
\end{minipage}%
\hspace{2.0cm}
\begin{minipage}[t]{0.3 \textwidth}
\centering
\subfigure[\hspace{0.2cm} $s_0/(m_0 \, r) = 10^{-1} \, , \ \al = -1$]{
\label{fig:P3-Vaidya-r=6-T=025-P=025-dMdx=1e-04-alpha=-1-mu=1e-1}
\rotatebox{0}{\includegraphics[width = 6.6cm, height = 5.0cm, scale = 1]{7d}}}
%\vspace{0.5cm}
\end{minipage}
\caption{\label{fig:P3-Vaidya-r=6-T=025-P=025-dMdx=1e-04} Azimuthal component $P^3(\tau)$ of the linear momentum for
$r = 6 M$ and $\hat{\th} = \hat{\ph} = \pi/4$ in the Vaidya background.
For Figs.~\ref{fig:P3-Vaidya-r=6-T=025-P=025-dMdx=1e-04-alpha=+1-mu=1e-2} and \ref{fig:P3-Vaidya-r=6-T=025-P=025-dMdx=1e-04-alpha=-1-mu=1e-2},
the change in magnitude is almost totally dominated by the first-order contribution in $\varepsilon$.
Figs.~\ref{fig:P3-Vaidya-r=6-T=025-P=025-dMdx=1e-04-alpha=+1-mu=1e-1} and \ref{fig:P3-Vaidya-r=6-T=025-P=025-dMdx=1e-04-alpha=-1-mu=1e-1},
in contrast, indicate that the second-order contribution in $\varepsilon$ dominates with a strongly unbounded increase in the amplitude
as $s_0/(m_0 \, r) = 10^{-1}$.}
\end{figure*}

\begin{figure*}
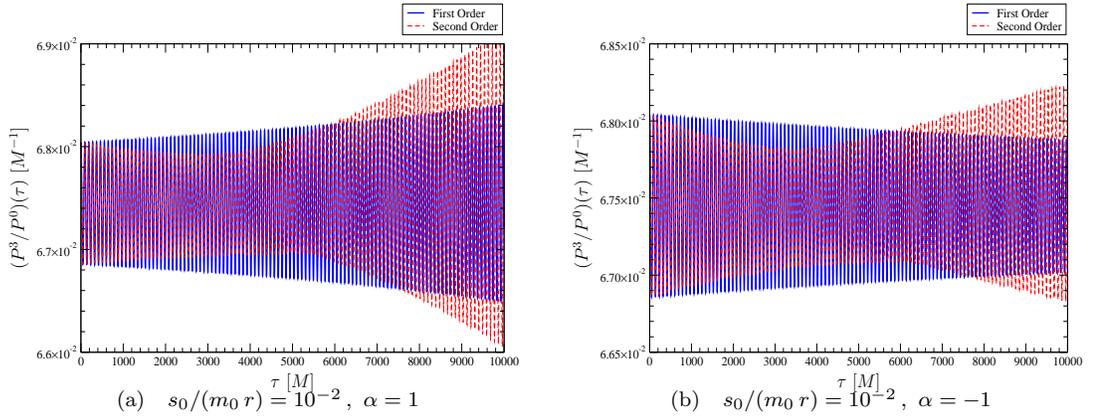

\psfrag{T}[tc][][1.8][0]{\Large $\tau \ [M]$}
\psfrag{P3/P0}[bc][][1.8][0]{\Large $(P^3/P^0)(\tau) \ [M^{-1}]$}
\begin{minipage}[t]{0.3 \textwidth}
\centering
\subfigure[\hspace{0.2cm} $s_0/(m_0 \, r) = 10^{-2} \, , \ \al = 1$]{
\label{fig:P3P0-Vaidya-r=6-T=025-P=025-dMdx=1e-04-alpha=+1-mu=1e-2}
\rotatebox{0}{\includegraphics[width = 6.6cm, height = 5.0cm, scale = 1]{8a}}}
%\vspace{0.5cm}
\end{minipage}%
\hspace{2.0cm}
\begin{minipage}[t]{0.3 \textwidth}
\centering
\subfigure[\hspace{0.2cm} $s_0/(m_0 \, r) = 10^{-2} \, , \ \al = -1$]{
\label{fig:P3P0-Vaidya-r=6-T=025-P=025-dMdx=1e-04-alpha=-1-mu=1e-2}
\rotatebox{0}{\includegraphics[width = 6.6cm, height = 5.0cm, scale = 1]{8b}}}
\end{minipage} \\
\hspace{2.0cm}
\caption{\label{fig:P3P0-Vaidya-r=6-T=025-P=025-dMdx=1e-04} Ratio of $P^3(\tau)$ to $P^0(\tau)$ in the Vaidya background for
$r = 6 M$ and $\hat{\th} = \hat{\ph} = \pi/4$.
Figs.~\ref{fig:P3P0-Vaidya-r=6-T=025-P=025-dMdx=1e-04-alpha=+1-mu=1e-2} and \ref{fig:P3P0-Vaidya-r=6-T=025-P=025-dMdx=1e-04-alpha=-1-mu=1e-2}
indicate that the second-order expression in $\varepsilon$ leads to a wider variation in the ratio than found in the
first-order expression only, irrespective of the given choice for $\al$.}
\end{figure*}

As in Paper I, examination of the linear momentum components $P^\mu (\tau)$, based on (\ref{P1-general}) and (\ref{P2-general}),
is in order.
This is given by Figures~\ref{fig:P1-Vaidya-r=6-T=025-P=025-dMdx=1e-04}--\ref{fig:P3-Vaidya-r=6-T=025-P=025-dMdx=1e-04}, which
show the radial, polar, and azimuthal components of the linear momentum in the Vaidya background, while the ratio
$P^3(\tau)/P^0(\tau)$ is displayed in Figure~\ref{fig:P3P0-Vaidya-r=6-T=025-P=025-dMdx=1e-04}.
Concerning the $O(\varepsilon^2)$ expression for $s_0/(m_0 \, r) = 10^{-2}$ and $\al = 1$,
Fig.~\ref{fig:P1-Vaidya-r=6-T=025-P=025-dMdx=1e-04-alpha=+1-mu=1e-2}
exhibits a slight contraction in the amplitude before outwardly expanding, with a similar behaviour shown in
Fig.~\ref{fig:P1-Vaidya-r=6-T=025-P=025-dMdx=1e-04-alpha=-1-mu=1e-2} for $\al = -1$.
Unlike the corresponding set of plots shown in Paper I, the difference between these two plots is virtually negligible.
When considering $s_0/(m_0 \, r) = 10^{-1}$ for $\al = 1$, the outward growth of the $O(\varepsilon^2)$ expression
for Fig.~\ref{fig:P1-Vaidya-r=6-T=025-P=025-dMdx=1e-04-alpha=+1-mu=1e-1} starts to become dominant just before $\tau = 1500 M$,
in accordance with the growth of $v(\tau)$ in Fig.~\ref{fig:v-Vaidya-r=6-T=025-P=025-dMdx=1e-04-alpha=+1-mu=1e-2}.
This is also true for Fig.~\ref{fig:P1-Vaidya-r=6-T=025-P=025-dMdx=1e-04-alpha=-1-mu=1e-1} when $\al = -1$ at around
$\tau = 2000 M$.

Figure~\ref{fig:P2-Vaidya-r=6-T=025-P=025-dMdx=1e-04} shows the polar component of the linear momentum, which indicates that while
the expression to first-order in $\varepsilon$ remains around zero on average, the expression to second-order is positive-valued.
For $s_0/(m_0 \, r) = 10^{-2}$, both Figs.~\ref{fig:P2-Vaidya-r=6-T=025-P=025-dMdx=1e-04-alpha=+1-mu=1e-2} and
\ref{fig:P2-Vaidya-r=6-T=025-P=025-dMdx=1e-04-alpha=-1-mu=1e-2} for $\al = 1$ and $\al = -1$, respectively, display a slight
positive-valued magnitude due to the $O(\varepsilon^2)$ expression for $P^2(\tau)$.
This behaves similarly to the result obtained in Paper I for the Kerr background, also indicating that
the spinning particle will no longer remain on the orbital plane after a sufficiently long time.
Furthermore, in terms of the choice for $\al$ there is essentially no difference in magnitude between these two plots
for $s_0/(m_0 \, r) = 10^{-2}$.
When considering $s_0/(m_0 \, r) = 10^{-1}$, as shown in Figs.~\ref{fig:P2-Vaidya-r=6-T=025-P=025-dMdx=1e-04-alpha=+1-mu=1e-1}
and \ref{fig:P2-Vaidya-r=6-T=025-P=025-dMdx=1e-04-alpha=-1-mu=1e-1}, the expression to second-order in $\varepsilon$
is more obviously non-zero compared to the first-order contribution, indicative of the orbital instabilities suggested by
Figs.~\ref{fig:v-Vaidya-r=6-T=025-P=025-dMdx=1e-04-alpha=+1-mu=1e-1} and \ref{fig:v-Vaidya-r=6-T=025-P=025-dMdx=1e-04-alpha=-1-mu=1e-1}.

The azimuthal component of the linear momentum is given by Figure~\ref{fig:P3-Vaidya-r=6-T=025-P=025-dMdx=1e-04},
where Figs.~\ref{fig:P3-Vaidya-r=6-T=025-P=025-dMdx=1e-04-alpha=+1-mu=1e-2}
and \ref{fig:P3-Vaidya-r=6-T=025-P=025-dMdx=1e-04-alpha=-1-mu=1e-2} refer to $\al = 1$ and $\al = -1$, respectively,
for $s_0/(m_0 \, r) = 10^{-2}$.
In both cases, the $O(\varepsilon^2)$ expression has very little impact on the overall plots.
Furthermore, it is evident that the last term in (\ref{P1-3}), which is linearly time-dependent, is most
likely responsible for the plots' slopes, due to an overall minus sign for this term.
When $s_0/(m_0 \, r) = 10^{-1}$, Figs.~\ref{fig:P3-Vaidya-r=6-T=025-P=025-dMdx=1e-04-alpha=+1-mu=1e-1}
and \ref{fig:P3-Vaidya-r=6-T=025-P=025-dMdx=1e-04-alpha=-1-mu=1e-1} indicate a strongly increasing amplitude
for the expression to second-order in $\varepsilon$, in similar fashion to the corresponding plots shown in Paper I.

The ratio $P^3(\tau)/P^0(\tau)$ is given by Figure~\ref{fig:P3P0-Vaidya-r=6-T=025-P=025-dMdx=1e-04} for $s_0/(m_0 \, r) = 10^{-2}$,
where Fig.~\ref{fig:P3P0-Vaidya-r=6-T=025-P=025-dMdx=1e-04-alpha=+1-mu=1e-2} describes $\al = 1$ and
Fig.~\ref{fig:P3P0-Vaidya-r=6-T=025-P=025-dMdx=1e-04-alpha=-1-mu=1e-2} refers to $\al = -1$.
Focussing on the expression to first-order in $\varepsilon$, it is interesting to note that the amplitude shows
a steady increase $(\al = 1)$ or decrease $(\al = -1)$, unlike the strictly constant amplitude for the corresponding
plots found in Paper I for the Kerr background.
This is due to the last term of (\ref{P1-3-P1-0}), which is linearly time-dependent with an overall positive sign.
Similarly to the plots of Paper I, when adding the second-order contribution,
the amplitude for the ratio slightly contracts before steadily growing in magnitude.

\begin{figure*}
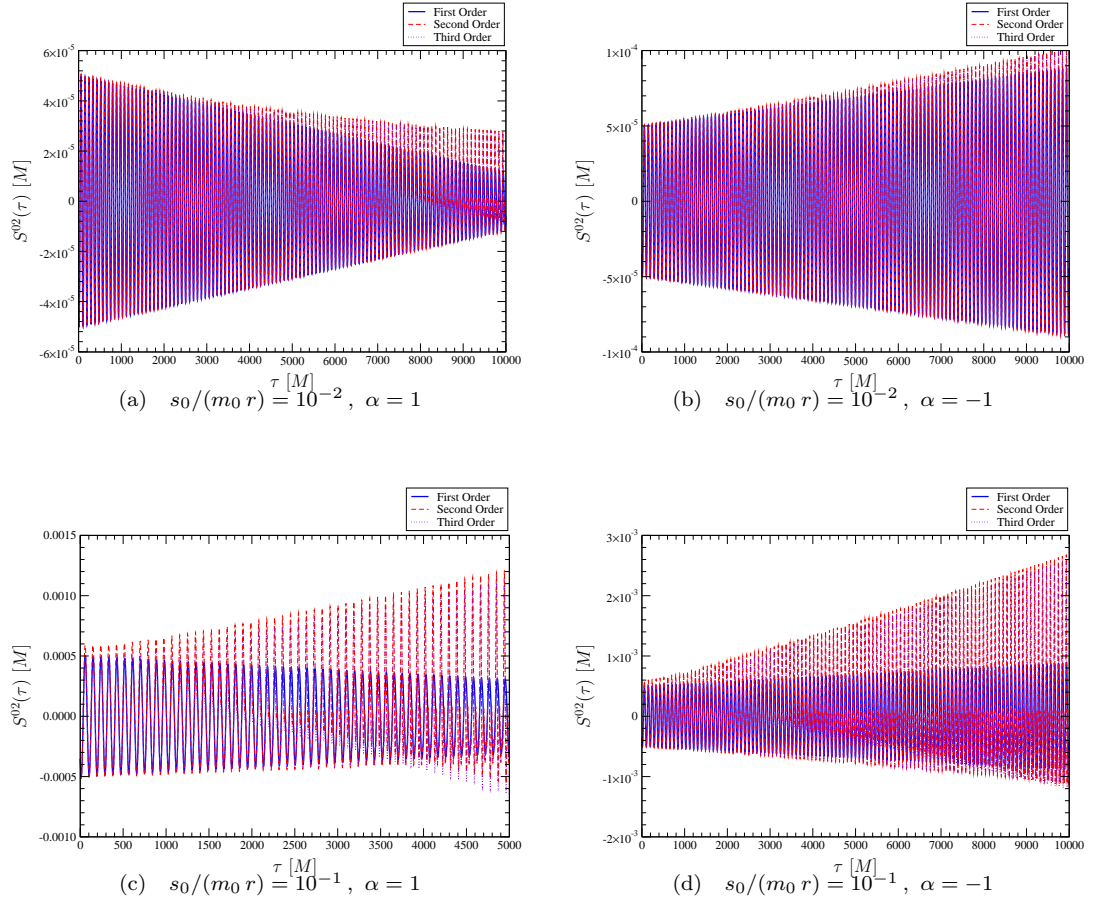

\psfrag{T}[tc][][1.8][0]{\Large $\tau \ [M]$}
\psfrag{S02}[bc][][1.8][0]{\Large $S^{02}(\tau) \ [M]$}
\begin{minipage}[t]{0.3 \textwidth}
\centering
\subfigure[\hspace{0.2cm} $s_0/(m_0 \, r) = 10^{-2} \, , \ \al = 1$]{
\label{fig:S02-Vaidya-r=6-T=025-P=025-dMdx=1e-04-alpha=+1-mu=1e-2}
\rotatebox{0}{\includegraphics[width = 6.6cm, height = 5.0cm, scale = 1]{9a}}}
%\vspace{0.5cm}
\end{minipage}%
\hspace{2.0cm}
\begin{minipage}[t]{0.3 \textwidth}
\centering
\subfigure[\hspace{0.2cm} $s_0/(m_0 \, r) = 10^{-2} \, , \ \al = -1$]{
\label{fig:S02-Vaidya-r=6-T=025-P=025-dMdx=1e-04-alpha=-1-mu=1e-2}
\rotatebox{0}{\includegraphics[width = 6.6cm, height = 5.0cm, scale = 1]{9b}}}
%\vspace{0.5cm}
\end{minipage} \\
\vspace{0.8cm}
\begin{minipage}[t]{0.3 \textwidth}
\centering
\subfigure[\hspace{0.2cm} $s_0/(m_0 \, r) = 10^{-1} \, , \ \al = 1$]{
\label{fig:S02-Vaidya-r=6-T=025-P=025-dMdx=1e-04-alpha=+1-mu=1e-1}
\rotatebox{0}{\includegraphics[width = 6.6cm, height = 5.0cm, scale = 1]{9c}}}
%\vspace{0.5cm}
\end{minipage}%
\hspace{2.0cm}
\begin{minipage}[t]{0.3 \textwidth}
\centering
\subfigure[\hspace{0.2cm} $s_0/(m_0 \, r) = 10^{-1} \, , \ \al = -1$]{
\label{fig:S02-Vaidya-r=6-T=025-P=025-dMdx=1e-04-alpha=-1-mu=1e-1}
\rotatebox{0}{\includegraphics[width = 6.6cm, height = 5.0cm, scale = 1]{9d}}}
%\vspace{0.5cm}
\end{minipage}
\caption{\label{fig:S02-Vaidya-r=6-T=025-P=025-dMdx=1e-04} The $S^{02}(\tau)$ component of the spin tensor for
$r = 6 M$ and $\hat{\th} = \hat{\ph} = \pi/4$ in the Vaidya background.
Fig.~\ref{fig:S02-Vaidya-r=6-T=025-P=025-dMdx=1e-04-alpha=+1-mu=1e-2} shows a modest decrease in amplitude for $s_0/(m_0 \, r) = 10^{-2}$
and $\al = 1$, with a corresponding modest increase in amplitude in Fig.~\ref{fig:S02-Vaidya-r=6-T=025-P=025-dMdx=1e-04-alpha=-1-mu=1e-2}
for $\al = -1$.
In contrast, both Figs.~\ref{fig:S02-Vaidya-r=6-T=025-P=025-dMdx=1e-04-alpha=+1-mu=1e-1} and
\ref{fig:S02-Vaidya-r=6-T=025-P=025-dMdx=1e-04-alpha=-1-mu=1e-1} show a significantly more pronounced amplitude increase for
$s_0/(m_0 \, r) = 10^{-1}$.}
\end{figure*}

As a final example, it is useful to examine one of the components of the spin tensor to illustrate its properties due
to spin-gravity interaction.
In keeping with Paper I for the sake of comparison, the $S^{02}(\tau)$ is chosen for study.
This is given by Figure~\ref{fig:S02-Vaidya-r=6-T=025-P=025-dMdx=1e-04}, expressed to third-order in $\varepsilon$.
Focussing on Figs.~\ref{fig:S02-Vaidya-r=6-T=025-P=025-dMdx=1e-04-alpha=+1-mu=1e-2} and
\ref{fig:S02-Vaidya-r=6-T=025-P=025-dMdx=1e-04-alpha=-1-mu=1e-2}, corresponding to $\al = 1$ and $\al = -1$ for
$s_0/(m_0 \, r) = 10^{-2}$, it is clear that the direction of radiation flow into or away from the black hole impacts upon
the amplitude of $S^{02}(\tau)$ about zero, and that the expression to third-order in $\varepsilon$ deviates
slightly from that due to the second-order contribution alone.
This is in contrast to the corresponding set of plots in Paper I, which show no significant difference between
the expressions to second- and third-order in $\varepsilon$.
When considering $s_0/(m_0 \, r) = 10^{-1}$, as shown in Figs.~\ref{fig:S02-Vaidya-r=6-T=025-P=025-dMdx=1e-04-alpha=+1-mu=1e-1} and
\ref{fig:S02-Vaidya-r=6-T=025-P=025-dMdx=1e-04-alpha=-1-mu=1e-1} for $\al = 1$ and $\al = -1$, respectively,
the large outward growth of the amplitude due to the expressions to second- and third-order in $\varepsilon$
is a further response to the orbital instability experienced by the spinning particle, as reflected in
Figure~\ref{fig:v-Vaidya-r=6-T=025-P=025-dMdx=1e-04}.

%%%%%%%%%%%%%%%%%%%%%

\section{Conclusion}
\label{sec:6}

This paper is an application of the generalized CMP approximation approach to the Mathisson-Papapetrou-Dixon equations
of motion of a spinning point particle in orbit around a spherical black hole in the presence of
radially inflowing and outflowing radiation, as described by the Vaidya metric.
When compared to a similar analysis performed for orbital motion around a Kerr black hole, as described in Paper I \cite{Singh2},
all relevant computations, including the ``radiative corrections'' to the particle's squared mass and spin magnitudes,
have nontrivial properties due to the explicit time-dependence of the space-time background.
It is somewhat ironic that, while the Vaidya metric is arguably much simpler in form than the Kerr metric,
its time-dependence leads to much more complicated mathematical structure in the generalized CMP approximation
than displayed in the previous application.
As with Paper I, some numerical analysis is performed to illustrate conditions for the emergence of
instabilities in the particle's orbit, with the suggestion that the M{\o}ller radius needs to remain positive-valued
in order to avoid the transition away from stable motion.

The next step in this exploration is to obtain the perturbed orbit from the results determined with the generalized
CMP approximation, while also incorporating the effects of gravitational radiation within the process.
As noted in Paper I, this generalization introduces conceptual and technical challenges that are still not
clearly understood at present.
This consideration will be deferred to a future publication once these challenges are overcome.
Another possibility is to explore a many-body interaction with spin incorporated through the generalized
CMP approximation.
To do this requires understanding the expected tidal and spin-spin interactions to be found when dealing
with such a problem, which have a separate set of conceptual and technical challenges to consider.
Nonetheless, both the work presented here and in Paper I illustrate the potential that comes from this line of research.

\begin{acknowledgments}
The author is thankful to Prof. Nader Mobed of the University of Regina for financial and moral support towards the
completion of this project.
\end{acknowledgments}

\begin{appendix}

\section{Solution for the First-Order Spin Shift}
\label{appendix:s1-bar^2}
\renewcommand{\theequation}{A.\arabic{equation}}
\setcounter{subsection}{0}
\setcounter{equation}{0}

To solve for $\bar{s}_1^2$ directly from (\ref{DS1/dt=0}) requires use of the
spin condition constraint equation (\ref{s.p=0}) for $j = 1$ \cite{Singh2}, which leads to
\be
A^\mu \, S^{(1)}_{\mu \nu} - B_\nu & = & 0 \, ,
\label{S1-4-equations}
\ee
where
\begin{subequations}
\label{A-B=}
\be
A^\mu & \equiv & P^\mu_{(0)} \, ,
\label{A-P}
\nl
\nn
B_\nu & \equiv & -P^\mu_{(1)} \, S^{(0)}_{\mu \nu} \, .
\label{B-P}
\ee
\end{subequations}
With (\ref{S1-4-equations}), the $S^{(1)}_{0j}$ components can be solved algebraically in terms of the
purely spatial components $S^{(1)}_{ij}$, %in the form
%%
%\begin{subequations}
%\label{S-0j=}
%\be
%S^{(1)}_{01} & = & {1 \over A^0} \lt[A^2 \, S^{(1)}_{12} - A^3 \, S^{(1)}_{31} + B_1\rt] \, ,
%\label{S-01}
%\nl
%S^{(1)}_{02} & = & {1 \over A^0} \lt[A^3 \, S^{(1)}_{23} - A^1 \, S^{(1)}_{12} + B_2\rt] \, ,
%\label{S-02}
%\nl
%S^{(1)}_{03} & = & {1 \over A^0} \lt[A^1 \, S^{(1)}_{31} - A^2 \, S^{(1)}_{23} + B_3\rt] \, .
%\label{S-03}
%\ee
%\end{subequations}
%%
while the three spatial components are determined as solutions to the matrix differential equation
\be
{D S^{(1)}_{ij} \over \d \tau} & = &
{\d S^{(1)}_{ij} \over \d \tau} + 2 \, u_{(0)}^\al \, \Gm^\bt{}_{\al[i} \, S^{(1)}_{j]\bt} \ = \ 0 \, .
\label{DS1/dt=0-ij}
\ee
In explicit component form, (\ref{DS1/dt=0-ij}) is equivalent to
\begin{subequations}
\label{dS1-matrix=}
\be
{\d S^{(1)}_{12}(\tau) \over \d \tau} + {1 \over 2} \, \al^{ij} \, S^{(1)}_{ij}(\tau) \ = \ \dl_{12}(\tau) \, ,
\label{dS1-12=}
\nl
{\d S^{(1)}_{23}(\tau) \over \d \tau} + {1 \over 2} \, \bt^{ij} \, S^{(1)}_{ij}(\tau) \ = \ \dl_{23}(\tau) \, ,
\label{dS1-23=}
\nl
{\d S^{(1)}_{31}(\tau) \over \d \tau} + {1 \over 2} \, \gm^{ij} \, S^{(1)}_{ij}(\tau) \ = \ \dl_{31}(\tau) \, ,
\label{dS1-31=}
\ee
\end{subequations}
where $\al^{ij}$, $\bt^{ij}$, and $\gm^{ij}$ are antisymmetric spatial tensors, which may be $\tau$-dependent for
a given choice of metric.
For the Vaidya metric given by (\ref{Vaidya-metric}), with (\ref{dM-approx}) chosen for $\Dl M$ and recalling (\ref{Q-cs}), it is shown that
\begin{widetext}
\begin{subequations}
\label{al-bt-gm-Vaidya=}
\be
\al^{12}(\tau) & \approx & \OmK \, \lt|\d \lt(\Dl M\rt) \over \d \xi \rt| \lt\{{2 \, \OmK \, \tau \over N \, A^3}
+ {\al \over N^3 \, A^2} \, {\lt[A^2 - 2 \, (r^2 \, \OmK^2) \rt] \over r \, \OmK} \, \lt[1 - \cos (2 \, \OmK \, \tau) \rt] \rt\} \, ,
\nl
\al^{23}(\tau) & \approx & {\OmK \, N \over r \, A^2} \lt[
1 + \al \, \lt|\d \lt(\Dl M\rt) \over \d \xi \rt| \lt\{ {(r \, \OmK) \over N^6} \, \sin (2 \, \OmK \, \tau)
- {1 \over 2 \, N^5} \lt[\lt(2 \, r \, \OmK - 1\rt) \sin (2 \, \OmK \, \tau) + 2 \, \OmK \, \tau \rt] \rt\} \rt] \, ,
\nl
\bt^{12}(\tau) & \approx & - {A^2 \over N} \, (r \, \OmK) \lt[1 %\rt.
%\nn
%&  &{} + \lt.
+ \al \, \lt|\d \lt(\Dl M\rt) \over \d \xi \rt|
\lt\{{(r \, \OmK) \over N^6} \, \sin (2 \, \OmK \, \tau)
- {1 \over 2 \, N^5} \, \lt[\lt(2 \, r \, \OmK - 1\rt) \sin (2 \, \OmK \, \tau) + 2 \, \OmK \, \tau \rt] \rt\} \rt] \, ,
\nn
\nl
\bt^{23}(\tau) & \approx & - 2 \, \OmK \lt|\d \lt(\Dl M\rt) \over \d \xi \rt| \lt\{{\OmK \, \tau \over N \, A^3}
- {\al \over N^3} \, {\lt[1 - \cos (2 \, \OmK \, \tau) \rt] \over r \, \OmK}  \rt\} \, ,
\nl
\gm^{31}(\tau) & \approx & {\al \over r} \, \lt|\d \lt(\Dl M\rt) \over \d \xi \rt| \lt\{
{1 \over N^3 \, A^2} \lt[A^2 - 2 \, (r^2 \, \OmK^2)\rt] \lt[1 - \cos (2 \, \OmK \, \tau) \rt] \rt\} \, ,
\nl
\al^{31} & = & \bt^{31} \ = \ \gm^{12} \ = \ \gm^{23} \ = \ 0 \, ,
\ee
\end{subequations}
\end{widetext}
and
\newpage
\begin{widetext}
\begin{subequations}
\label{dl-Vaidya=}
\be
\dl_{12}(\tau) & = & -{1 \over 4} \, {m_0 \, r \over N^2 \, A} \,
\lt(s_0 \over m_0 \, r \rt)^2  \, (r^4 \, \OmK^4) \lt[3 \, Q_{\rm s}^- (\OmK \, \tau \, ,  2 \, \hat{\th} \, , \hat{\ph}) \rt.
\nn
&  &{} - {\al \over A} \, \lt|\d \lt(\Dl M\rt) \over \d \xi \rt|
\lt\{ {1 \over N^6} \, (r^2 \, \OmK^2) \, \lt[Q_{\rm c}^- (3 \, \OmK \, \tau \, ,  2 \, \hat{\th} \, , \hat{\ph})
- 3 \, Q_{\rm c}^- (\OmK \, \tau \, ,  2 \, \hat{\th} \, , -\hat{\ph}) \rt] \rt.
\nn
&  &{} - \lt. \lt.
{6 \over N^5} \, (r \, \OmK) \, Q_{\rm s}^- (\OmK \, \tau \, ,  2 \, \hat{\th} \, , \hat{\ph}) \, (\OmK \, \tau) \rt\} \rt] \, ,
\nl
\dl_{23}(\tau) & = & -{3 \over 2} \, {m_0 \, r^2 \over N^2 \, A^2} \, \lt|\d \lt(\Dl M\rt) \over \d \xi \rt| \,
\lt(s_0 \over m_0 \, r \rt)^2 \, (r^2 \, \OmK^2) \, Q_{\rm s}^- (\OmK \, \tau \, ,  2 \, \hat{\th} \, , \hat{\ph}) \, (\OmK \, \tau) \, ,
\nl
\dl_{31}(\tau) & = & {1 \over 8} \, {m_0 \, r \over N^3} \,
\lt(s_0 \over m_0 \, r \rt)^2 \, (r^4 \, \OmK^4) \lt[ 3 \, Q_{\rm s}^+ (2 \, \OmK \, \tau \, ,  2 \, \hat{\th} \, , 2 \, \hat{\ph})
- 3 \, \sin (2 \, \OmK \, \tau - 2 \, \hat{\ph})  \rt.
\nn
&  &{} - {\al \over A} \, \lt|\d \lt(\Dl M\rt) \over \d \xi \rt|
\lt\{{1 \over N^6} \, (r^2 \, \OmK^2) \lt[Q_{\rm c}^+ (4 \, \OmK \, \tau \, ,  2 \, \hat{\th} \, , 2 \, \hat{\ph})
- 2 \, \cos (4 \, \OmK \, \tau - 2 \, \hat{\ph})
\rt. \rt.
\nn
&  &{}
+ \lt. 12 \, Q_{\rm c}^+ (2 \, \OmK \, \tau \, ,  2 \, \hat{\th} \, , 0) + 8 \, \cos (2 \, \OmK \, \tau)
+ 3 \, Q_{\rm c}^+ (0 \, ,  2 \, \hat{\th} \, , 2 \, \hat{\ph}) - 6 \, \cos (2 \, \hat{\th}) \rt]
\nn
&  &{} + \lt. \lt. \, {6 \over N^5} \, (r \, \OmK) \lt[Q_{\rm s}^+ (2 \, \OmK \, \tau \, ,  2 \, \hat{\th} \, , 2 \, \hat{\ph})
- 2 \, \sin (2 \, \OmK \, \tau - 2 \, \hat{\ph}) \rt] (\OmK \, \tau) \rt\} \rt] \, ,
\ee
\end{subequations}
\end{widetext}
which are substituted into (\ref{dS1-matrix=}) to solve for $S^{(1)}_{\mu \nu}(\tau)$.

\section{Orthonormal Tetrad Frame}
\label{appendix:tetrad-frame}
\renewcommand{\theequation}{B.\arabic{equation}}
\setcounter{subsection}{0}
\setcounter{equation}{0}

This Appendix outlines the derivation of the orthonormal tetrad frame $\lm^\mu{}_{\Cal}$ for orbital motion
in the Vaidya space-time background, following the approach given elsewhere \cite{Chicone1,Mashhoon2}, with the assumption
that $\Dl M/M_0~\ll~1$.
To proceed, recall from (\ref{Vaidya-metric}) the Vaidya metric $g_{\mu \nu}$ in $\lt(t, r, \th, \ph\rt)$ co-ordinates and
consider the orthonormal tetrad frame $\Lm^\mu{}_{\Cal}$ corresponding to fundamental static observers in the Vaidya background,
subject to
\be
\eta_{\Cal \Cbt} & = & g_{\mu \nu} \, \Lm^\mu{}_{\Cal} \, \Lm^\nu{}_{\Cbt} \, .
\label{orthonormal-tetrad-condition}
\ee
The tetrad set for static observers is assumed to take the form
\be
\Lm^\mu{}_{\hat{0}} & = & \lt(Z_0, 0, 0, 0 \rt),
\label{tetrad-0-static}
\nl
\Lm^\mu{}_{\hat{1}} & = & \lt(Z_1, Z, 0, 0\rt),
\label{tetrad-1-static}
\nl
\Lm^\mu{}_{\hat{2}} & = & \lt(0, 0, {1 \over r}, 0\rt),
\label{tetrad-2-static}
\nl
\Lm^\mu{}_{\hat{3}} & = & \lt(0, 0, 0, {1 \over r \, \sin \th} \rt),
\label{tetrad-3-static}
\ee
where satisfying (\ref{orthonormal-tetrad-condition}) leads to
\be
Z & = & \lt(A^2 - {2 \Dl M \over r} \rt)^{1/2} \, ,
\label{Z=}
\nl
Z_0 & = & Z^{-1} \, ,
\label{Z0=}
\nl
Z_1 & = & {2 \, \al \over A^2 \, Z} \, {\Dl M \over r} \, .
\label{Z1=}
\ee

The main idea is to Lorentz boost $\Lm^\mu{}_{\Cal} \rightarrow \tilde{\Lm}^\mu{}_{\Cal}$ with speed $\tilde{\bt}$
to its location on the orbit, set $\tilde{\Lm}^\mu{}_{\ze} = \lm^\mu{}_{\ze}$, and determine $\lm^\mu{}_{\hat{\jmath}}$
accordingly to accommodate the parallel transport condition $D \lm^\mu{}_{\Cal}/ \d \tau = 0$.
Because $M$ changes with proper time $\tau$ along the null co-ordinate $\xi$ according to (\ref{xi}), it follows that the frame
will be boosted along at least the azimuthal and radial directions.
This implies that the orbit is strictly no longer circular and introduces some complications in determining $\lm^\mu{}_{\ze}$.
However, it is shown in Appendix~\ref{appendix:zeroth-component} that, for $\th = \pi/2$,
\be
\lm^\mu{}_{\ze} & = & {1 \over K} \lt({E \over A^2} + \Dl u^0 \, , \Dl u^1 \, , 0 \, , {1 \over r} \lt({L \over r} + r \, \Dl u^3\rt) \rt),
\nn
\label{lm0=}
\ee
where
\be
K & = & \lt[Z^2 \lt({E \over A^2} + \Dl u^0\rt)^2
- {4 \, \al \over A^2} \, {\Dl M \over r} \lt({E \over A^2} + \Dl u^0\rt) \Dl u^1 \rt.
\nn
&  &{} - \lt. {1 \over A^4} \lt(A^2 + {2 \Dl M \over r}\rt) \lt(\Dl u^1\rt)^2 - \lt({L \over r} + r \, \Dl u^3\rt)^2 \rt]^{1/2} \,
\nn
\label{K}
\ee
is a normalization condition for $\lm^\mu{}_{\ze}$, and $\Dl u^0$, $\Dl u^1$, and $\Dl u^3$
are contributions to the tetrad frame's overall four-velocity determined in Appendix~\ref{appendix:zeroth-component}.
It is straightforward to confirm that $K \rightarrow 1$ and
$\lm^\mu{}_{\ze} \rightarrow \lm^\mu{}_{\hat{0} \, (\rm Sch)}$ when $\Dl M \rightarrow 0$, as expected.

Applying a Lorentz transformation to $\Lm^\mu{}_{\Cal}$ and given (\ref{lm0=}), it is true that
\be
\tilde{\Lm}^\mu{}_{\ze} & = & \tilde{\gm} \lt[\Lm^\mu{}_{\ze} + \tilde{\bt} \lt(\cos \tilde{\al} \, \Lm^\mu{}_{\tr}
+ \sin \tilde{\al} \, \Lm^\mu{}_{\on}\rt) \rt], \qquad
\label{Lm0-boost=}
\nl
\tilde{\Lm}^\mu{}_{\on} & = & \cos \tilde{\al} \, \Lm^\mu{}_{\on} - \sin \tilde{\al} \, \Lm^\mu{}_{\tr} \, ,
\label{Lm1-boost=}
\nl
\tilde{\Lm}^\mu{}_{\tw} & = & \Lm^\mu{}_{\tw} \, ,
\label{Lm2-boost=}
\nl
\tilde{\Lm}^\mu{}_{\tr} & = & \tilde{\gm} \lt[\lt(\cos \tilde{\al} \, \Lm^\mu{}_{\tr}
+ \sin \tilde{\al} \, \Lm^\mu{}_{\on}\rt) + \tilde{\bt} \, \Lm^\mu{}_{\rm \ze}\rt],
\label{Lm3-boost=}
\ee
where $\tilde{\gm} = 1/\sqrt{1 - \tilde{\bt}^2}$ is the Lorentz factor.
It follows that identification of (\ref{Lm0-boost=}) with (\ref{lm0=}) leads to
\be
\tilde{\gm} & = & {Z \over K} \lt[{E \over A^2} + \Dl u^0
- {2 \, \al \over A^2 \, Z^2} \, {\Dl M \over r} \, \Dl u^1 \rt],
\label{gm=}
\nl
\tilde{\bt} & = & {1 \over K \, \tilde{\gm}} \lt[{\lt(\Dl u^1\rt)^2 \over Z^2} + \lt({L \over r} + r \, \Dl u^3\rt)^2\rt]^{1/2} \, ,
\qquad
\label{bt=}
\nl
\tilde{\al} & = & \tan^{-1} \lt[\lt({L \over r} + r \, \Dl u^3\rt)^{-1} \, {\Dl u^1 \over Z} \rt].
\label{al=}
\ee

Finally, the spatial triad need to be rotated back by $\OmK \, \tau $ to reflect the parallel propagation
of the tetrad along the orbit \cite{Chicone1}, such that
\be
\lm^\mu{}_{\on} & = & \tilde{\Lm}^\mu{}_{\on} \, \cos \lt(\OmK \, \tau \rt) - \tilde{\Lm}^\mu{}_{\tr} \, \sin \lt(\OmK \, \tau \rt),
\label{lm1-formal}
\nl
\lm^\mu{}_{\tw} & = & \tilde{\Lm}^\mu{}_{\tw} \, ,
\label{lm2-formal}
\nl
\lm^\mu{}_{\tr} & = & \tilde{\Lm}^\mu{}_{\on} \, \sin \lt(\OmK \, \tau \rt) + \tilde{\Lm}^\mu{}_{\tr} \, \cos \lt(\OmK \, \tau \rt).
\label{lm3-formal}
\ee
This leads to the final expression for the orthonormal tetrad frame
\be
\lm^\mu{}_{\ze} & = & {1 \over K} \lt({E \over A^2} + \Dl u^0 \, , \, \Dl u^1 \, , \, 0 \, , \, {1 \over r} \lt({L \over r} + r \, \Dl u^3\rt) \rt),
\nn
\label{lm0-final}
\nl
\lm^\mu{}_{\on} & = & \lt({1 \over Z} \lt[-\tilde{\gm} \, \tilde{\bt} \, \sin \lt(\OmK \, \tau \rt)
+ {2 \, \al \over A^2} \, {\Dl M \over r} \, F_{\rm c}^+(\OmK \, \tau , \tilde{\al})\rt],  \rt.
\nn
&  &{} \lt. Z \, F_{\rm c}^+(\OmK \, \tau , \tilde{\al}) \, , \, 0 \, , \, {1 \over r} \, F_{\rm s}^-(\OmK \, \tau , \tilde{\al}) \rt),
\label{lm1-final}
\nl
\lm^\mu{}_{\tw} & = & \lt(0 \, , \, 0 \, , \, {1 \over r} \, , \, 0 \rt),
\label{lm2-final}
\nl
\lm^\mu{}_{\tr} & = & \lt({1 \over Z} \lt[\tilde{\gm} \, \tilde{\bt} \, \cos \lt(\OmK \, \tau \rt)
+ {2 \, \al \over A^2} \, {\Dl M \over r} \, F_{\rm s}^+(\OmK \, \tau , \tilde{\al})\rt],  \rt.
\nn
&  &{} \lt. Z \, F_{\rm s}^+(\OmK \, \tau , \tilde{\al}) \, , \, 0 \, , \, -{1 \over r} \, F_{\rm c}^-(\OmK \, \tau , \tilde{\al}) \rt),
\label{lm3-final}
\ee
where
\be
F_{\rm c}^\pm(\OmK \, \tau , \tilde{\al}) & = & {1 \over 2} \lt(1 - \tilde{\gm}\rt) \cos \lt(\OmK \, \tau  - \tilde{\al}\rt)
\nn
&  &{}
\pm {1 \over 2} \lt(1 + \tilde{\gm}\rt) \cos \lt(\OmK \, \tau  + \tilde{\al}\rt), \qquad
\label{Fc=}
\nl
F_{\rm s}^\pm(\OmK \, \tau , \tilde{\al}) & = & {1 \over 2} \lt(1 - \tilde{\gm}\rt) \sin \lt(\OmK \, \tau  - \tilde{\al}\rt)
\nn
&  &{}
\pm {1 \over 2} \lt(1 + \tilde{\gm}\rt) \sin \lt(\OmK \, \tau  + \tilde{\al}\rt).
\label{Fs=}
\ee
As a final consistency check, it is straightforward to verify that
$\lm^\mu{}_{\tr} \lt(\OmK \, \tau  + {\pi \over 2}\rt) = \lm^\mu{}_{\on} \lt(\OmK \, \tau \rt)$, and that
(\ref{lm0-final})--(\ref{lm3-final}) reduce to (\ref{tetrad-0-sch})--(\ref{tetrad-3-sch}) in the limit as $\Dl M \rightarrow 0$.

\section{Zeroth Component of the Tetrad Frame}
\label{appendix:zeroth-component}
\renewcommand{\theequation}{C.\arabic{equation}}
\setcounter{subsection}{0}
\setcounter{equation}{0}

This Appendix outlines the method to determine the zeroth component $\lm^\mu{}_{\ze}$ of the orthonormal tetrad frame
in the Vaidya space-time background.
Suppose that $\lm^\mu{}_{\ze} = \bar{\Lm}^\mu{}_{\ze}/K$, where
\be
\bar{\Lm}^\mu{}_{\ze} & = & \lm^\mu{}_{\hat{0} \, (\rm Sch)} + \Dl \lm^\mu{}_{\hat{0}} \, ,
\label{tetrad-0-bar}
\nl
\Dl \lm^\mu{}_{\hat{0}} & = & \lt(\Dl u^0, \Dl u^1, \Dl u^2, \Dl u^3 \rt),
\label{Delta-tetrad-0}
\ee
and
\be
K & \equiv & \lt(-g_{\mu \nu} \, \bar{\Lm}^\mu{}_{\ze} \, \bar{\Lm}^\nu{}_{\ze} \rt)^{1/2} \, .
\label{K=}
\ee
From (\ref{Vaidya-metric}), it is possible to identify the metric connection as
\be
\Gm^\mu{}_{\al \bt} & = & \Gm^\mu{}_{\al \bt \, (\rm Sch)} + \Dl \Gm^\mu{}_{\al \bt} \, ,
\label{metric-connection}
\ee
where $\Dl \Gm^\mu{}_{\al \bt}$ represents the contributions dependent on $\Dl M$.
Then it follows from requiring $D \lm^\mu{}_{\ze}/ \d \tau = 0$ that
\be
{\d \over \d \tau} \lt(\Dl \lm^\mu{}_{\ze}\rt) + \tilde{P}^\mu{}_\al \lt(\Dl \lm^\al{}_{\ze}\rt) & = & \tilde{Q}^\mu \, ,
\label{parallel-transport-Dl-lm}
\ee
where
\be
\tilde{P}^\mu{}_\al & = & 2 \, \Gm^\mu{}_{\al \bt \, (\rm Sch)} \, \lm^\bt{}_{\hat{0} \, (\rm Sch)} \, ,
\label{P=}
\nl
\tilde{Q}^\mu & = & - \lt(\Dl \Gm^\mu{}_{\al \bt}\rt) \, \lm^\al{}_{\hat{0} \, (\rm Sch)} \, \lm^\bt{}_{\hat{0} \, (\rm Sch)} \, .
\label{Q=}
\ee
After specifying $\th = \pi/2$ for the orbital plane, it is shown that $\d \lt(\Dl \lm^2{}_{\ze}\rt)/ \d \tau = \tilde{Q}^2 = 0$,
which implies that $\Dl \lm^2{}_{\ze}$ is a constant that can be set to zero.
This leads to the column vector differential equation
\begin{widetext}
\be
{\d \over \d \tau}
\lt(\begin{array}{c}
\Dl \lm^0{}_{\ze} \\ \\
\Dl \lm^1{}_{\ze} \\ \\
\Dl \lm^3{}_{\ze}
\end{array} \rt) +
\lt(\begin{array}{ccc}
0 & {2E \, M_0 \over r^2 \, A^4} & 0 \\ \\
{2E \, M_0 \over r^2} & 0 & - {2 A^2 \, L \over r} \\ \\
0 & {2L \over r^3} & 0
\end{array} \rt)
\lt(\begin{array}{c}
\Dl \lm^0{}_{\ze} \\ \\
\Dl \lm^1{}_{\ze} \\ \\
\Dl \lm^3{}_{\ze}
\end{array} \rt)
& = &
\lt(\begin{array}{c}
\tilde{Q}^0 \\ \\
\tilde{Q}^1 \\ \\
\tilde{Q}^3
\end{array} \rt)
\label{parallel-transport-matrix}
\ee
\end{widetext}
to solve, where
\be
\tilde{Q}^0 & = & {\al \, E^2 \over r A^6} \lt[\lt|\d \lt(\Dl M\rt) \over \d \xi \rt| - 2 \lt({\Dl M \over r}\rt)^2 \rt],
\label{Q0=}
\nl
\tilde{Q}^1 & = & -{2 L^2 \over r^3} \, {\Dl M \over r} \, ,
\label{Q1=}
\nl
\tilde{Q}^3 & = & 0 \, .
\label{Q3=}
\ee

In matrix notation, (\ref{parallel-transport-matrix}) is represented as
\be
{\d \over \d \tau} \lt(\Dl \vec{\lm}{}_{\ze}\rt) + {\cal \tilde{P}} \, \lt(\Dl \vec{\lm}{}_{\ze}\rt) & = & \vec{\cal \tilde{Q}} \, .
\label{parallel-transport-Dl-lm-vector}
\ee
The normal mode expression for $\Dl \vec{\lm}{}_{\ze}$ is determined by finding an invertible matrix ${\cal C}$ that is constant in $\tau$,
such that for $\Dl \vec{\lm}{}_{\ze}' = {\cal C}^{-1} \, \Dl \vec{\lm}{}_{\ze}$ and $\vec{\cal \tilde{Q}}' = {\cal C}^{-1} \, \vec{\cal \tilde{Q}}$,
\be
{\d \over \d \tau} \lt(\Dl \vec{\lm}{}_{\ze}'\rt) + {\cal \tilde{P}}' \, \lt(\Dl \vec{\lm}{}_{\ze}'\rt) & = & \vec{\cal \tilde{Q}}' \, ,
\label{parallel-transport-Dl-lm-vector-diag}
\ee
where ${\cal \tilde{P}}' = {\cal C}^{-1} \, {\cal \tilde{P}} \, {\cal C}$ is diagonalized.
It follows that
\be
{\cal \tilde{P}}' & = & \lt(\begin{array}{ccc}
0 & 0 & 0 \\ \\
0 & 2 \, i \lt(r \OmK\rt) & 0 \\ \\
0 & 0 & -2 \, i \lt(r \OmK\rt)
\end{array} \rt)
\label{P'=}
\ee
for
\begin{subequations}
\label{C-C-inv}
\be
{\cal C} & = & \lt(\begin{array}{ccc}
{A^2 \, L \over E \lt(r \OmK\rt)^2} & -i \, {E \lt(r \OmK\rt) \over A^4} & i \, {E \lt(r \OmK\rt) \over A^4} \\ \\
0 & 1 & 1 \\ \\
1 & -i \, {L \over r^2 \lt(r \OmK\rt)} & i \, {L \over r^2 \lt(r \OmK\rt)}
\end{array} \rt), \qquad
\nl
\nn
\nn
{\cal C}^{-1} & = &
\lt(\begin{array}{ccc}
{E \, L \over r^2} & 0 & -{E^2 \lt(r \OmK\rt) \over A^4} \\ \\
-i \, {E \lt(r \OmK\rt) \over 2} & {1 \over 2} & i \, {A^2 \, L \over 2 \lt(r \OmK\rt)} \\ \\
i \, {E \lt(r \OmK\rt) \over 2} & {1 \over 2} & -i \, {A^2 \, L \over 2 \lt(r \OmK\rt)}
\end{array} \rt),
\ee
\end{subequations}
leading to
\begin{widetext}
\be
{\d \over \d \tau}
\lt(\begin{array}{c}
\Dl \lm'^0{}_{\ze} \\ \\
\Dl \lm'^1{}_{\ze} \\ \\
\Dl \lm'^3{}_{\ze}
\end{array} \rt) +
\lt(\begin{array}{ccc}
0 & 0 & 0 \\ \\
0 & 2 \, i \lt(r \OmK\rt) & 0 \\ \\
0 & 0 & -2 \, i \lt(r \OmK\rt)
\end{array} \rt)
\lt(\begin{array}{c}
\Dl \lm'^0{}_{\ze} \\ \\
\Dl \lm'^1{}_{\ze} \\ \\
\Dl \lm'^3{}_{\ze}
\end{array} \rt)
& = &
\lt(\begin{array}{c}
\tilde{Q}'^0 \\ \\
\tilde{Q}'^1 \\ \\
\tilde{Q}'^3
\end{array} \rt),
\label{parallel-transport-matrix-diag}
\ee
\end{widetext}
where
\be
\tilde{Q}'^0 & = & \lt({E \, L \over r^2}\rt) \tilde{Q}^0 \, ,
\label{Q'0=}
\nl
\tilde{Q}'^1 & = & {1 \over 2} \lt[\tilde{Q}^1 - i \, E \lt(r \OmK\rt) \, \tilde{Q}^0\rt] \, ,
\label{Q'1=}
\nl
\tilde{Q}'^3 & = & \lt(\tilde{Q}'^1\rt)^* \, .
\label{Q'3=}
\ee
The solution to (\ref{parallel-transport-matrix-diag}) is then
\be
\Dl \lm'^0{}_{\ze} & = & \int_0^\tau \tilde{Q}'^0 (\tau') \, \d \tau' \, ,
\label{lm'0=}
\nl
\Dl \lm'^1{}_{\ze} & = & e^{-2 \, i \lt(\OmK \, \tau \rt)} \int_0^\tau e^{2 \, i \lt(\OmK \, \tau '\rt)} \, \tilde{Q}'^1 (\tau') \, \d \tau' \, ,
\qquad
\label{lm'1=}
\nl
\Dl \lm'^3{}_{\ze} & = & \lt(\Dl \lm'^1{}_{\ze}\rt)^* \, ,
\label{lm'3=}
\ee
which from $\Dl \vec{\lm}{}_{\ze} = {\cal C} \, \Dl \vec{\lm}'_{\ze}$ and (\ref{Delta-tetrad-0}) leads to
\be
\Dl u^0 & = & {A^2 \, L \over E \lt(r \OmK\rt)^2} \, \lt(\Dl \lm'^0{}_{\ze}\rt)
+ {2 E \lt(r \OmK\rt) \over A^4} \, {\rm Im} \lt(\Dl \lm'^1{}_{\ze}\rt) \, ,
\nn
\label{u0=}
\nl
\Dl u^1 & = & 2 \, {\rm Re} \lt(\Dl \lm'^1{}_{\ze}\rt) \, ,
\label{u1=}
\nl
\Dl u^2 & = & 0 \, ,
\label{u2=}
\nl
\Dl u^3 & = & \Dl \lm'^0{}_{\ze} + {2 L \over r^2 \lt(r \OmK\rt)} \, {\rm Im} \lt(\Dl \lm'^1{}_{\ze}\rt) \, .
\label{u3=}
\ee
%

%To eventually obtain $\lm^\mu{}_{\Cal}$ from $\Lm^\mu{}_{\Cal}$ requires first
\section{Fermi-Frame Riemann Tensor Components}
\label{appendix:riemann-frame-components}
\renewcommand{\theequation}{D.\arabic{equation}}
\setcounter{subsection}{0}
\setcounter{equation}{0}

Given that the nonzero Riemann tensor components for the Vaidya metric are
\be
R_{0101} & = & -{2 \over r^3} \lt(M_0 + \Dl M\rt),
\nl
R_{0202} & = & {Z^2 \over r} \, \lt(M_0 + \Dl M\rt) + \lt|\d \lt(\Dl M\rt) \over \d \xi \rt|,
\nl
R_{0212} & = & {\al \over A^2} \lt[-2\lt(M_0 + \Dl M\rt) {\Dl M \over r^2} + \lt|\d \lt(\Dl M\rt) \over \d \xi \rt| \rt],
\nl
R_{0303} & = & R_{0202} \, \sin^2 \th \, ,
\nl
R_{0313} & = & R_{0212} \, \sin^2 \th \, ,
\nl
R_{1212} & = & {1 \over A^4} \lt[-{1 \over r} \lt(A^2 + {2 \Dl M \over r} \rt)\lt(M_0 + \Dl M\rt) + \lt|\d \lt(\Dl M\rt) \over \d \xi \rt| \rt],
\nn
\nl
R_{1313} & = & R_{1212} \, \sin^2 \th \, ,
\nl
R_{2323} & = & 2 \, r \lt(M_0 + \Dl M\rt) \sin^2 \th \, ,
\ee
the nonzero components of the Riemann curvature tensor ${}^F{}R_{\Cmu \Cnu \Cal \Cbt}$ in the Fermi frame are listed as follows:
\newpage
\newpage
\begin{widetext}
\be
{}^F{}R_{\ze \on \ze \on} & = & {\tilde{\bt}^2 \, \tilde{\gm}^2 \over K^2 \, Z^2} \lt[\lt(\Dl u^1\rt)^2 R_{0101}
+ \lt({L \over r^2} + \Dl u^3 \rt)^2 R_{0303} \rt] \sin^2 \lt(\OmK \, \tau \rt)
\nn
\nn
\nn
&  &{} + {2 \, \tilde{\bt} \, \tilde{\gm} \over K^2} \lt\{ \lt[
\lt({E \over A^2} + \Dl u^0 - {2 \, \al \over A^2 \, Z^2} \, {\Dl M \over r} \, \Dl u^1 \rt) \Dl u^1 \, R_{0101} \rt. \rt.
\nn
\nn
\nn
&  &{} - \lt.
\lt({L \over r^2} + \Dl u^3 \rt)^2 \lt(R_{0313} + {2 \, \al \over A^2 \, Z^2} \, {\Dl M \over r} \, R_{0303} \rt) \rt]F_{\rm c}^+\lt(\OmK \, \tau , \tilde{\al}\rt)
\nn
\nn
\nn
&  &{} + \lt. {1 \over Z \, r} \lt({L \over r^2} + \Dl u^3 \rt) \lt[\Dl u^1 \, R_{0313} + \lt({E \over A^2} + \Dl u^0\rt) R_{0303} \rt]
F_{\rm s}^-\lt(\OmK \, \tau , \tilde{\al}\rt) \rt\}\sin \lt(\OmK \, \tau \rt)
\nn
\nn
\nn
&  &{} + {Z^2 \over K^2} \lt\{\lt({E \over A^2} + \Dl u^0 - {2 \, \al \over A^2 \, Z^2} \, {\Dl M \over r} \, \Dl u^1 \rt)^2 R_{0101} \rt.
\nn
\nn
\nn
&  &{} + \lt. \lt({L \over r^2} + \Dl u^3 \rt)^2 \lt[R_{1313}
+ {4 \over A^2 \, Z^2} \lt(\al \, R_{0313} + {1 \over A^2 \, Z^2} \, {\Dl M \over r} \, R_{0303} \rt) {\Dl M \over r} \rt]
\rt\}\lt[F_{\rm c}^+\lt(\OmK \, \tau , \tilde{\al}\rt)\rt]^2
\nn
\nn
\nn
&  &{} - {2 \, Z \over K^2 \, r} \lt({L \over r^2} + \Dl u^3 \rt) \lt[\Dl u^1 \, R_{1313}
+ {2 \, \al \over A^2 \, Z^2} \lt({E \over A^2} + \Dl u^0\rt) {\Dl M \over r} \, R_{0303} \rt.
\nn
\nn
\nn
&  &{} \lt. + \lt({E \over A^2} + \Dl u^0 + {2 \, \al \over A^2 \, Z^2} \, {\Dl M \over r} \, \Dl u^1 \rt) R_{0313} \rt]
F_{\rm s}^-\lt(\OmK \, \tau , \tilde{\al}\rt) \, F_{\rm c}^+\lt(\OmK \, \tau , \tilde{\al}\rt)
\nn
\nn
\nn
&  &{} + {1 \over K^2 \, r^2} \lt\{\lt(\Dl u^1\rt)^2 R_{1313} + \lt({E \over A^2} + \Dl u^0 \rt) \lt[\lt({E \over A^2} + \Dl u^0 \rt) R_{0303}
+ 2 \, \Dl u^1 \, R_{0313} \rt] \rt\} \lt[F_{\rm s}^-\lt(\OmK \, \tau , \tilde{\al}\rt)\rt]^2 \, , \qquad
\ee
\be
R_{\ze \on \ze \tr} & = &
-{\tilde{\bt}^2 \, \tilde{\gm}^2 \over K^2 \, Z^2} \lt[\lt(\Dl u^1\rt)^2 R_{0101}
+ \lt({L \over r^2} + \Dl u^3 \rt)^2 R_{0303} \rt] \sin^2 \lt(\OmK \, \tau \rt) \cos \lt(\OmK \, \tau \rt)
\nn
\nn
\nn
&  &{} - {\tilde{\bt} \, \tilde{\gm} \over K^2} \lt\{ \lt[
\lt({E \over A^2} + \Dl u^0 - {2 \, \al \over A^2 \, Z^2} \, {\Dl M \over r} \, \Dl u^1 \rt) \Dl u^1 \, R_{0101} \rt. \rt.
\nn
\nn
\nn
&  &{} - \lt.
\lt({L \over r^2} + \Dl u^3 \rt)^2 \lt(R_{0313} + {2 \, \al \over A^2 \, Z^2} \, {\Dl M \over r} \, R_{0303} \rt) \rt]
F_{\rm c}^+\lt(2 \, \OmK \, \tau , \tilde{\al}\rt)
\nn
\nn
\nn
&  &{} + \lt. {1 \over Z \, r} \lt({L \over r^2} + \Dl u^3 \rt) \lt[\Dl u^1 \, R_{0313} + \lt({E \over A^2} + \Dl u^0\rt) R_{0303} \rt]
F_{\rm s}^-\lt(2 \, \OmK \, \tau , \tilde{\al}\rt) \rt\}
\nn
\nn
\nn
&  &{} + {Z^2 \over K^2} \lt\{\lt[\lt({E \over A^2} + \Dl u^0 \rt)^2
- {4 \over A^2 \, Z^2} \lt[\al \lt({E \over A^2} + \Dl u^0 \rt) - {1 \over A^2 \, Z^2} \, {\Dl M \over r} \, \Dl u^1 \rt] {\Dl M \over r} \, \Dl u^1  \rt]R_{0101} \rt.
\nn
\nn
\nn
&  &{} + \lt. \lt({L \over r^2} + \Dl u^3 \rt)^2 \lt[R_{1313}
+ {4 \over A^2 \, Z^2} \lt(\al \, R_{0313} + {1 \over A^2 \, Z^2} \, {\Dl M \over r} \, R_{0303} \rt) {\Dl M \over r} \rt]
\rt\} F_{\rm c}^+\lt(\OmK \, \tau , \tilde{\al}\rt) \, F_{\rm s}^+\lt(\OmK \, \tau , \tilde{\al}\rt)
\nn
\nn
\nn
&  &{} + {Z \over K^2 \, r} \lt({L \over r^2} + \Dl u^3 \rt) \lt[\Dl u^1 \, R_{1313}
+ {2 \, \al \over A^2 \, Z^2} \lt({E \over A^2} + \Dl u^0\rt) {\Dl M \over r} \, R_{0303} \rt.
\nn
\nn
\nn
&  &{} \lt. + \lt({E \over A^2} + \Dl u^0 + {2 \, \al \over A^2 \, Z^2} \, {\Dl M \over r} \, \Dl u^1 \rt) R_{0313} \rt]
\lt[F_{\rm c}^-\lt(\OmK \, \tau , \tilde{\al}\rt) F_{\rm c}^+\lt(\OmK \, \tau , \tilde{\al}\rt) - F_{\rm s}^-\lt(\OmK \, \tau , \tilde{\al}\rt) F_{\rm s}^+\lt(\OmK \, \tau , \tilde{\al}\rt)\rt]
\nn
\nn
\nn
&  &{} - {1 \over K^2 \, r^2} \lt\{\lt(\Dl u^1\rt)^2 R_{1313} + \lt({E \over A^2} + \Dl u^0 \rt) \lt[\lt({E \over A^2} + \Dl u^0 \rt) R_{0303}
+ 2 \, \Dl u^1 \, R_{0313} \rt] \rt\}
\nn
\nn
\nn
&  &{} \times F_{\rm s}^-\lt(\OmK \, \tau , \tilde{\al}\rt) \, F_{\rm c}^-\lt(\OmK \, \tau , \tilde{\al}\rt) \, ,
\ee
\be
R_{\ze \on \on \tr} & = &
-{\tilde{\bt}^2 \, \tilde{\gm}^2 \over K \, Z} \lt\{ \Dl u^1 \, R_{0101} \, F_{\rm c}^+\lt(0, \tilde{\al}\rt)
+ {1 \over Z \, r} \lt({L \over r^2} + \Dl u^3 \rt) R_{0303} \, F_{\rm s}^-\lt(0, \tilde{\al}\rt) \rt\} \sin \lt(\OmK \, \tau \rt)
\nn
\nn
\nn
&  &{} - {\tilde{\bt} \, \tilde{\gm} \, Z \over K} \lt({E \over A^2} + \Dl u^0 - {2 \, \al \over A^2 \, Z^2} \rt) R_{0101}
\lt[ \lt[F_{\rm c}^+\lt(\OmK \, \tau , \tilde{\al}\rt)\rt]^2 \cos \lt(\OmK \, \tau \rt) + F_{\rm s}^+\lt(\OmK \, \tau , \tilde{\al}\rt) \, \sin \lt(\OmK \, \tau \rt) \rt]
\nn
\nn
\nn
&  &{} - {\tilde{\bt} \, \tilde{\gm} \over K \, Z \, r^2} \lt[\lt({E \over A^2} + \Dl u^0\rt) R_{0303} + \Dl u^1 \, R_{0313} \rt]
F_{\rm s}^-\lt(0, \tilde{\al}\rt) \, F_{\rm s}^-\lt(\OmK \, \tau , \tilde{\al}\rt)
\nn
\nn
\nn
&  &{} + {\tilde{\bt} \, \tilde{\gm} \over K \, r} \lt({L \over r^2} + \Dl u^3 \rt) \lt(R_{0313} + {2 \, \al \over A^2 \, Z^2} \, {\Dl M \over r} \, R_{0303} \rt)
\lt[F_{\rm c}^-\lt(2 \, \OmK \, \tau , \tilde{\al}\rt) \, F_{\rm s}^-\lt(\OmK \, \tau , \tilde{\al}\rt) \rt.
\nn
\nn
\nn
&  &{} - \lt. 2 \, F_{\rm c}^-\lt(\OmK \, \tau , \tilde{\al}\rt) \, F_{\rm c}^+\lt(\OmK \, \tau , \tilde{\al}\rt) \, \sin \lt(\OmK \, \tau \rt) \rt]
\nn
\nn
\nn
&  &{} + {Z \over K \, r} \lt\{Z \lt({L \over r^2} + \Dl u^3 \rt) \lt[R_{1313}
+ {4 \over A^2 \, Z^2} \lt(\al \, R_{0313} + {1 \over A^2 \, Z^2} \, {\Dl M \over r} \, R_{0303} \rt) {\Dl M \over r} \rt] F_{\rm c}^+\lt(\OmK \, \tau , \tilde{\al}\rt) \rt.
\nn
\nn
\nn
&  &{} - {1 \over r} \lt[\Dl u^1 \, R_{1313} + {2 \, \al \over A^2 \, Z^2} \lt({E \over A^2} + \Dl u^0 \rt) {\Dl M \over r} \, R_{0303}
+ \lt({E \over A^2} + \Dl u^0 + {2 \, \al \over A^2 \, Z^2} \, {\Dl M \over r} \, \Dl u^1 \rt) R_{0313} \rt]
\nn
\nn
\nn
&  &{} \times \lt. F_{\rm s}^-\lt(\OmK \, \tau , \tilde{\al}\rt) \rt\}
\lt[F_{\rm c}^-\lt(\OmK \, \tau , \tilde{\al}\rt) F_{\rm c}^+\lt(\OmK \, \tau , \tilde{\al}\rt) + F_{\rm s}^-\lt(\OmK \, \tau , \tilde{\al}\rt) F_{\rm s}^+\lt(\OmK \, \tau , \tilde{\al}\rt)\rt] \, ,
\nl
\nn
\nn
R_{\ze \tw \ze \tw} & = &
{1 \over K^2 \, r^2} \lt\{\lt(\Dl u^1\rt)^2 R_{1212} + \lt({E \over A^2} + \Dl u^0 \rt) \lt[\lt({E \over A^2} + \Dl u^0 \rt) R_{0202} + 2 \, \Dl u^1 \, R_{0212} \rt]
+ \lt({L \over r^2} + \Dl u^3 \rt) R_{2323}  \rt\} \, , \qquad
\nl
\nn
\nn
R_{\ze \tw \on \tw} & = &
- {\tilde{\bt} \, \tilde{\gm} \over K \, Z \, r^2} \lt[\lt({E \over A^2} + \Dl u^0 \rt) R_{0202} + \Dl u^1 \, R_{0212} \rt] \sin \lt(\OmK \, \tau \rt)
+ {1 \over K \, r^3} \lt({L \over r^2} + \Dl u^3 \rt) R_{2323} \, F_{\rm s}^-\lt(\OmK \, \tau , \tilde{\al}\rt)
\nn
\nn
\nn
&  &{} + {Z \over K \, r^2} \lt[\Dl u^1 \, R_{1212} + {2 \, \al \over A^2 \, Z^2} \lt({E \over A^2} + \Dl u^0 \rt) {\Dl M \over r} \, R_{0202}
+ \lt({E \over A^2} + \Dl u^0 + {2 \, \al \over A^2 \, Z^2} \, {\Dl M \over r} \, \Dl u^1 \rt) R_{0212} \rt]
\nn
\nn
\nn
&  &{} \times F_{\rm c}^+\lt(\OmK \, \tau , \tilde{\al}\rt) \, ,
\nl
\nn
\nn
R_{\ze \tw \tw \tr} & = &
- {\tilde{\bt} \, \tilde{\gm} \over K \, Z \, r^2} \lt[\lt({E \over A^2} + \Dl u^0 \rt) R_{0202} + \Dl u^1 \, R_{0212} \rt] \cos \lt(\OmK \, \tau \rt)
+ {1 \over K \, r^3} \lt({L \over r^2} + \Dl u^3 \rt) R_{2323} \, F_{\rm c}^-\lt(\OmK \, \tau , \tilde{\al}\rt)
\nn
\nn
\nn
&  &{} - {Z \over K \, r^2} \lt[\Dl u^1 \, R_{1212} + {2 \, \al \over A^2 \, Z^2} \lt({E \over A^2} + \Dl u^0 \rt) {\Dl M \over r} \, R_{0202}
+ \lt({E \over A^2} + \Dl u^0 + {2 \, \al \over A^2 \, Z^2} \, {\Dl M \over r} \, \Dl u^1 \rt) R_{0212} \rt]
\nn
\nn
\nn
&  &{} \times F_{\rm s}^+\lt(\OmK \, \tau , \tilde{\al}\rt) \, ,
\ee
\be
R_{\ze \tr \ze \tr} & = &
{\tilde{\bt}^2 \, \tilde{\gm}^2 \over K^2 \, Z^2} \lt[\lt(\Dl u^1\rt)^2 R_{0101}
+ \lt({L \over r^2} + \Dl u^3 \rt)^2 R_{0303} \rt] \cos^2 \lt(\OmK \, \tau \rt)
\nn
\nn
\nn
&  &{} - {2 \, \tilde{\bt} \, \tilde{\gm} \over K^2} \lt\{ \lt[
\lt({E \over A^2} + \Dl u^0 - {2 \, \al \over A^2 \, Z^2} \, {\Dl M \over r} \, \Dl u^1 \rt) \Dl u^1 \, R_{0101} \rt. \rt.
\nn
\nn
\nn
&  &{} - \lt.
\lt({L \over r^2} + \Dl u^3 \rt)^2 \lt(R_{0313} + {2 \, \al \over A^2 \, Z^2} \, {\Dl M \over r} \, R_{0303} \rt) \rt]
F_{\rm s}^+\lt(\OmK \, \tau , \tilde{\al}\rt)
\nn
\nn
\nn
&  &{} - \lt. {1 \over Z \, r} \lt({L \over r^2} + \Dl u^3 \rt) \lt[\Dl u^1 \, R_{0313} + \lt({E \over A^2} + \Dl u^0\rt) R_{0303} \rt]
F_{\rm c}^-\lt(2 \, \OmK \, \tau , \tilde{\al}\rt) \rt\} \cos \lt(\OmK \, \tau \rt)
\nn
\nn
\nn
&  &{} + {Z^2 \over K^2} \lt\{\lt({E \over A^2} + \Dl u^0 - {2 \, \al \over A^2 \, Z^2} \, {\Dl M \over r} \, \Dl u^1 \rt)^2 R_{0101} \rt.
\nn
\nn
\nn
&  &{} + \lt. \lt({L \over r^2} + \Dl u^3 \rt)^2 \lt[R_{1313}
+ {4 \over A^2 \, Z^2} \lt(\al \, R_{0313} + {1 \over A^2 \, Z^2} \, {\Dl M \over r} \, R_{0303} \rt) {\Dl M \over r} \rt]
\rt\}\lt[F_{\rm s}^+\lt(\OmK \, \tau , \tilde{\al}\rt)\rt]^2
\nn
\nn
\nn
&  &{} + {2 \, Z \over K^2 \, r} \lt({L \over r^2} + \Dl u^3 \rt) \lt[\Dl u^1 \, R_{1313}
+ {2 \, \al \over A^2 \, Z^2} \lt({E \over A^2} + \Dl u^0\rt) {\Dl M \over r} \, R_{0303} \rt.
\nn
\nn
\nn
&  &{} \lt. + \lt({E \over A^2} + \Dl u^0 + {2 \, \al \over A^2 \, Z^2} \, {\Dl M \over r} \, \Dl u^1 \rt) R_{0313} \rt]
F_{\rm c}^-\lt(\OmK \, \tau , \tilde{\al}\rt) \, F_{\rm s}^+\lt(\OmK \, \tau , \tilde{\al}\rt)
\nn
\nn
\nn
&  &{} + {1 \over K^2 \, r^2} \lt\{\lt(\Dl u^1\rt)^2 R_{1313} + \lt({E \over A^2} + \Dl u^0 \rt) \lt[\lt({E \over A^2} + \Dl u^0 \rt) R_{0303}
+ 2 \, \Dl u^1 \, R_{0313} \rt] \rt\} \lt[F_{\rm c}^-\lt(\OmK \, \tau , \tilde{\al}\rt)\rt]^2 \, , \qquad
\ee
\be
R_{\ze \tr \on \tr} & = &
{\tilde{\bt}^2 \, \tilde{\gm}^2 \over K \, Z} \lt\{ \Dl u^1 \, R_{0101} \, F_{\rm c}^+\lt(0, \tilde{\al}\rt)
+ {1 \over Z \, r} \lt({L \over r^2} + \Dl u^3 \rt) R_{0303} \, F_{\rm s}^-\lt(0, \tilde{\al}\rt) \rt\} \cos \lt(\OmK \, \tau \rt)
\nn
\nn
\nn
&  &{} - {\tilde{\bt} \, \tilde{\gm} \, Z \over K} \lt({E \over A^2} + \Dl u^0 - {2 \, \al \over A^2 Z^2} \rt) R_{0101}
\lt[ \lt[F_{\rm s}^+\lt(\OmK \, \tau , \tilde{\al}\rt)\rt]^2 \sin \lt(\OmK \, \tau \rt) + F_{\rm s}^+\lt(\OmK \, \tau , \tilde{\al}\rt) \, \cos \lt(\OmK \, \tau \rt) \rt]
\nn
\nn
\nn
&  &{} + {\tilde{\bt} \, \tilde{\gm} \over K \, Z \, r^2} \lt[\lt({E \over A^2} + \Dl u^0\rt) R_{0303} + \Dl u^1 \, R_{0313} \rt]
F_{\rm s}^-\lt(0, \tilde{\al}\rt) \, F_{\rm c}^-\lt(\OmK \, \tau , \tilde{\al}\rt)
\nn
\nn
\nn
&  &{} + {\tilde{\bt} \, \tilde{\gm} \over K \, r} \lt({L \over r^2} + \Dl u^3 \rt) \lt(R_{0313} + {2 \, \al \over A^2 \, Z^2} \, {\Dl M \over r} \, R_{0303} \rt)
\lt[F_{\rm c}^+\lt(2 \, \OmK \, \tau , \tilde{\al}\rt) \, F_{\rm c}^-\lt(\OmK \, \tau , \tilde{\al}\rt) \rt.
\nn
\nn
\nn
&  &{} + \lt. 2 \, F_{\rm s}^-\lt(\OmK \, \tau , \tilde{\al}\rt) \, F_{\rm s}^+\lt(\OmK \, \tau , \tilde{\al}\rt) \, \cos \lt(\OmK \, \tau \rt) \rt]
\nn
\nn
\nn
&  &{} + {Z \over K \, r} \lt\{Z \lt({L \over r^2} + \Dl u^3 \rt) \lt[R_{1313}
+ {4 \over A^2 \, Z^2} \lt(\al \, R_{0313} + {1 \over A^2 \, Z^2} \, {\Dl M \over r} \, R_{0303} \rt) {\Dl M \over r} \rt] F_{\rm s}^+\lt(\OmK \, \tau , \tilde{\al}\rt) \rt.
\nn
\nn
\nn
&  &{} + {1 \over r} \lt[\Dl u^1 \, R_{1313} + {2 \, \al \over A^2 \, Z^2} \lt({E \over A^2} + \Dl u^0 \rt) {\Dl M \over r} \, R_{0303}
+ \lt({E \over A^2} + \Dl u^0 + {2 \, \al \over A^2 \, Z^2} \, {\Dl M \over r} \, \Dl u^1 \rt) R_{0313} \rt]
\nn
\nn
\nn
&  &{} \times \lt. F_{\rm c}^-\lt(\OmK \, \tau , \tilde{\al}\rt) \rt\}
\lt[F_{\rm c}^-\lt(\OmK \, \tau , \tilde{\al}\rt) F_{\rm c}^+\lt(\OmK \, \tau , \tilde{\al}\rt) + F_{\rm s}^-\lt(\OmK \, \tau , \tilde{\al}\rt) F_{\rm s}^+\lt(\OmK \, \tau , \tilde{\al}\rt)\rt] \, ,
\nl
\nn
\nn
R_{\on \tw \on \tw} & = &
{\tilde{\bt}^2 \, \tilde{\gm}^2 \over Z^2 \, r^2} \, R_{0202} \, \sin^2 \lt(\OmK \, \tau \rt)
- {2 \, \tilde{\bt} \, \tilde{\gm} \over r^2} \lt(R_{0212} + {2 \, \al \over A^2 \, Z^2} \, {\Dl M \over r} \, R_{0202} \rt)
F_{\rm c}^+\lt(\OmK \, \tau , \tilde{\al}\rt) \, \sin \lt(\OmK \, \tau \rt)
\nn
\nn
\nn
&  &{} + {Z^2 \over r^2} \lt[R_{1212}
+ {4 \over A^2 \, Z^2} \lt(\al \, R_{0212} + {1 \over A^2 \, Z^2} \, {\Dl M \over r} \, R_{0202} \rt) {\Dl M \over r} \rt] \lt[F_{\rm c}^+\lt(\OmK \, \tau , \tilde{\al}\rt)\rt]^2
\nn
\nn
\nn
&  &{} + {1 \over r^4} \, R_{2323} \lt[F_{\rm s}^+\lt(\OmK \, \tau , \tilde{\al}\rt)\rt]^2 \, ,
\nl
\nn
\nn
R_{\on \tw \tw \tr} & = &
{\tilde{\bt}^2 \, \tilde{\gm}^2 \over Z^2 \, r^2} \, R_{0202} \, \sin \lt(\OmK \, \tau \rt) \, \cos \lt(\OmK \, \tau \rt)
- {\tilde{\bt} \, \tilde{\gm} \over r^2} \lt(R_{0212} + {2 \, \al \over A^2 \, Z^2} \, {\Dl M \over r} \, R_{0202} \rt)
F_{\rm c}^+\lt(2 \, \OmK \, \tau , \tilde{\al}\rt)
\nn
\nn
\nn
&  &{} - {Z^2 \over r^2} \lt[R_{1212}
+ {4 \over A^2 \, Z^2} \lt(\al \, R_{0212} + {1 \over A^2 \, Z^2} \, {\Dl M \over r} \, R_{0202} \rt) {\Dl M \over r} \rt]
F_{\rm s}^+\lt(\OmK \, \tau , \tilde{\al}\rt) \, F_{\rm c}^+\lt(\OmK \, \tau , \tilde{\al}\rt)
\nn
\nn
\nn
&  &{} + {1 \over r^4} \, R_{2323} \, F_{\rm s}^-\lt(\OmK \, \tau , \tilde{\al}\rt) \, F_{\rm c}^-\lt(\OmK \, \tau , \tilde{\al}\rt) \, ,
\ee
\be
R_{\on \tr \on \tr} & = &
\tilde{\bt}^2 \, \tilde{\gm}^2 \lt\{
R_{0101} \lt[\lt(\lt[F_{\rm c}^+\lt(\OmK \, \tau , \tilde{\al}\rt)\rt]^2 \cos^2 \lt(\OmK \, \tau \rt)
+ \lt[F_{\rm s}^+\lt(\OmK \, \tau , \tilde{\al}\rt)\rt]^2 \sin^2 \lt(\OmK \, \tau \rt) \rt) \rt. \rt.
\nn
\nn
\nn
&  &{} + \lt. 2 \, F_{\rm s}^+\lt(\OmK \, \tau , \tilde{\al}\rt) \, F_{\rm c}^+\lt(\OmK \, \tau , \tilde{\al}\rt) \, \sin \lt(\OmK \, \tau \rt) \rt]
\nn
\nn
\nn
&  &{} + {1 \over Z^2 \, r^2} \, R_{0303} \lt[\lt(\lt[F_{\rm c}^-\lt(\OmK \, \tau , \tilde{\al}\rt)\rt]^2 \sin^2 \lt(\OmK \, \tau \rt)
+ \lt[F_{\rm s}^-\lt(\OmK \, \tau , \tilde{\al}\rt)\rt]^2 \cos^2 \lt(\OmK \, \tau \rt) \rt) \rt.
\nn
\nn
\nn
&  &{} - \lt. \lt. 2 \, F_{\rm s}^-\lt(\OmK \, \tau , \tilde{\al}\rt) \, F_{\rm c}^-\lt(\OmK \, \tau , \tilde{\al}\rt) \, \sin \lt(\OmK \, \tau \rt) \rt] \rt\}
\nn
\nn
\nn
&  &{} + {2 \, \tilde{\bt} \, \tilde{\gm} \over r^2} \lt(R_{0313} + {2 \, \al \over A^2 \, Z^2} \, {\Dl M \over r} \, R_{0303} \rt) F_{\rm s}^-\lt(0, \tilde{\al}\rt)
\lt[F_{\rm c}^-\lt(\OmK \, \tau , \tilde{\al}\rt) F_{\rm c}^+\lt(\OmK \, \tau , \tilde{\al}\rt) + F_{\rm s}^-\lt(\OmK \, \tau , \tilde{\al}\rt) F_{\rm s}^+\lt(\OmK \, \tau , \tilde{\al}\rt)\rt]
\nn
\nn
\nn
&  &{} + {Z^2 \over r^2} \lt[R_{1313}
+ {4 \over A^2 \, Z^2} \lt(\al \, R_{0313} + {1 \over A^2 \, Z^2} \, {\Dl M \over r} \, R_{0303} \rt) {\Dl M \over r} \rt]
\nn
\nn
\nn
&  &{} \times
\lt[F_{\rm c}^-\lt(\OmK \, \tau , \tilde{\al}\rt) F_{\rm c}^+\lt(\OmK \, \tau , \tilde{\al}\rt) + F_{\rm s}^-\lt(\OmK \, \tau , \tilde{\al}\rt) F_{\rm s}^+\lt(\OmK \, \tau , \tilde{\al}\rt)\rt]^2
\, ,
\nl
\nn
\nn
R_{\tw \tr \tw \tr} & = &
{\tilde{\bt}^2 \, \tilde{\gm}^2 \over Z^2 \, r^2} \, R_{0202} \, \cos^2 \lt(\OmK \, \tau \rt)
+ {2 \, \tilde{\bt} \, \tilde{\gm} \over r^2} \lt(R_{0212} + {2 \, \al \over A^2 \, Z^2} \, {\Dl M \over r} \, R_{0202} \rt)
F_{\rm s}^+\lt(\OmK \, \tau , \tilde{\al}\rt) \, \cos \lt(\OmK \, \tau \rt)
\nn
\nn
\nn
&  &{} + {Z^2 \over r^2} \lt[R_{1212}
+ {4 \over A^2 \, Z^2} \lt(\al \, R_{0212} + {1 \over A^2 \, Z^2} \, {\Dl M \over r} \, R_{0202} \rt) {\Dl M \over r} \rt]
\lt[F_{\rm s}^+\lt(\OmK \, \tau , \tilde{\al}\rt)\rt]^2
\nn
\nn
\nn
&  &{} + {1 \over r^4} \, R_{2323} \, \lt[F_{\rm c}^-\lt(\OmK \, \tau , \tilde{\al}\rt)\rt]^2 \, .
\ee
\end{widetext}

\newpage

\section{Selected Plots for the Generalized CMP Approximation of the MPD Equations}
\label{appendix:plots}
\renewcommand{\theequation}{E.\arabic{equation}}
\setcounter{subsection}{0}
\setcounter{equation}{0}

\end{appendix}

\end{document}